\documentclass[12pt,a4paper,aps,prx,superscriptaddress,showpacs,preprint]{revtex4-1}
\usepackage{graphicx}
\usepackage{color}
\usepackage{latexsym}
\usepackage{amsmath}
\usepackage{amssymb}

\newcommand{\ie}{\textit{i.e.}}
\newcommand{\eg}{\textit{e.g.}}
\newcommand{\EQ}{Eq.~}

\newcommand{\FIG}{Fig.~}
\newcommand{\FIGS}{Figs.~}

\newcommand{\SEC}{Sec.~}

\renewcommand{\Pr}{{P}}

\begin{document}

\title{Predictability of conversation partners}
	
\author{Taro Takaguchi}
\affiliation{Department of Mathematical Informatics, The University of Tokyo, 7-3-1 Hongo, Bunkyo, Tokyo 113-8656, Japan}
\author{Mitsuhiro Nakamura}
\affiliation{Department of Mathematical Informatics, The University of Tokyo, 7-3-1 Hongo, Bunkyo, Tokyo 113-8656, Japan}
\author{Nobuo Sato}
\affiliation{Central Research Laboratory, Hitachi, Ltd., 1-280 Higashi-Koigakubo, Kokubunji-shi, Tokyo, Japan}
\author{Kazuo Yano}
\affiliation{Central Research Laboratory, Hitachi, Ltd., 1-280 Higashi-Koigakubo, Kokubunji-shi, Tokyo, Japan}
\author{Naoki Masuda}
\affiliation{Department of Mathematical Informatics, The University of Tokyo, 7-3-1 Hongo, Bunkyo, Tokyo 113-8656, Japan}
\affiliation{PRESTO, Japan Science and Technology Agency, 4-1-8 Honcho, Kawaguchi, Saitama 332-0012, Japan}

\begin{abstract}
Recent developments in sensing technologies have enabled us to examine the nature of human social behavior in greater detail.
By applying an information theoretic method to the spatiotemporal data of cell-phone locations,
[C. Song~\textit{et al.}~Science~{\bf 327},~1018~(2010)] found that human mobility patterns are remarkably predictable. 
Inspired by their work, we address a similar predictability question in a different kind of human social activity: conversation events.
The predictability in the sequence of one's conversation partners is defined as
the degree to which one's next conversation partner can be predicted given the current partner.
We quantify this predictability by using the mutual information.
We examine the predictability of conversation events for each individual using the longitudinal data of face-to-face interactions
collected from two company offices in Japan.
Each subject wears a name tag equipped with an infrared sensor node,
and conversation events are marked when signals are exchanged between sensor nodes in close proximity.
We find that the conversation events are predictable to a certain extent; knowing the current partner decreases the uncertainty about the next partner by $28.4\%$ on average.
Much of the predictability is explained by long-tailed distributions of interevent intervals.
However, a predictability also exists in the data, apart from the contribution of their long-tailed nature.
In addition, an individual's predictability is correlated with the position of the individual in the static social network derived from the data.
Individuals confined in a community -- in the sense of an abundance of surrounding triangles -- tend to have low predictability,
and those bridging different communities tend to have high predictability.
\end{abstract}


\maketitle

\section{Introduction}\label{sec:introduction}
Recently, interest in the statistical and dynamical features of human social behavior has been growing, enabled by the development of new devices that allow tracking of social data in real time, with increasing precision and duration~\cite{AVazquez2006mba,Gonzalez2008,Rybski2009slo,Iribarren2009ioh,Cattuto2010dop,Isella2010,Tang2010,Wu2010,Isella2011}.
A remarkable recent finding from the analysis of spatiotemporal data on cell-phone locations is
that human mobility patterns are highly predictable~\cite{Gonzalez2008,Song2010,Song2010a},
a finding that is in contrast to the traditional view.
For instance, in epidemic models that take the mobility of subjects into account,
subjects are usually assumed to perform a conventional random walk from one location to another~\cite{Anderson1991,Colizza2006}.
However, actual traveling patterns of humans often deviate from such random walk models,
and  the displacement distribution follows a power law~\cite{Brockmann2006,Gonzalez2008}.
Furthermore, the statistics of the next location of the individual is
affected not only by the current location, but also by the history of the traveling pattern,
resulting in approximately 90\% predictability of the mobility patterns~\cite{Song2010}.

In this study, we address a similar predictability question
for a different component of human social behavior: conversation events.
Conversation events mediate the spreading and routing of diverse contents such as new ideas, opinions, and infectious diseases
in social networks~\cite{Castellano2009spo,Barrat2008dpo_book}. 
In models describing these phenomena, it is a norm that each individual possesses a dynamically changing state
(\eg, opinion A or opinion B in opinion dynamics, and susceptible or infected state in epidemic dynamics).
The law of transition from one state to another is usually assumed to be Markovian,
\ie, independent of the history of the process.
The Markovian property, which is a type of unpredictability, is an assumption for simulating such dynamics
based on a static social network~\cite{Castellano2009spo,Barrat2008dpo_book}. 

However, the plausibility of this assumption is unclear.
Imagine the office that you share with other colleagues in your company.
When you have a question about a project, you may talk to your boss.
After this conversation event, you may tend to talk to a particular individual to communicate the instruction of the boss.
During lunchtime, you may chat with your close colleagues
in a particular order that you do not perceive.
How predictable is your choice of your next conversation partner given the current partner?

We examine the predictability of conversation events
using two sets of longitudinal data collected from company offices in Japan.
We use the information about the timing and duration of conversations between each pair of individuals, 
but do not use \textit{a priori} knowledge about status or other social attributes of individuals.
Our data are unique in that they are collected from a relatively high number of individuals (\ie, approximately 200 individuals) over a long recording period (\ie, approximately three months).
We examine the sequence of conversation events for each individual. 
We find that a conversation event has notable deterministic components.
In other words, the uncertainty about the next partner that you talk with decreases by $28.4\%$ on average, given the identity of the partner you are currently talking with (see \SEC\ref{sec:Predictability of partner sequences}).

It should be noted that our approach is related to, but different from, the studies of power-law interval distributions in conversation events. 
The interval between successive conversation events for an individual or a given pair of individuals often follows a power law~\cite{Eckmann2004eod,Barabasi2005,AVazquez2006mba,Candia2008,Malmgren2008,Iribarren2009ioh,Cattuto2010dop,Isella2010,Wu2010}.
Modeling studies have revealed implications of these empirical results in contagions~\cite{AVazquez2007ion,Iribarren2009ioh,Karrer2010mpa,Isella2010,Karsai2011,Isella2011,Min2011,Miritello2011} 
and opinion formation~\cite{Takaguchi2010,Fernandez-Gracia2011}.
In contrast to conventional models in which the Poisson interval distribution is assumed,
these results indicate that the next conversation time given the previous one is relatively predictable in
that a conversation event in the recent past is a precursor to a burst of events in the near future.
We argue that the bursty nature of the point process largely contributes to the predictability of conversation events.

We also show that the degree of predictability depends on individuals.
Individuals located inside a network community, \ie, a dense subnetwork loosely connected to other parts of the entire network~\cite{Fortunato2010}, quantified in this study via strong links and the clustering coefficient, behave relatively randomly.
On the other hand, individuals that connect different communities by weak links tend to have a high predictability.

\section{Data and Methods}\label{sec:methods}

We analyze two sets of face-to-face interaction logs obtained from different company offices
using the Business Microscope system developed by Hitachi, Ltd., Japan~\cite{Wakisaka2009,Yano2009}.
The data were collected by World Signal Center, Hitachi, Ltd., Japan.
Data set $D_1$ consists of recordings from $N=163$ individuals for 73 days.
Data set $D_2$ consists of recordings from $N=211$ individuals for 120 days.
Each subject wears a name tag strapped around the neck and placed at the chest,
and each name tag contains an infrared module.
The infrared modules can communicate with each other if they are less than 3 meters apart.
An infrared module only senses the modules situated within a $120^\circ$ circular sector in front of the name tag,
and the system detects conversation events only when two individuals are facing each other.
Communication between modules includes exchanging the owners' IDs every 10 sec.
We regard two individuals to be involved in a conversation event if their infrared modules communicate with each other
at least once in 1 min.
In other words,  the time resolution of the system is equal to one minute.
The list of conversation partners and time stamps is stored in the name tag of each individual
and sent to the central database on a daily basis. 
The data transfer occurs when the individual leaves work and puts the name tag on a gateway device connected to the individual's computer~\cite{Wakisaka2009,Yano2009}.
Each data set contains a list of conversation events, as shown in \FIG\ref{fig:sequence-generation}.
A conversation event is specified by the IDs of the two individuals talking with each other,
the date and time at which the dialogue starts, and the duration of the dialogue.
We are not concerned with the content of the dialogue.
Data sets $D_1$ and $D_2$ contain $51,879$ and $125,345$ events, respectively.

We investigate the predictability of each individual's conversation patterns. 
Our preliminary data analysis revealed that the timing of conversation events lacks sufficient temporal correlation
and is unpredictable.
Therefore, we neglect the timing of conversation events in the data unless otherwise stated
and focus on the partner sequence defined as follows.
To generate the partner sequence of individual~1,
we first sift out all the conversation events that involve individual~1 from the entire data set
(\FIG\ref{fig:sequence-generation}(b)).
Next, we ignore the time stamp and duration of the conversation events.
The remaining data define the partner sequence,
\ie, the chronologically ordered sequence of the IDs of
the conversation partners for individual~1~(\FIG\ref{fig:sequence-generation}(c)).
When multiple conversation events involving individual~1 are initiated in the same minute,
we determine their order at random.

To evaluate the predictability of the partner sequence,
we calculate three entropy measures,
inspired by those used for the analysis of human mobility patterns~\cite{Song2010}.
First, we define the random entropy for individual~$i$ as
\begin{equation}
H^0_i \equiv \log_2 k_i,
\label{eq:H0}
\end{equation}
where $k_i$ represents the number of $i$'s partners for the entire recording.
If $i$ chooses the partner with equal probability $1/k_i$ from all the $i$'s acquaintances in each conversation event,
$H_i^0$ quantifies the degree of randomness.
Second, we define the uncorrelated entropy as
\begin{equation}
H^1_i \equiv  -\sum_{j \in {\cal N}_i} \Pr_i(j) \log_2 \Pr_i(j),
\end{equation}
where ${\cal N}_i$ is the set of $i$'s partners containing $k_i$ elements.
$\Pr_i(j)$ represents the probability that individual~$i$ talks with individual~$j$ in a conversation event for $i$;
the normalization is given by $\sum_{j \in {\cal N}_i} \Pr_i(j) = 1$.
Compared to $H_i^0$, $H^1_i$ accounts for
the heterogeneity among $\Pr_i(j)$~$(j\neq i)$.
Third, we define the conditional entropy as
\begin{equation}
H^2_i \equiv - \sum_{j \in {\cal N}_i} \Pr_i(j) \sum_{\ell \in {\cal N}_i} \Pr_i(\ell | j) \log_2 \Pr_i(\ell |  j ),  
\label{eq:def_H2}
\end{equation}
where $\Pr_i(\ell | j)$ represents the conditional probability that individual~$i$ talks with individual~$\ell$
immediately after talking with individual~$j$.
$H^2_i$ measures
the second-order correlation in the partner sequence of $i$.
For each individual, $0 \leq H^2_i \leq H^1_i \leq H^0_i$ is satisfied.
We quantify the predictability of the partner sequence
by the mutual information as follows:	
\begin{equation}
I_i \equiv H^1_i - H^2_i = \sum_{j, \ell \in {\cal N}_i} \Pr_i(\ell, j) \log_2 \frac{\Pr_i(\ell, j)}{\Pr_i(\ell) \Pr_i(j)},
\label{eq:MI}
\end{equation}
where $\Pr_i(\ell, j)$ represents the joint probability that individual~$i$ talks with individual~$\ell$
immediately after talking with individual~$j$.
For each individual, $0 \leq I_i \leq H^1_i$ is satisfied.
$I_i$ quantifies the predictability of the partner sequence; it is equal to the amount of 
the information about the next partner that is earned by knowing the current partner.
When the partner sequence lacks a second-order correlation such that $H^1_i = H^2_i$,
$I_i$ takes the minimum value~$0$. 
In this case, knowing the current partner does not help predict the next partner at all.
When the partner sequence is completely deterministic,
\ie, the next partner is completely predicted from the current partner
such that $H^2_i = 0$,
$I_i$ takes the maximum value $H^1_i$.

Although our primary interest in this study is the temporal properties of partner sequences,
we also analyze the conversation networks (CNs) $G_1$ and $G_2$
constructed by aggregating all the conversation events in $D_1$ and $D_2$, respectively, over the entire recording.
In a CN, the node represents an	individual, and the weight of the link, denoted as $w_{ij}$,
represents the number of conversation events between individuals~$i$ and $j$
during the entire recording period.
By the definition of the conversation event,
$w_{ij} = w_{ji}$ $( i,j = 1,2, \cdots, N)$ holds true; the CN is an undirected network.
The degree $k_i$ of individual~$i$ is equal to the number of $j$'s for which $w_{ij}>0$.

\section{Results}
\subsection{Properties of the CN}

We found that both CNs, $G_1$ and $G_2$, are composed of a single connected component.
The CN $G_1$ is visualized in \FIG\ref{fig:graph_00}; we will analyze the relation between the CNs and the predictability in \SEC\ref{sec:variation}.
The clustering coefficient~\cite{Watts1998} of the unweighted versions of $G_1$ and $G_2$
is equal to $0.646$ and $0.611$, respectively.
The Pearson assortativity coefficient~\cite{Newman2002} of the degree of $G_1$ and $G_2$
is equal to $0.169$ and $0.296$, respectively.
Therefore, the CNs have typical properties of social networks~\cite{Newman2003},
\ie, high clustering and positive assortativity.
  
For the two CNs, we measure the distributions of degree, node strength, and link weight.
The node strength $s_i$ is the sum of link weights connecting to node~$i$~\cite{Yook2001,Barrat2004},
\ie, the total number of conversation events for individual~$i$, defined as
\begin{equation}
s_i \equiv \sum_{j \in {\cal N}_i} w_{ij}.
\end{equation}
The mean and standard deviation of $k_i$ of $G_1$ and $G_2$
are equal to $26.07 \pm 11.01$ and $69.56 \pm 29.47$~(mean $\pm$ standard deviation), respectively.
Because two individuals are adjacent if there is at least one conversation event for a few months,
the mean $k_i$ of both networks is relatively large.
$s_i$ of $G_1$ and $G_2$
is equal to $636.6 \pm 516.7$ and $1188.1 \pm 622.1$, respectively.
$w_{ij}$ of $G_1$ and $G_2$
is equal to $24.41 \pm 53.69$ and $17.08 \pm 45.77$, respectively.
The cumulative distribution of the three quantities are shown in \FIG\ref{fig:statdist_CN}.

\subsection{Predictability of partner sequences}\label{sec:Predictability of partner sequences}

We examine the predictability of partner sequences using the entropy measures. 
Because the estimation of entropy is notoriously biased when the data size is small,
we discard individuals with less than 100 conversation events (\ie, $s_i<100$).
There remain 146 and 210 individuals in data sets $D_1$ and $D_2$, respectively after the thresholding. 
Because the results for the two datasets are similar, 
we report the results for $D_1$ in the following.
The results for $D_2$ are given in Appendix~A.

The histograms of the three types of entropies for partner sequences are shown in \FIG\ref{fig:hist_H}(a).
For all the individuals, $H^1_i$ is at least $9.94\%$ smaller than $H^0_i$.
This implies that individuals exhibit a preference when selecting partners from their neighbors in the CN.

The values of $H_i^1$ and $H_i^2$ for each individual are shown in \FIG\ref{fig:hist_H}(b).
The mutual information $I_i = H^1_i - H^2_i$ is positive for all the individuals regardless of the value of $H_i^1$.
In general, the finite size effect decreases $H^1_i$ and $H^2_i$ by different amounts
such that the estimated $I_i$ is generally inherited with a positive bias~\cite{Panzeri2007a}.
For our data, the positive values of $I_i$ are not an artifact caused by the small data size.
Through a bootstrap test (see Appendix~B for details), we confirmed that the empirical values of $I_i$ are significantly (at $1\%$ level) 
larger than the values obtained from the bootstrap samples.
In short, the bootstrap samples are randomized partner sequences
that destroy temporal correlation in the data but preserve the original $H_i^1$ and 
account for the portion of $I_i$ derived from the finite size effect.
It should also be noted that we determined the order of partners at random 
when conversation events with different partners initiate in the same minute.
This randomization does not make $I_i$ larger
because it conserves $H_i^1$ and makes $H_i^2$ larger than the true value.
In fact, the Pearson correlation coefficient between $I_i$
and the fraction of such overlapping conversation events for individual~$i$~($1 \leq i \leq N$, $s_i \geq 100$)
is slightly negative~(\ie, $-0.0811$).
In summary, the information about the current conversation partner
gives the information about the next partner; $H_i^2$ is, on average, $28.4\%$ smaller than $H_i^1$.

The predictability present in the data is mainly explained by the bursty activity patterns, \ie, long-tailed distributions of the interevent intervals, that have been observed for various data~\cite{Eckmann2004eod,Barabasi2005,AVazquez2006mba,Candia2008,Malmgren2008,Iribarren2009ioh,Cattuto2010dop,Isella2010,Wu2010}.
Our data also possess this feature (see Appendix~C for details).
Because the interevent interval for a given pair of individuals obeys a long-tailed distribution,
individual~$i$ tends to talk with individual~$j$ again within a short period from their previous conversation.
In the remainder of this section, we show that the predictability is mainly caused by the bursty activity patterns (\FIG\ref{fig:IEI-SEtest}(a)) and that predictability also exists in the data even if we omit the bursts from the data (\FIG\ref{fig:IEI-SEtest}(b)).

We examine the contribution of the bursty activity pattern to the predictability 
by calculating the mutual information~$I_i^{\rm burst}$ of the randomized partner sequence.
The randomization of the interevent intervals between each pair of individuals is realized
by swapping interevent intervals of the original data within each day in a completely random order 
(see Appendix~D for the precise methods).
Because of the computational cost of the randomization procedure,
we obtain the mean and standard deviation of $I_i^{\rm burst}$ from 100 randomized partner sequences,
instead of estimating the confidential interval of $I_i^{\rm burst}$.
The mean $I_i^{\rm burst}$ accounts for $79.5\%$ of the original $I_i$ on average~(\FIG\ref{fig:IEI-SEtest}(a)).
Because the randomization procedure preserves the interevent interval distribution,
\FIG\ref{fig:IEI-SEtest}(a) suggests that a large $I_i$ is mainly attributed to the bursty activity patterns.
It should be noted that $I_i^{\rm burst}$ is large
partly because the randomizing procedure conserves the timings of the first and last conversation events of each pair on any day. 
Therefore, we may be overestimating the contribution of burstiness to $I_i$.

The predictability is not solely determined by the bursty activity patterns.
To clarify this point, we calculate the mutual information~$I_i^{\rm merge}$ of the modified partner sequence 
generated by merging the consecutive conversation events with the same partner in the original partner sequence into one event.  
This merging procedure allows us to eliminate the contribution of the bursty activity pattern to the predictability.
For example, if individual~$i$ talks with individual~$j$ 3 times without being interrupted by other partners,
we merge the three conversation events into one.
The values of $I_i^{\rm merge}$ are shown in \FIG\ref{fig:IEI-SEtest}(b).
To confirm that the positive values of $I_i^{\rm merge}$ are not an artifact caused by the small data size,
we carry out a bootstrap test for $I_i^{\rm merge}$ similar to that for $I_i$.
By definition, no partner ID appears successively in the merged partner sequence.  
Therefore, we generate the bootstrap sample of the merged partner sequence
by sampling from the merged sequence with replacement
under the condition that the same partner is not consecutively chosen~(see Appendix~D for details).
$I_i^{\rm merge}$ is significantly larger than the values obtained from the bootstrap samples.
Therefore, the original partner sequence possesses some predictability even after removing bursts originating from the bursty nature.

\subsection{Variation among the predictabilities of individuals}\label{sec:variation}

The predictability, quantified by $I_i$, depends on individuals. 
In this section, we investigate the relationship between the predictability of individuals and the properties of nodes in the CN.
The results shown in this section are summarized as follows.
First, $I_i$ is negatively correlated with node strength $s_i$ and with mean node weight defined as $\overline{w}_i \equiv \sum_{j \in {\cal N}_i} w_{ij} / k_{i}$ (\FIG\ref{fig:corr_MI}).
Second, the CN possesses the ``strength of weak ties'' structure (\FIG\ref{fig:overlap_MI.CC}(a)).
Third, the individuals bridging different communities with weak links tend to have large $I_i$,
and those concealed in a single community and surrounded by strong links tend to have small $I_i$ (\FIG\ref{fig:overlap_MI.CC}(b)).

One may speculate that $I_i$ is strongly affected by the node degree $k_i$ because $H^0_i = \log_2 k_i$
and $H_i^1$ and $H_i^2$ comprise many terms if $k_i$ is large.
However,  $k_i$ and $I_i$ are uncorrelated, as shown in \FIG\ref{fig:corr_MI}(a).
We found that $I_i$ is negatively correlated with $s_i$ (\FIG\ref{fig:corr_MI}(b))
and with $\overline{w}_i$ (\FIG\ref{fig:corr_MI}(c)).
Using the bootstrap test, we verified that the negative correlation
shown in \FIG\ref{fig:corr_MI}(b) and \ref{fig:corr_MI}(c)
is not because of the finite sampling size~(see Appendix~B for details).
The correlation shown in \FIG\ref{fig:corr_MI} and the following results do not qualitatively change
if we use the normalized mutual information~\cite{Strehl2003} $I_i / H_i^1$~(see Appendix~E).
We also verified that alternatively defining the link weight by the total duration of the conversation events for each pair, instead of the total number of the conversation events, does not qualitatively change the results described in this section (see Appendix F for details). 

For a fixed $k_i$, both $s_i$ and $\overline{w}_i$ decrease with the number of weak links
 (\ie, the links with small weight) connected to individual~$i$.
This fact leads us to hypothesize that individuals surrounded by weak links select partners
in a relatively deterministic order.
According to Granovetter's theory of the strength of weak ties,
weak links tend to interconnect different communities in a social network
and bring valuable external information to both end nodes,
while strong links tend to be intracommunity links~\cite{Granovetter1973}.
Therefore,
the individuals bridging different communities with weak links may have large values of $I_i$.

We first verify the strength of weak ties hypothesis in the CN.
The network visualized in \FIG\ref{fig:graph_00} appears to be consistent with the hypothesis;
weak links tend to connect communities composed of strong links.
To quantify the extent to which a link is engaged in intracommunity connection, 
we measure the relative neighborhood overlap of a link~\cite{Onnela2007}, defined as
\begin{equation}
O_{ij} = \frac{\left|{\cal N}_i \cap {\cal N}_j \right|}{\left|{\cal N}_i \cup {\cal N}_j \right| - 2},
\end{equation}
where $\left| \cdot \right|$ denotes the number of elements in the set.
When $O_{ij} = 0$, individuals $i$ and $j$ do not have a common neighbor
and the link~$(i,j)$ is considered to connect different communities.
When $O_{ij} = 1$, individuals $i$ and $j$ share all of the neighbors
and the link~$(i,j)$ is confined in a community. 
The strength of weak ties hypothesis suggests
that $O_{ij}$ is positively correlated with $w_{ij}$~\cite{Onnela2007}.
In \FIG\ref{fig:overlap_MI.CC}(a),
$O_{ij}$ averaged over the links with weights smaller than $w$,
denoted as $\langle O \rangle_w$, is plotted against the fraction of links with weights smaller than $w$,
denoted as $P_{\rm cum}(w)$.
Because $\langle O \rangle_w$ monotonically increases with $P_{\rm cum}(w)$,
the CN possesses the strength of weak ties property,
as in the case of mobile communication networks~\cite{Onnela2007}.

Because weak links are associated with a large $I_i$~(\FIG\ref{fig:corr_MI}(c))
and intercommunity links (\FIG\ref{fig:overlap_MI.CC}(a)),
individuals with a large $I_i$ are expected to bridge different communities
and those with a small $I_i$ are expected to be shielded inside a community.
This concept is consistent with the visual inspection of \FIG\ref{fig:graph_00}.
To verify this point,
we show that $I_i$ is negatively correlated with a calibrated clustering coefficient in the following (\FIG\ref{fig:overlap_MI.CC}(b)).
Note that, when the clustering coefficient is large, the individual tends to be inside a community
quantified by the abundance of triangles~\cite{Palla2005}.
When it is small, the individual tends to connect different communities~\cite{Eckmann2002,Ravasz2003}.

The clustering coefficient for each node is defined by
$C_i =$ (number of triangles including individual~i)/$\left[k_i(k_i-1)/2\right]$
$(0 \leq C_i \leq 1, i=1,2, \cdots, N)$~\cite{Watts1998}.
In \FIG\ref{fig:overlap_MI.CC}(b), the Pearson correlation coefficient between $I_i$ and $C_i(w_{\rm thr})$ is plotted against $w_{\rm thr}$,
where $C_i(w_{\rm thr})$ is the local clustering coefficient $C_i$ for the subgraph of the CN generated by eliminating the links with weights smaller than $w_{\rm thr}$.
We opted to use $C_i(w_{\rm thr})$ instead of the weighted clustering coefficient defined for weighted networks~\cite{Barrat2004,Onnela2005} because the latter quantity is, by definition, strongly correlated with $s_i$ and $\overline{w}_i$;
we already discussed the negative correlation between $I_i$ and $s_i$ and between $I_i$ and $\overline{w}_i$
in \FIG\ref{fig:corr_MI}(b) and \ref{fig:corr_MI}(c), respectively.
For $w_{\rm thr}=1$, $I_i$ and $C_i(w_{\rm thr})$ are almost uncorrelated.
This is because almost all the individuals have a large $C_i$ regardless of $I_i$
in the original CN $G_1$~(refer to \FIG\ref{fig:graph_00} for a visual confirmation of this statement).
For $2 \leq w_{\rm thr} \leq 100$,
$I_i$ and $C_i(w_{\rm thr})$ are negatively correlated~(squares in \FIG\ref{fig:overlap_MI.CC}(b)).
Therefore, an individual with a large $I_i$ tends to bridge different communities as quantified by the clustering coefficient.
An individual with a small $I_i$ tends to be confined within communities.  
The circles in \FIG\ref{fig:overlap_MI.CC}(b) represent the partial correlation coefficient
between $I_i$ and $C_i(w_{\rm thr})$, with $k_i(w_{\rm thr})$ and $s_i(w_{\rm thr})$ fixed.
Here, $k_i(w_{\rm thr})$ and $s_i(w_{\rm thr})$ are, respectively,  the degree and strength of individual~$i$,
calculated after eliminating the links with weights smaller than $w_{\rm thr}$.
Because the Pearson and partial correlation coefficients behave similarly,
the negative correlation between $I_i$ and $C_i(w_{\rm thr})$
is not ascribed to the negative correlation between $I_i$ and $s_i$~(\FIG\ref{fig:corr_MI}(b))
or between $I_i$ and $\overline{w}_i$~(\FIG\ref{fig:corr_MI}(c)).

In closing this section,
we stress the robustness of our results against observation failures. 
The wearable tag used in our measurement fails to detect a conversation event
if the tag is sealed behind obstacles such as a desk or partition.
For example, suppose that two individuals chat for five minutes
and either of their tags is just under a desk and is undetected in the third minute.
Then, the single conversation event is split into two spurious conversation events, each lasting for two minutes.
To examine the robustness of our results against such observation failures,
we repeat the same set of analyses after filling short intervals between successive conversations
between the same pair of individuals.
If individual~$i$ has two successive conversation events with individual~$j$
and the interval between the two events is smaller than or equal to $m$ minutes,
we merge the two events into one.
The original partner sequence corresponds to $m=0$.
The number of conversation events decreases with $m$.
The interpolation reduces $w_{ij}$, $s_i$, and $\overline{w}_i$ and conserves $k_i$, $H^0_i$, and $C_i$.
We confirmed that our findings are reproduced when we interpolate the original data
with $m=1$ and $m=5$~(see Appendix~G for details).

\section{Discussion}

We have shown that sequences of conversation events have
deterministic components. The entropy in the distribution of the
conversation partners of an individual decreases by, on average, $28.4\%$
for data set $D_1$ and $34.8\%$ for data set $D_2$, if we know the current partner.
Much of the predictability of conversation events results from the bursty activity patterns.
In general, daily and weekly rhythms of human activity can cause bursty activity patterns~\cite{Malmgren2008}.
During the night and weekend, the individuals are out of the office. Therefore, interevent intervals are usually longer than those within working hours.
Nevertheless, we consider that the effects of such long interevent intervals on the predictability of conversation partners are small.
This is because the fraction of long interevent intervals, \ie, those over five hours, for example, is relatively small, occupying 4.31\% in $D_1$ and 2.95\% in $D_2$.
In addition, there is no particular reason to believe that the last conversation partner in a day and the first partner in the next day are specifically correlated. In this study, we did not correct for the effect of the night and weekend.

The degree of predictability depends on individuals.
In particular, we have shown that individuals connecting different communities in
conversation networks behave relatively deterministically.
We quantified the degree to which an individual is confined in communities by the clustering coefficient.
In the context of an overlapping community structure,
individuals connect different communities when they belong to multiple overlapping communities~\cite{Palla2005}.
Such individuals tend to be surrounded by many triangles if we define the community by 3-cliques (\ie, triangles).
This apparently contradicts our results.
This contradiction comes from the difference in what we mean by connecting different communities.
We regard individuals as bridging different communities when they are not strongly bound to any community 
and they have links to different communities.
In this sense, nodes with small clustering coefficient values connect different communities
in networks with hierarchal structure~\cite{Ravasz2002,Ravasz2003}.
In general, links bridging different communities have large betweenness centrality values~\cite{Girvan2002}.
The clustering coefficient of a node tends to decease with the betweenness centrality~\cite{Goh2003}.
This lends more support to our view that individuals with small clustering coefficient values tend to connect different network communities.
It should be noted that the strength of weak ties property of the CN and the relationship between $I_i$ and the individual's position in the CN are preserved, if we define the link weight by the total duration of the conversation events for each pair (see Appendix F).

We do not have an access to the contents of dialogs for ethical reasons.
Therefore, the understanding of the reason for the correlation between the individual's position and predictability is limited. 
Nevertheless, individuals that own many weak links and connect distinct groups may mediate
information flows necessary to coordinate tasks involving these groups (\eg, project groups in a company). 
Such individuals may control the information flow between the groups in a rigid manner to yield a large $I_i$.
In contrast, individuals with few weak links may enjoy casual (and perhaps creative) conversations
within their own groups to choose the partners in a random manner.
Such individuals may tend to have a small $I_i$.
It should be noted that our data were obtained in company offices.
Roles or formal positions of individuals in the company may affect $I_i$ and the local abundance of weak links surrounding the individuals.
 
Song~\textit{et al.} discovered a remarkable predictability in the mobility patterns of humans~\cite{Song2010}.
In terms of the analysis tools, our methods are similar to theirs.
We have applied the entropy measures and the concept of predictability to different types of data sets.
In our data, the physical location of individuals is irrelevant;
individuals work in offices in the companies. 
It should be noted that although we have not implemented the prediction algorithm,
the predictability of the data is implied by the large mutual
information that we observed. 
This logic parallels that made for human mobility patterns~\cite{Song2010}.

\section*{Acknowledgments}
T.T. acknowledges the support provided through Grant-in-Aid for Scientific Research (No. 10J06281) from JSPS, Japan.
M.N. acknowledges the support provided through Grant-in-Aid for Scientific Research (No. 10J08999) from JSPS, Japan.
N.M. acknowledges the support provided through Grants-in-Aid for Scientific Research (No. 20760258 and No. 23681033) from MEXT, Japan.

\section*{Appendix}
\subsection*{A. Results for data set $D_2$} 

We obtained qualitatively the same results for $D_2$ as those for $D_1$.
The results for $D_2$ are shown in \FIGS\ref{fig:hist_H_D2}, \ref{fig:IEI-SEtest_D2}, \ref{fig:corr_MI_D2}, and \ref{fig:overlap_MI.CC_D2},
which correspond to \FIGS\ref{fig:hist_H}, \ref{fig:IEI-SEtest}, \ref{fig:corr_MI}, and \ref{fig:overlap_MI.CC} in the main text, respectively.

\subsection*{B. Details of the bootstrap test}\label{sec:boottest_MI}

To confirm that the large value of the empirically obtained $I_i$ is not because of the small data size,
we carry out a bootstrap test as follows.
First, we make a bootstrap sample of a partner sequence with length $s_i$
by resampling partners' IDs
from the empirical partner sequence of individual~$i$ without replacement~(\ie, shuffling).
Then, we use \EQ\eqref{eq:MI} to calculate the mutual information $\hat{I}_i$ for the bootstrap sample.
By resampling 5,000 bootstrap partner sequences,
we construct the distribution of $\hat{I}_i$, which we denote as $p(\hat{I}_i)$.
On the basis of $p(\hat{I}_i)$, we carry out a hypothesis test for $I_i$.
The null hypothesis of the test is that
$I_i$ is positive just because of the small data size.
The alternative hypothesis is that
$I_i$ is larger than the value expected for unstructured data of a small size.
We set the significance level of the test to $1\%$.
Consequently, the critical region of the null hypothesis is the half-open interval
above the 99~percentile point of $p(\hat{I}_i)$.
In \FIG\ref{fig:boottestMI}, the results of the bootstrap test are summarized.
Apparently, $I_i$ is above the 99~percentile point (\ie, the upper end of the each error bar).
In fact, for all the individuals in $D_1$ and $D_2$, except individual~14 in $D_1$ and 149 in $D_2$,
the null hypothesis is rejected with a $1\%$~significance level.
	
\subsection*{C. Long-tailed behavior of interevent intervals}
Human activity patterns are characterized
by long-tailed distributions of the interevent intervals~\cite{Eckmann2004eod,Barabasi2005,AVazquez2006mba,Candia2008,Malmgren2008,Iribarren2009ioh,Cattuto2010dop,Isella2010,Wu2010}, a feature that is shared by our data.
We define the interevent interval~$\tau$
as the interval between the initiation time of two successive conversation events involving a given individual.
The unit of $\tau$ is a minute, corresponding to the time resolution of the recording.
As shown in \FIG\ref{fig:taus}(a), the distribution of $\tau$, denoted by $p(\tau)$,
for a typical individual in $D_1$ is long-tailed. 
The tail of the empirical data (solid line) is much fatter than 
that of the exponential distribution whose mean is equal to that of the empirical data (dashed line).
The histogram of the coefficient of variation (CV) of $p(\tau)$ on the basis of all the individuals in $D_1$ and the same histogram for $D_2$
are shown in \FIG\ref{fig:taus}(b).
The value of CV is equal to the ratio of the standard deviation to the mean and is equal to unity for exponential distribution.
Figure~\ref{fig:taus}(b) indicates that the CV of $p(\tau)$ is much larger than unity for all the individuals.

\subsection*{D. Components of the predictability of conversation events}
A possible mechanism governing the predictability of the conversation events is the bursty activity patterns.
To examine the effect of the long-tailed behavior of $p(\tau)$ on the predictability, we carry out a statistical test based on the shuffling of $I_i$ as follows.
Consider the sequence of conversation events of focal individual~$i$ with individual~$j$.
If $i$ and $j$ talk four times in a given day and the interevent intervals are equal to $\tau_1$, $\tau_2$, and  $\tau_3$ in the chronological order,
we randomize their order.
For example, the interevent intervals in the shuffled data are ordered as $\tau_2$, $\tau_1$, and $\tau_3$.
We carry out the same randomization for each day and each partner~$j$.
Then, we combine the randomized sequences (\ie, point processes) for different $j$'s into the one point process from which we read out the randomized partner sequence for $i$.
We define $I_i^{\rm burst}$ as the mutual information for this randomized partner sequence.
In \FIG\ref{fig:IEI-SEtest}(a), the mean and standard deviation of $I_i^{\rm burst}$ obtained from 100 randomized partner sequences are shown for different individuals in $D_1$.
The empirical values of $I_i$~(circles) are significantly larger than $I_i^{\rm burst}$ for most individuals.
However, $I_i^{\rm burst}$ consistently occupies a large fraction of $I_i$ and increases with $I_i$. 
Therefore, the burstiness is a major cause of the predictability regardless of the value of $I_i$.

The burstiness is not the only contributor to the predictability.
To show this,
we examine the reduced partner sequence generated by merging all the consecutive events with the same partner into one event.
For example,
the original partner sequence~$\left\{ 2,3,3,6,4,4,3,3,3,2,6,2 \right\}$ yields the merged partner sequence~$\left\{ 2,3,6,4,3,2,6,2 \right\}$.
We calculate the mutual information in the merged partner sequence, denoted by $I_i^{\rm merge}$.
$I_i^{\rm merge}$ measures the predictability of conversation events that does not result from the burstiness.
We do not directly compare $I_i^{\rm merge}$ with the original $I_i$ because the merging procedure shortens
the length of the partner sequence and the amount of mutual information generally depends on the length of a sequence~\cite{Strehl2003}.
Instead, we carry out a bootstrap test for $I_i^{\rm merge}$.
By definition, the partner changes every time in the merged partner sequence.
We obtain bootstrap samples respecting this property as follows.
The frequency with which partner~$j$ appears in the merged partner sequence of individual~$j$ is denoted by $\Pr^{\rm merge}_i(j)$.
We select the first partner of $i$, denoted by $\ell$, randomly according to $\Pr^{\rm merge}_i(j)$.
The second partner is selected according to $\Pr^{\rm merge}_i(j) / (1- \Pr^{\rm merge}_i(\ell))$,
where $j \neq \ell$.
We repeat the same procedure until the generated sequence becomes as long as the merged partner sequence.
Figure~\ref{fig:IEI-SEtest}(b) summarizes the results of the bootstrap test for $I_i^{\rm merge}$.
$I_i^{\rm merge}$ is consistently larger than the values expected for the bootstrap samples for all the individuals.
Therefore, the partner sequence is predictable to some extent even without the effect of the bursty activity patterns.

\subsection*{E. Use of normalized mutual information}
In the field of cluster partitioning, the normalized mutual information $\overline{I}_i \equiv I_i / H_i^1$ is used
to quantify the accuracy of partitioning methods,
because the relationship $0 \leq \overline{I}_i \leq 1$ is convenient for comparing different methods~\cite{Strehl2003}.
Our main results are qualitatively the same if we replace $I_i$ by $\overline{I}_i$~(\FIG\ref{fig:normMI}).  

\subsection*{F. Alternative definition of the link weight based on the duration of conversation}

In the main text, we defined the link weight by the total number of conversation events for each pair.
An alternative definition is given by the total duration of the conversation events for each pair.
This alternative definition changes $w_{ij}$, $s_i$, and $\overline{w}_i$ and conserves $k_i$, $H_i^{0,1,2}$, and $I_i$.
For the CN where the link weight is defined by the total duration,
we repeat the same set of analyses as that conducted in \SEC\ref{sec:variation}.
As shown in \FIG\ref{fig:duration_weight}, the change in the definition of the link weight does not affect our main results.
We observed a negative correlation between $I_i$ and $s_i$ (\FIG\ref{fig:duration_weight}(a)), that between $I_i$ and $\overline{w}_i$ (\FIG\ref{fig:duration_weight}(b)), the ``strength of weak ties'' property (\FIG\ref{fig:duration_weight}(c)), and a negative correlation between $I_i$ and $C_i(w_{\rm thr})$ (\FIG\ref{fig:duration_weight}(d)).

\subsection*{G. Robustness against observation failures}

To examine the robustness of our results against observation failures,
we analyze the data sets after interpolating short intervals
between successive conversations between the same pairs of individuals.
Suppose that individuals~$i$ and $j$ talk with each other twice and that $i$ does not talk with anybody else
between the two conversation events with $j$.
We merge the two conversation events into one
if the difference between the ending time of the first event and the starting time of the second event is less than or equal to $m$ minutes.

In \FIG\ref{fig:compare_m1}, $\tilde{s}_i$, $\tilde{H}^1_i$, $\tilde{H}^2_i$, and $\tilde{I}_i$,
which are the quantities calculated for the data obtained with $m=1$, are compared
with $s_i$, $H^1_i$, $H^2_i$, and $I_i$, respectively.
As expected, $\tilde{s}_i$ is smaller than $s_i$, and 
$\tilde{H}^1_i$ and $\tilde{H}^2_i$ are generally larger than $H^1_i$ and $H^2_i$, respectively.
As shown in \FIG\ref{fig:confirm_m1}, the important properties of the data sets are not changed by
the interpolation with $m=1$.
In other words,
a negative correlation between $\tilde{I}_i$ and $\tilde{s}_i$ (\FIG\ref{fig:confirm_m1}(a))
and that between $\tilde{I}_i$ and $\tilde{\overline{w}}_i$ (\FIG\ref{fig:confirm_m1}(b)),
the strength of weak ties property~(\FIG\ref{fig:confirm_m1}(c)), 
and a negative correlation between $\tilde{I}_i$ and $C_i(w_{\rm thr})$~(\FIG\ref{fig:confirm_m1}(d)) are observed.
The results are qualitatively the same for $m=5$, as shown in \FIGS\ref{fig:compare_m5} and \ref{fig:confirm_m5}.


\setcounter{figure}{0}
\renewcommand{\thefigure}{\arabic{figure}}
\renewcommand{\thesection}{}

\clearpage
\begin{figure}
\centering
\includegraphics[width=0.9\hsize]{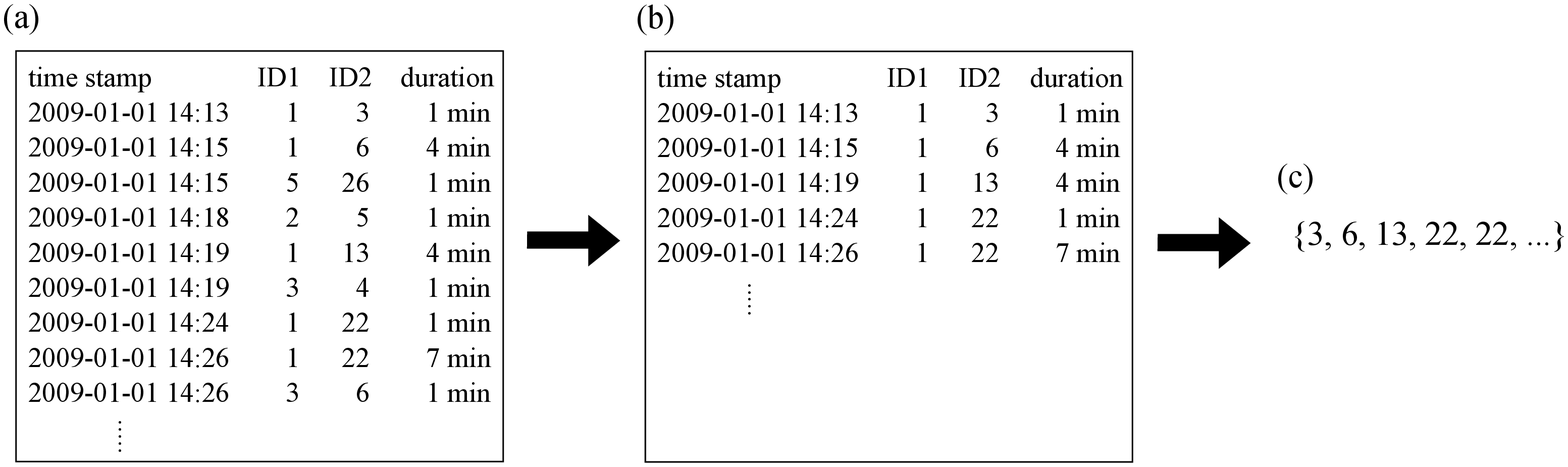}
\caption{
Procedure for generating the partner sequence of individual~1.
(a) Original data set. (b) List of conversation events that involve individual~1.
(c) Partner sequence of individual~1.
The data set shown in (a) is an artificial one, and is provided for the purpose of explanation.
}
\label{fig:sequence-generation}
\end{figure}

\clearpage
\begin{figure}[h]
\centering
\includegraphics[width=\hsize]{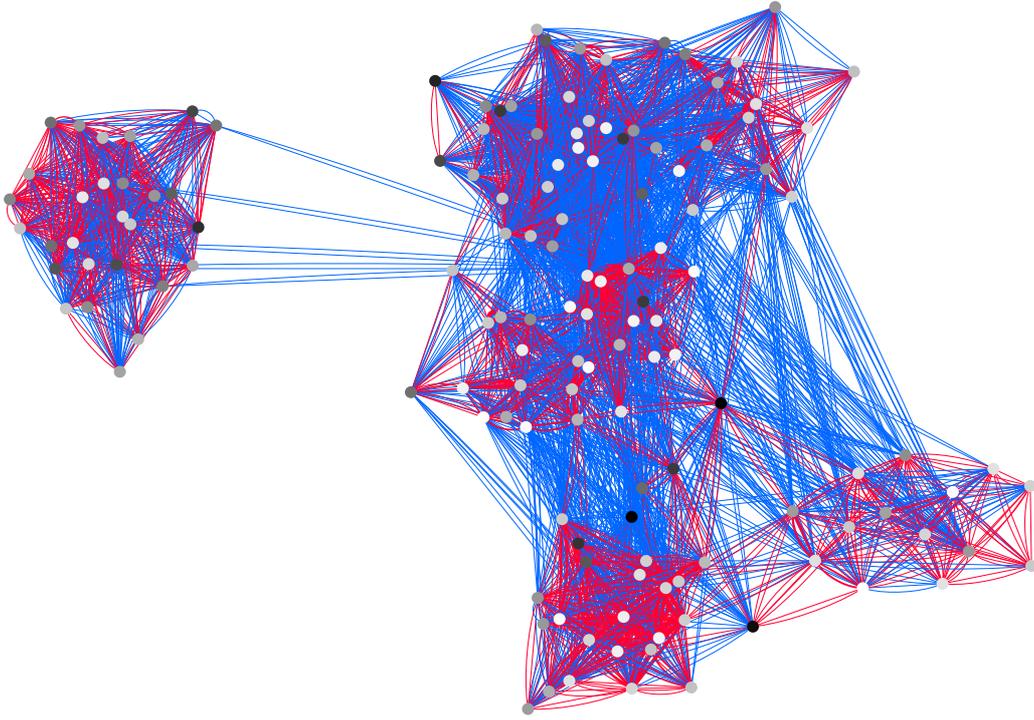}
\caption{
Visualization of CN $G_1$.
For clarity, only the nodes with strengths larger than 100 and the links among them are drawn.
The darkness of the node color represents the value of $I_i$; a darker node has a larger $I_i$.
The thickness of the link is proportional to its weight.
The links with weights larger than or equal to  (smaller than) the median value (\ie, 5) are drawn by red (blue) lines.}
\label{fig:graph_00}
\end{figure}

\clearpage
\begin{figure}[h]
\centering
\includegraphics[width=0.45\hsize]{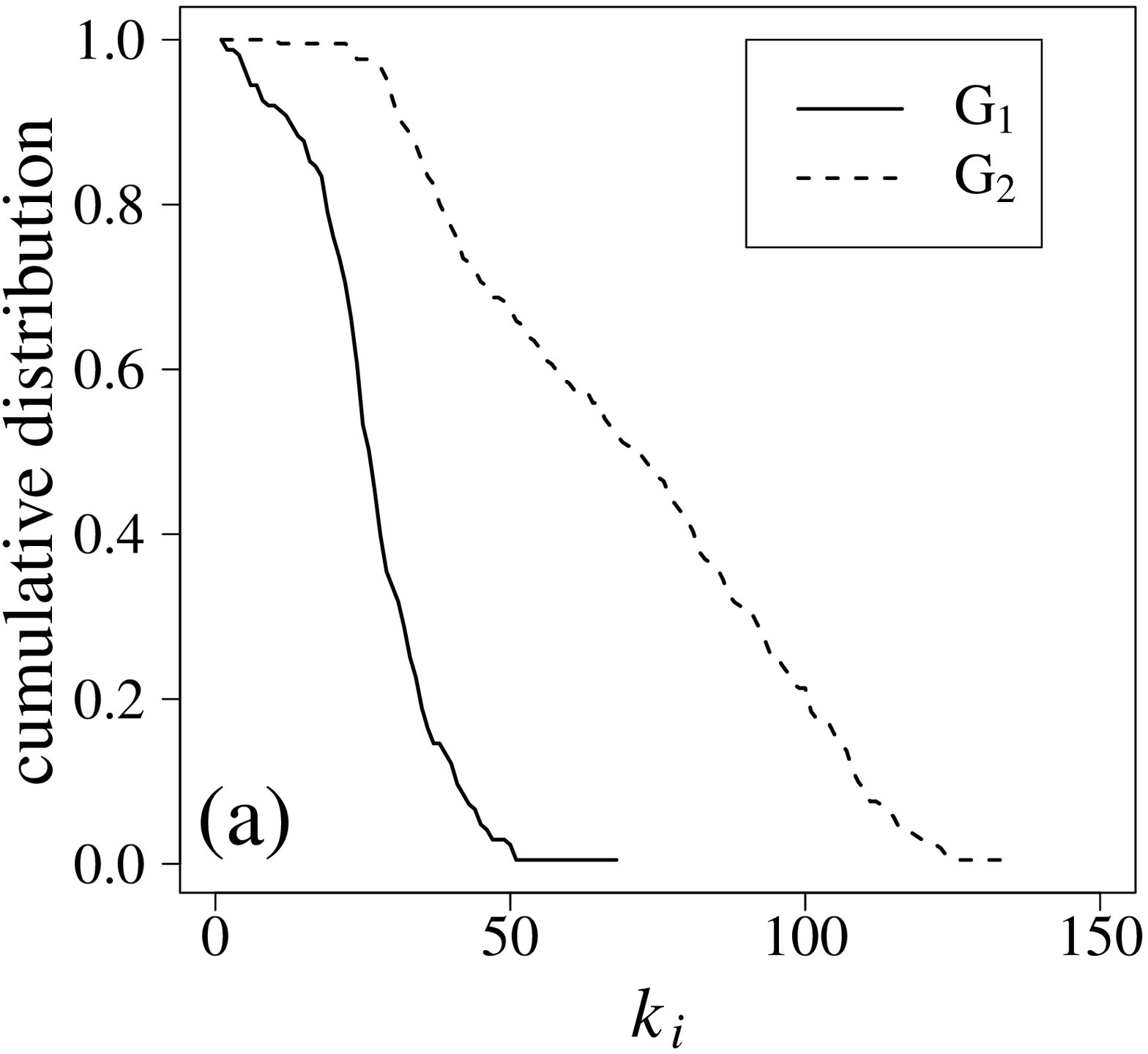}
\includegraphics[width=0.45\hsize]{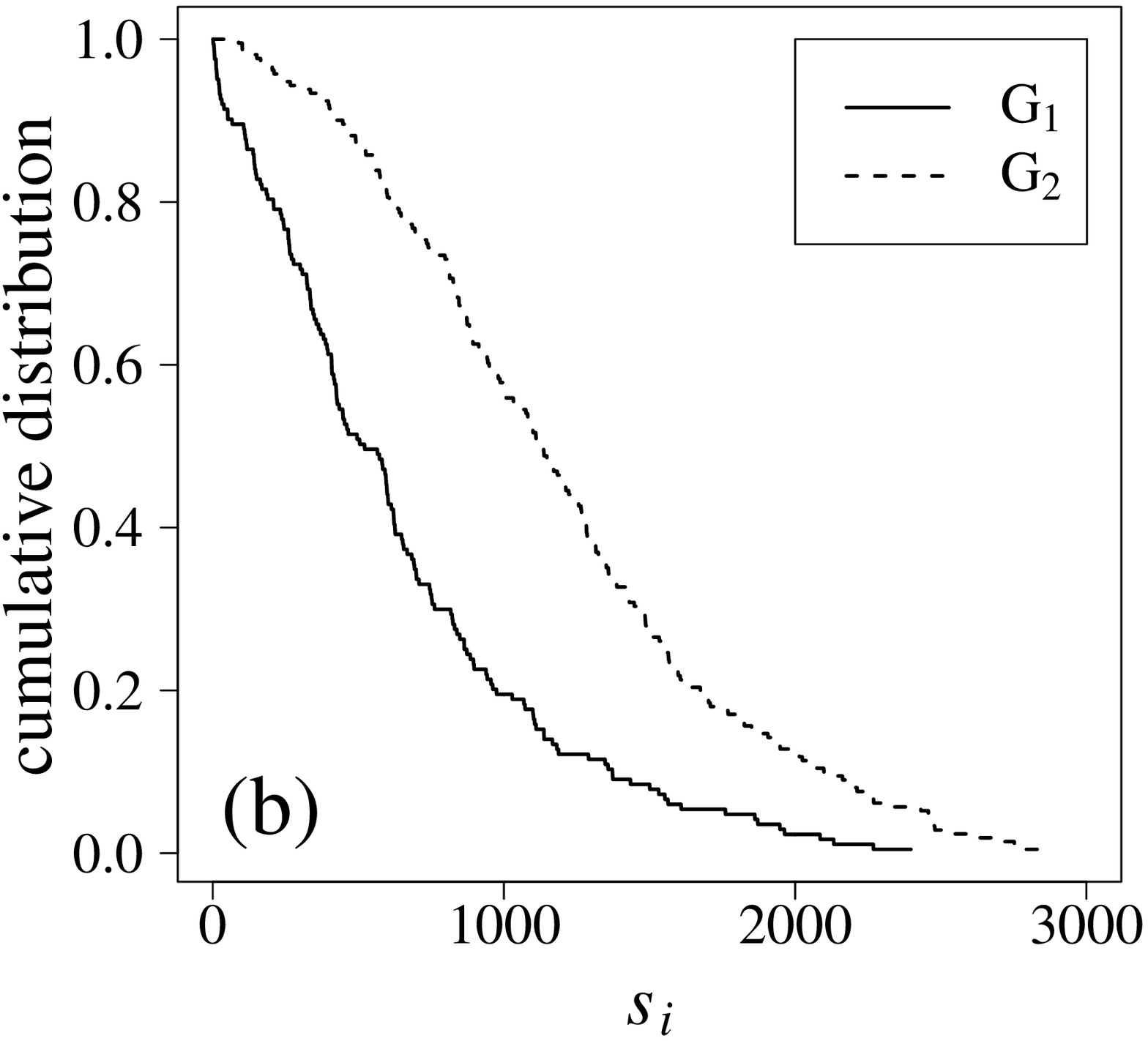}\\
\includegraphics[width=0.45\hsize]{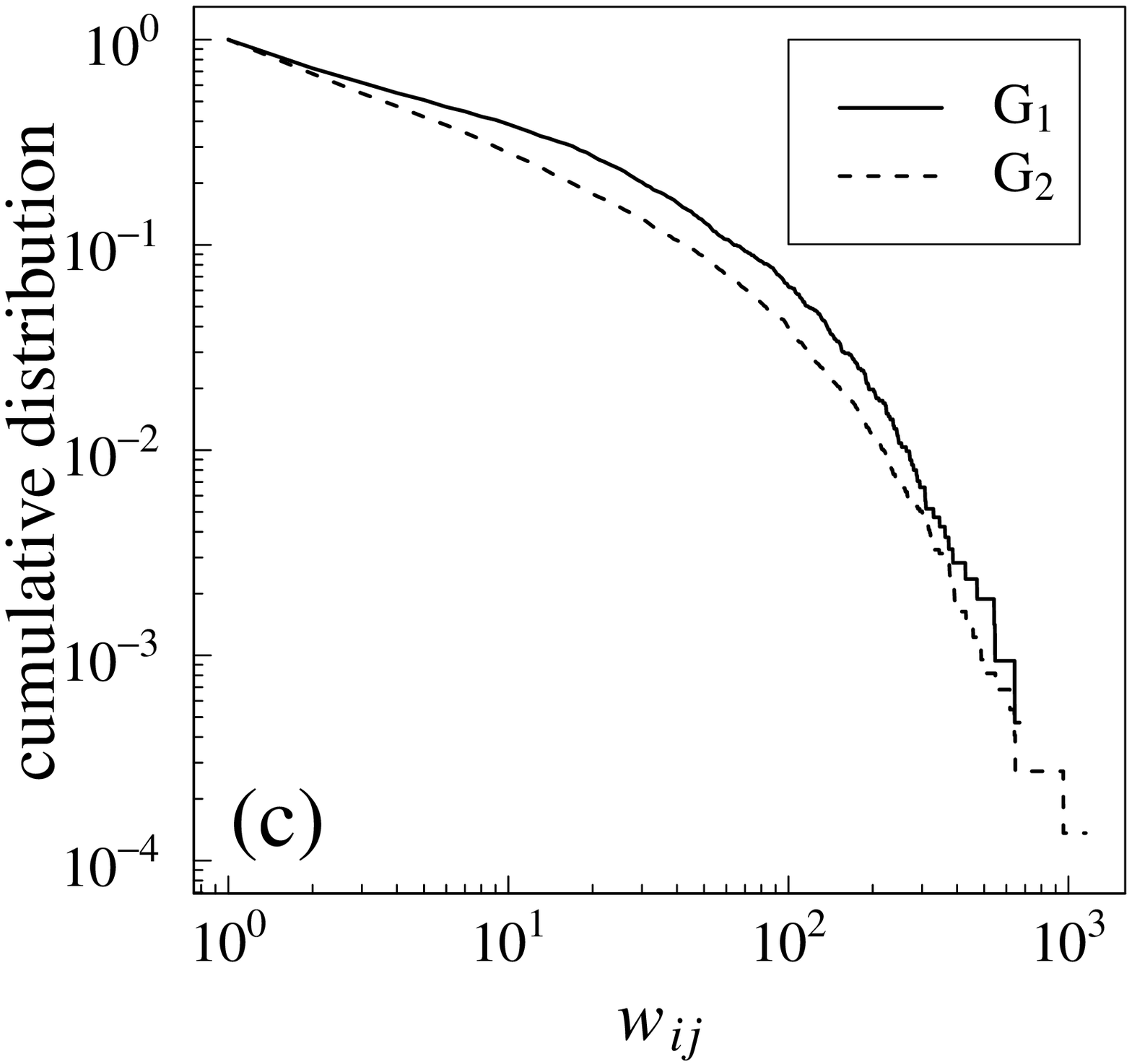}
\caption{
Cumulative distribution of (a) degree, (b) node strength, and (c) link weight of the CNs.
}
\label{fig:statdist_CN}
\end{figure}

\clearpage
\begin{figure}
\centering
\includegraphics[width=0.45\hsize]{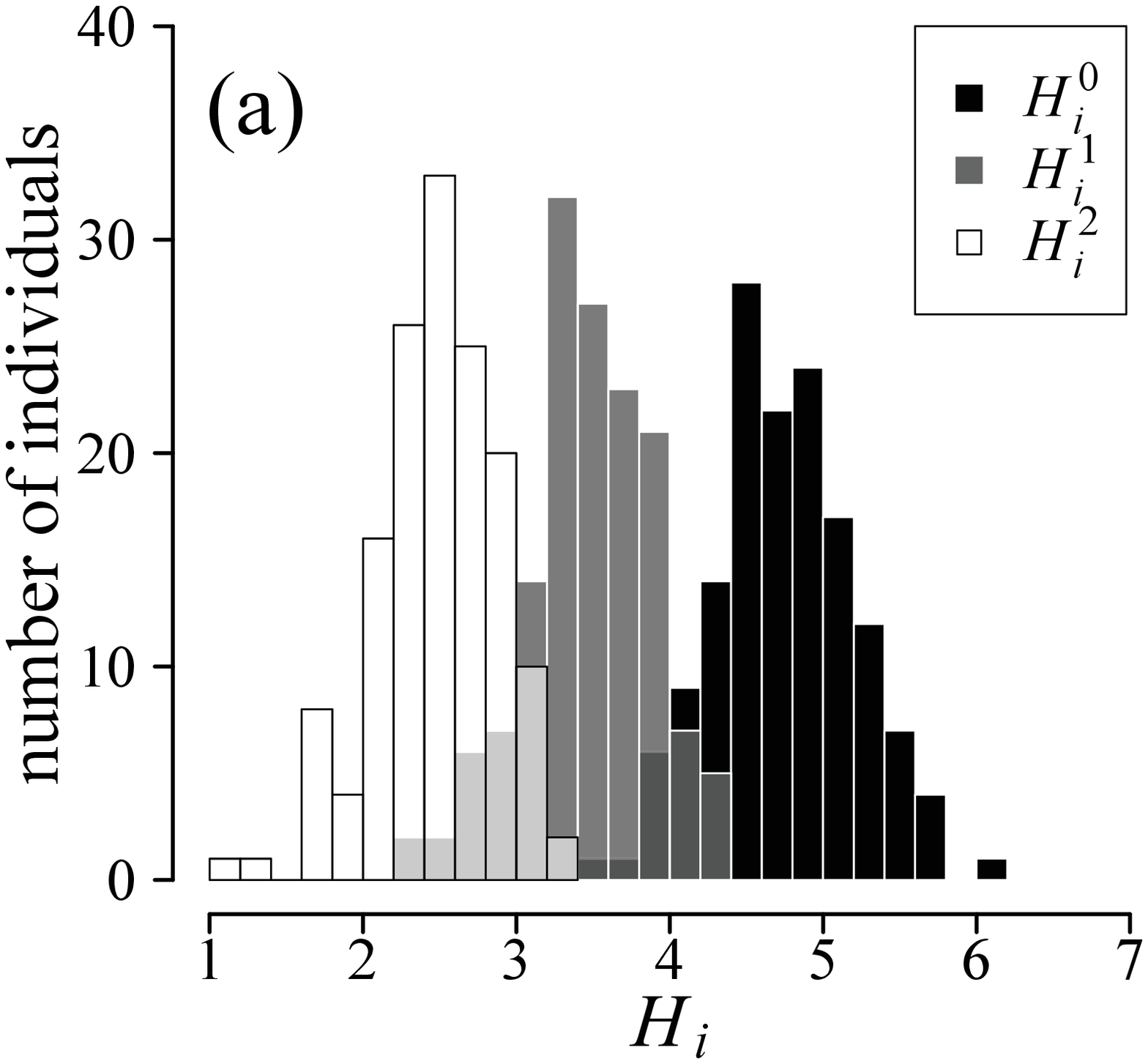}
\includegraphics[width=0.45\hsize]{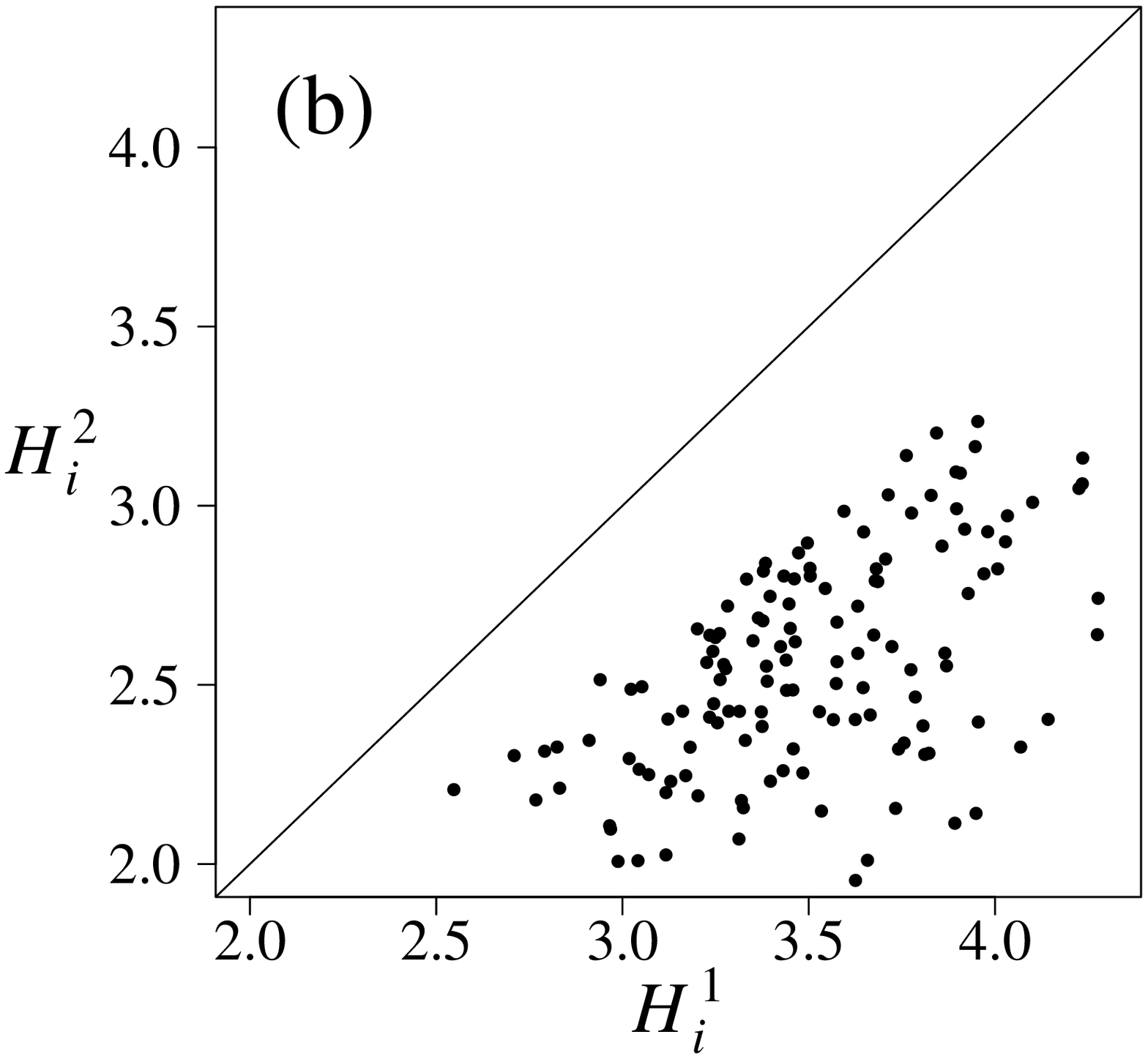}
\caption{
(a) Histograms of the entropies for $D_1$.
(b) Relationship between $H^1_i$ and $H^2_i$ in $D_1$. The solid line represents $H^1_i = H^2_i$.
}
\label{fig:hist_H}
\end{figure}

\clearpage
\begin{figure}[h]
\centering
\includegraphics[width=0.9\hsize]{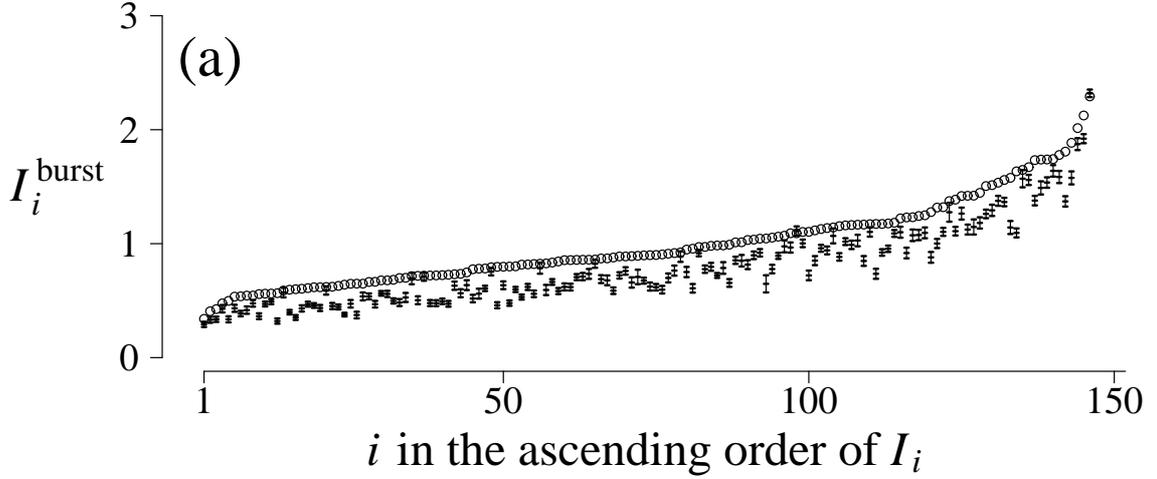}\\
\includegraphics[width=0.9\hsize]{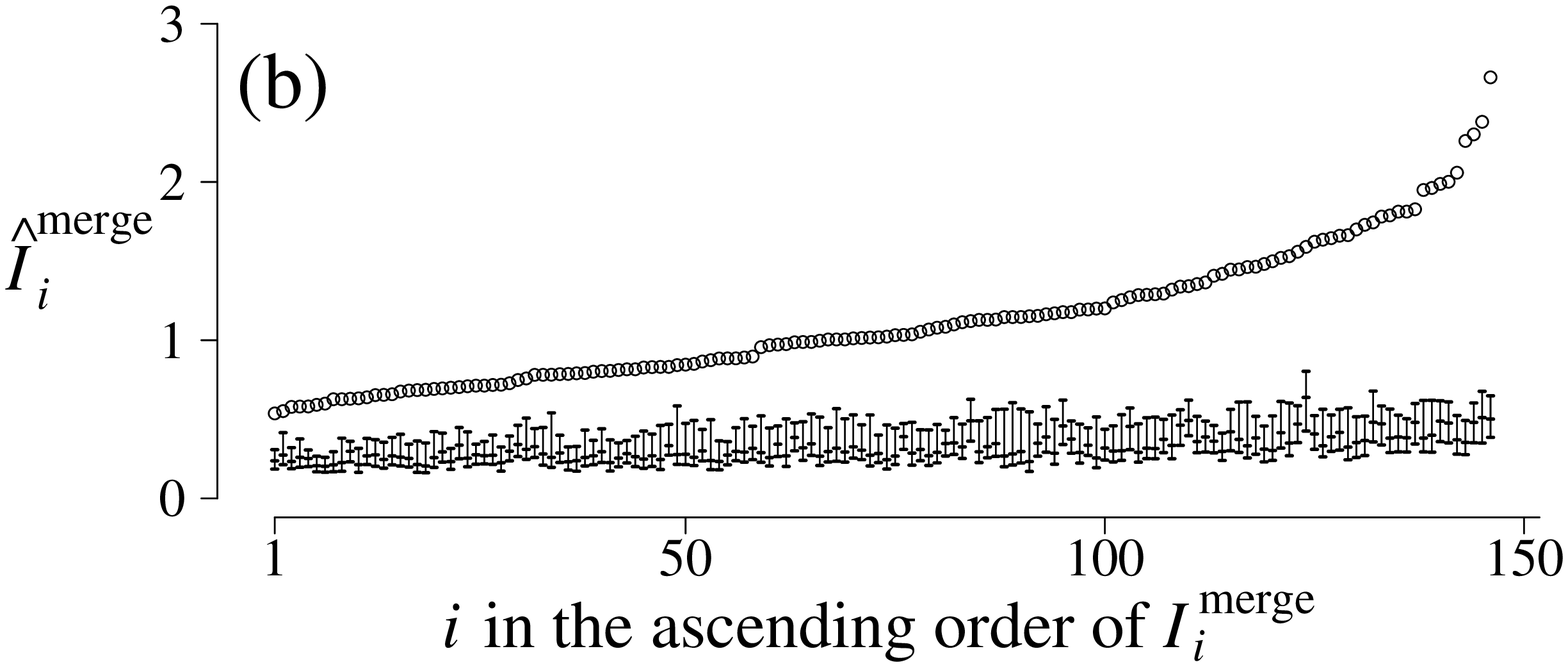}
\caption{
Results of the bootstrap test for $D_1$. The circles represent $I_i$ and $I_i^{\rm merge}$ in (a) and (b), respectively.
The error bars represent the statistics for the bootstrap samples.
(a) Results of the shuffling test.
$I_i$ and the error bars are plotted in the ascending order of $I_i$.
The error bars indicate 1 standard deviation around the mean of $I_i^{\rm burst}$,
which was obtained from 100 shuffled partner sequences.
The ticks at the middle of the error bars indicate the mean.
(b) Results of the merging test.
$I_i^{\rm merge}$ and the confidential intervals (error bars) are plotted 
in the ascending order of $I_i^{\rm merge}$.
The lower and upper ends of the error bars represent $0$ and $99$ percentile points, respectively.
The ticks at the middle of the error bars indicate the mean.
}
\label{fig:IEI-SEtest}
\end{figure}

\clearpage
\begin{figure}[t]
\centering
\includegraphics[width=0.45\hsize]{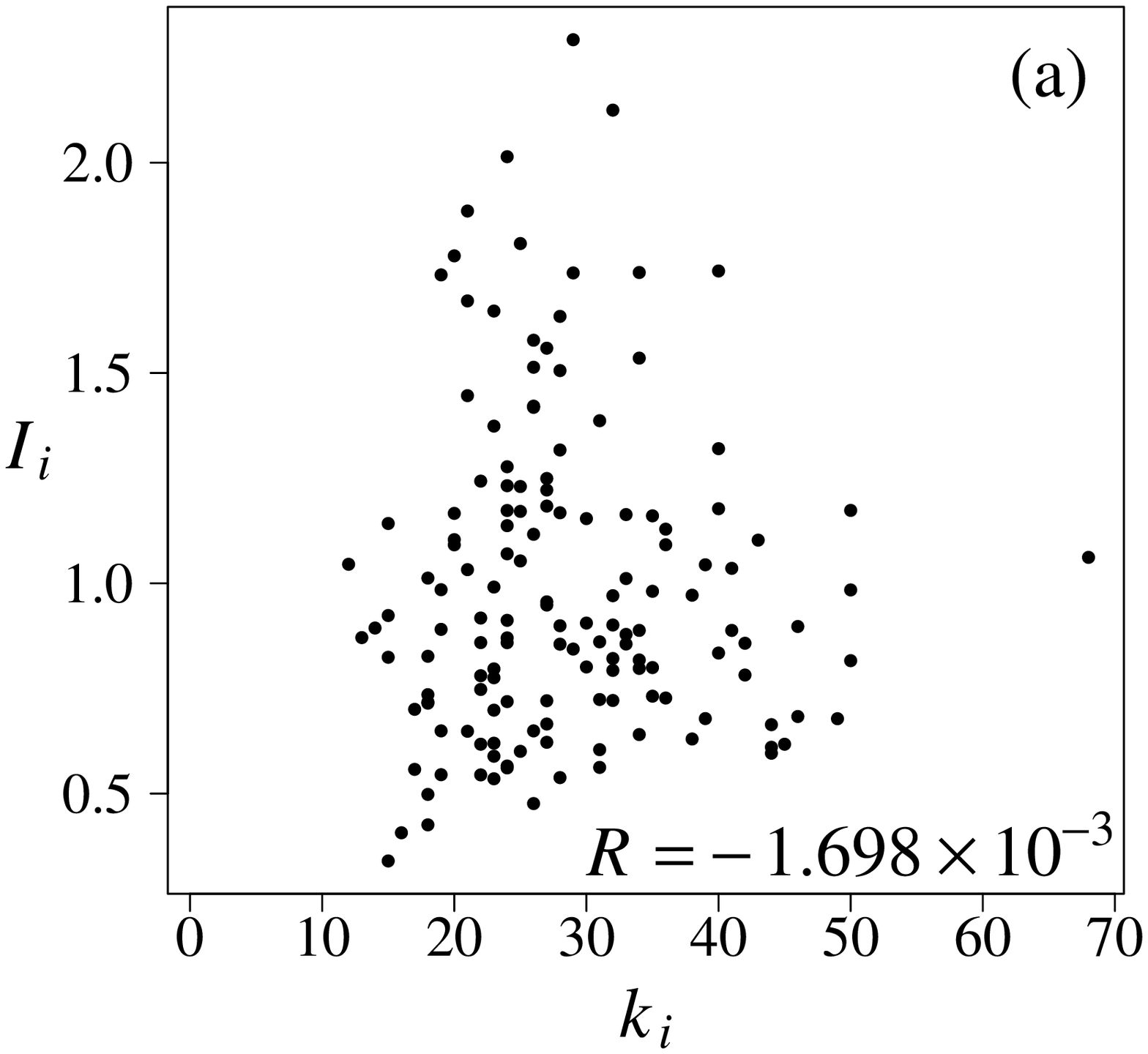}
\includegraphics[width=0.45\hsize]{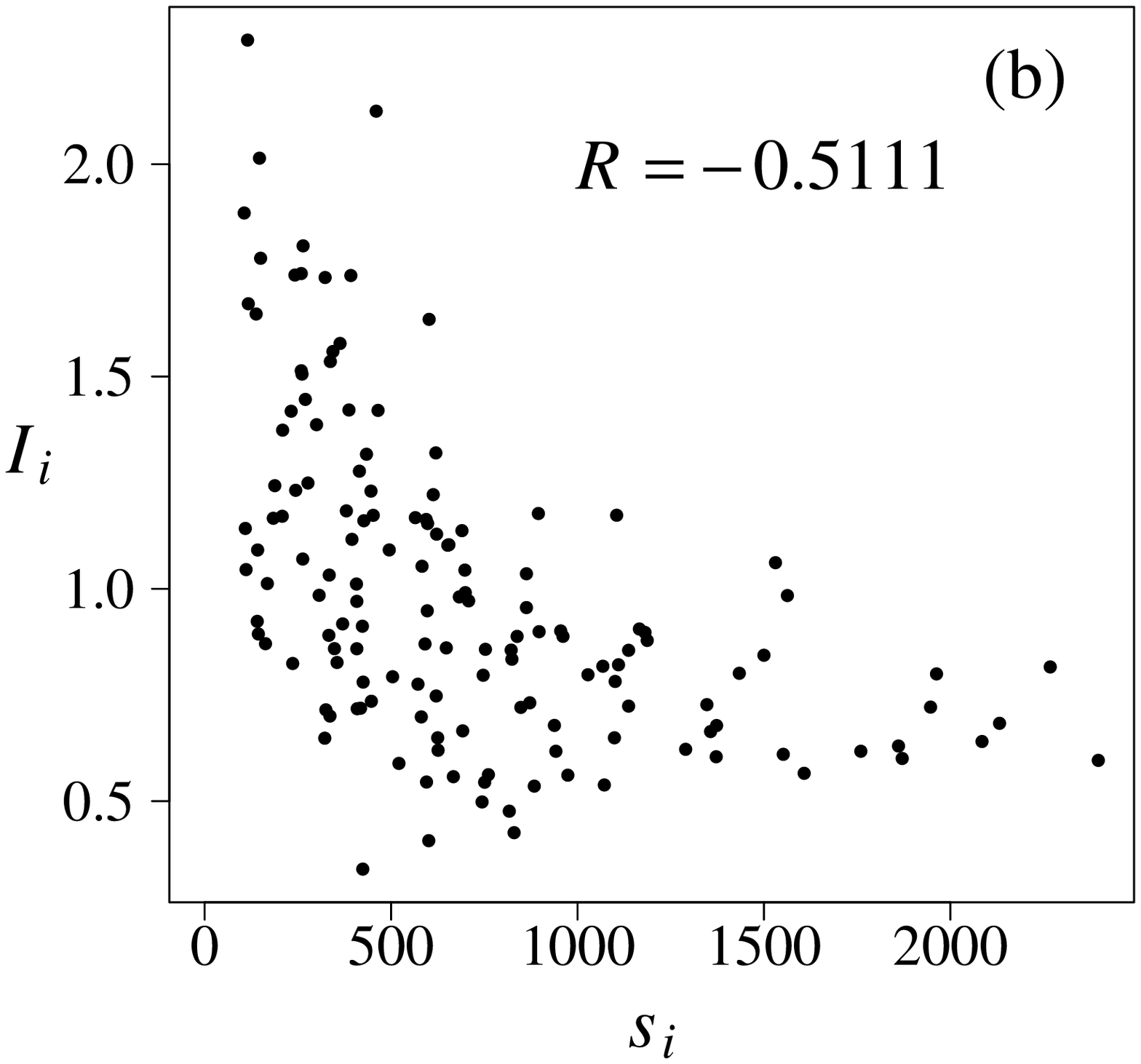}\\
\includegraphics[width=0.45\hsize]{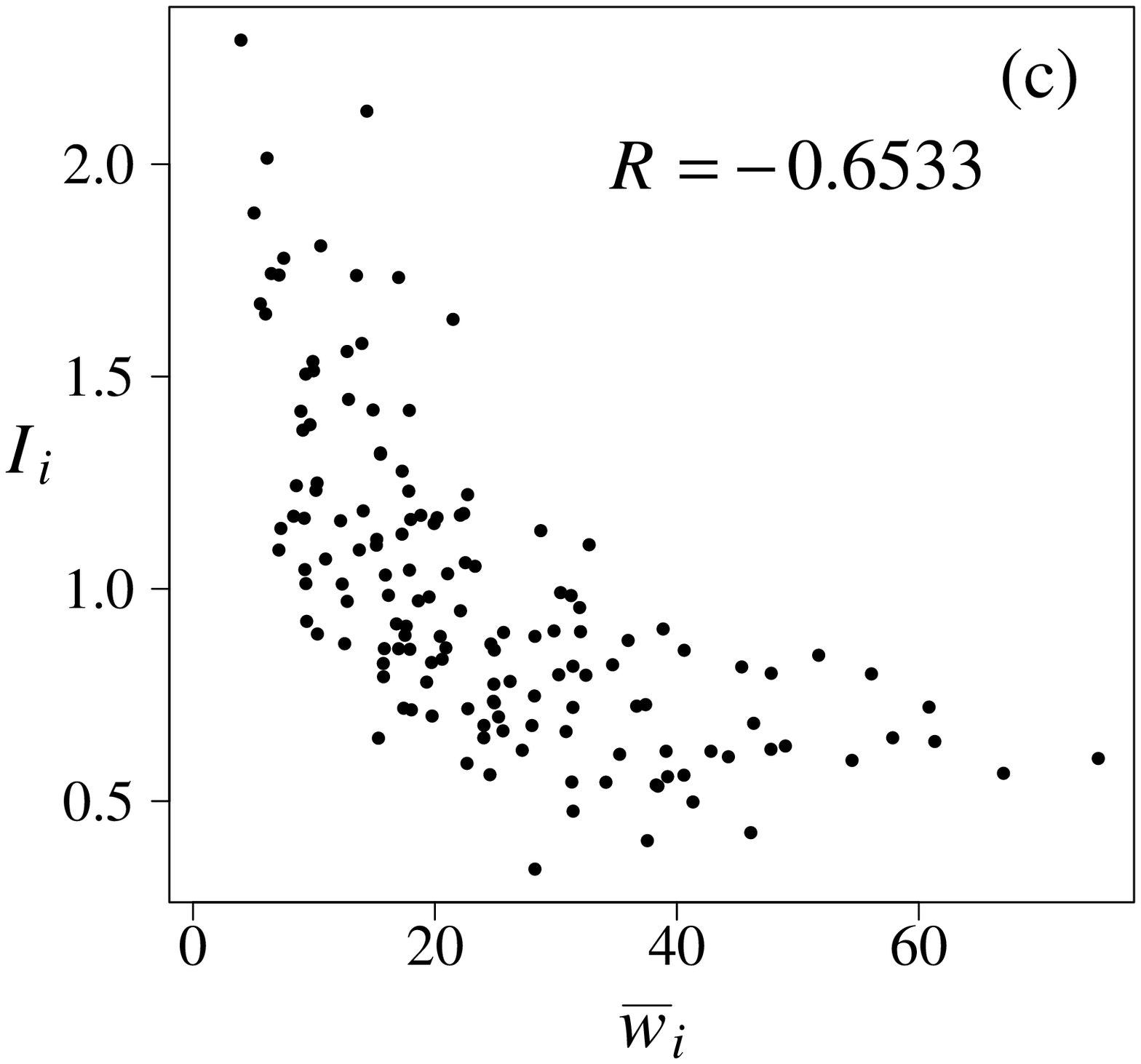}
\caption{
Mutual information $I_i$ is plotted against (a) degree $k_i$, (b) node strength $s_i$,
and (c) average node weight $\overline{w}_i$, for $D_1$.
The Pearson correlation coefficient $R$ between the plotted quantities is also shown.
}
\label{fig:corr_MI}
\end{figure}

\begin{figure}[h]
\centering
\includegraphics[width=0.45\hsize]{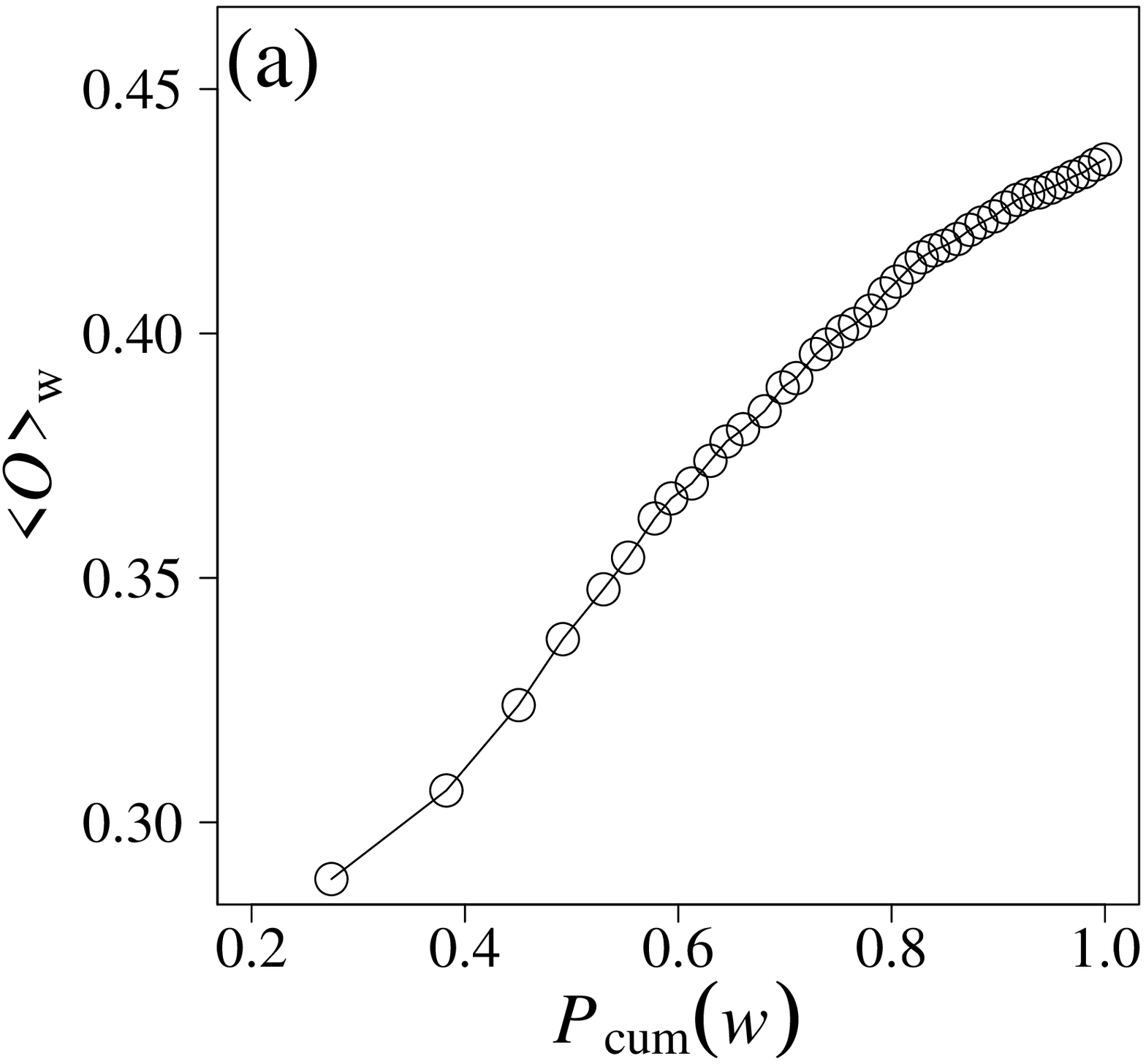}
\includegraphics[width=0.45\hsize]{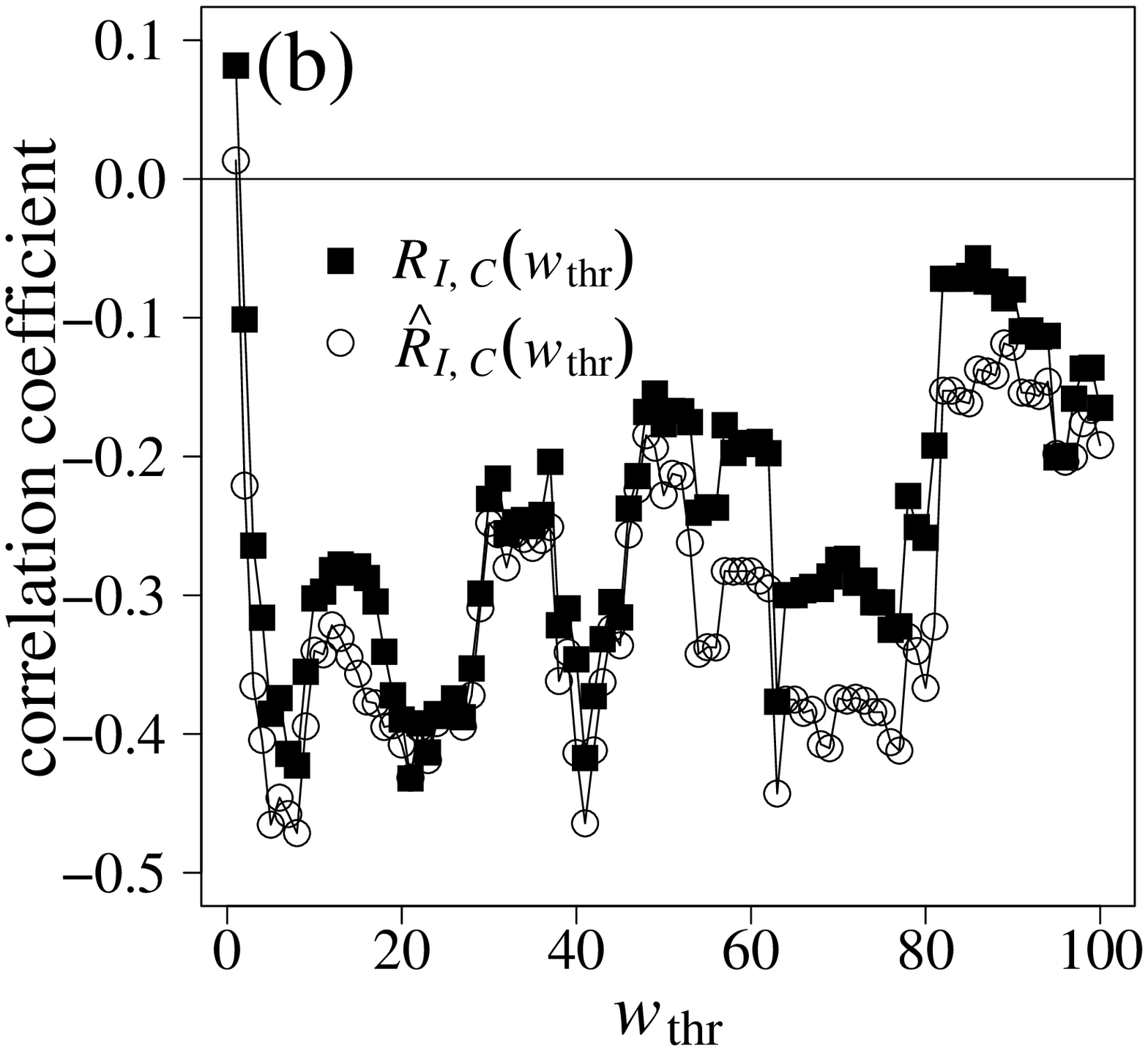}
\caption{
(a) Averaged neighborhood overlap $\langle O \rangle_w$ as a function of the fraction of links with weights smaller than $w$ for $D_1$.
(b) Pearson correlation coefficient between $I_i$ and $C_i(w_{\rm thr})$ (squares) and 
the partial correlation coefficient between them with $k_i(w_{\rm thr})$ and $s_i(w_{\rm thr})$ fixed (circles), for $D_1$.
The horizontal line represents zero correlation.
}
\label{fig:overlap_MI.CC}
\end{figure}

\setcounter{figure}{0}
\renewcommand{\thefigure}{A\arabic{figure}}

\clearpage
\begin{figure}
\centering
\includegraphics[width=0.45\hsize]{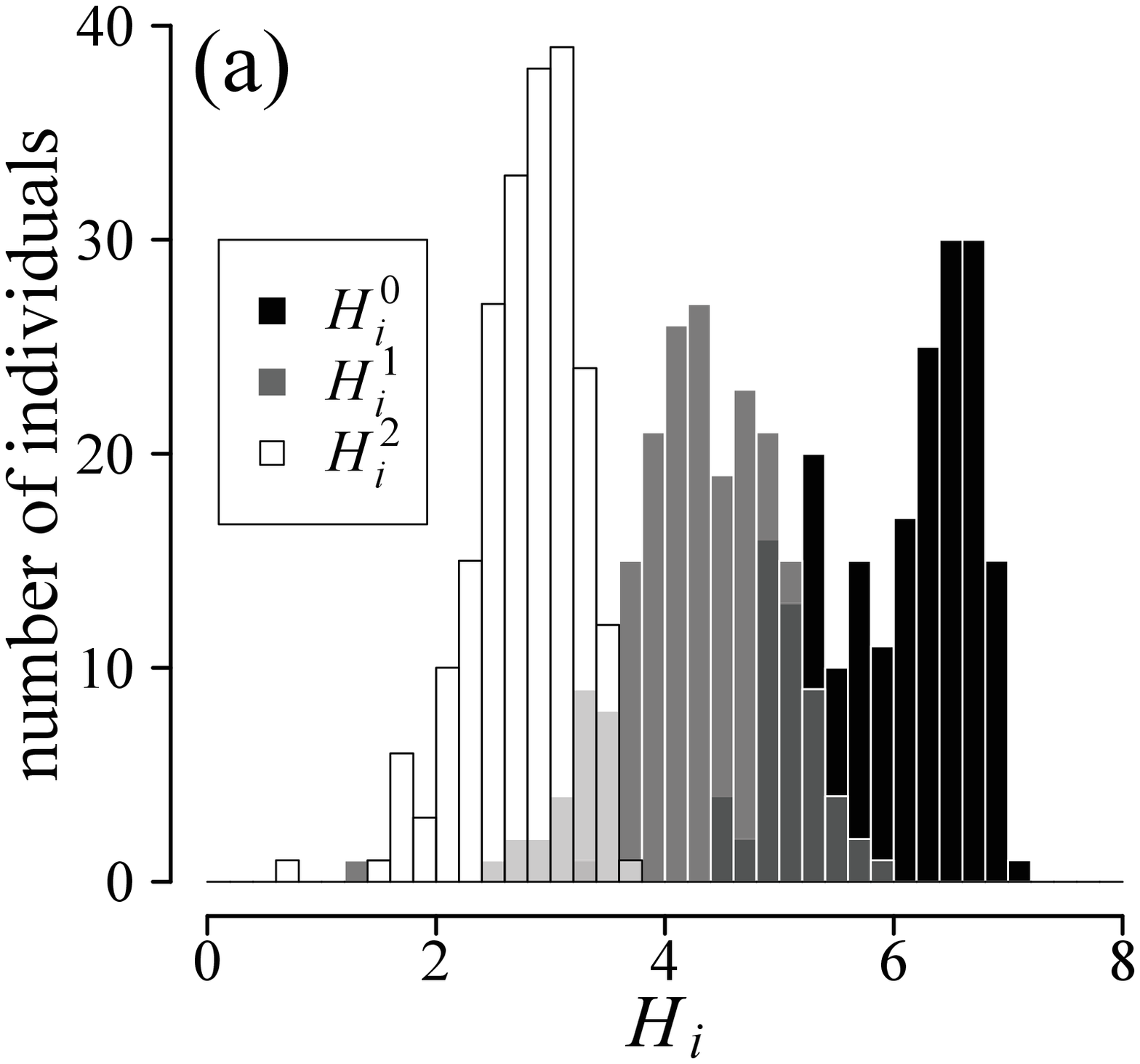}
\includegraphics[width=0.45\hsize]{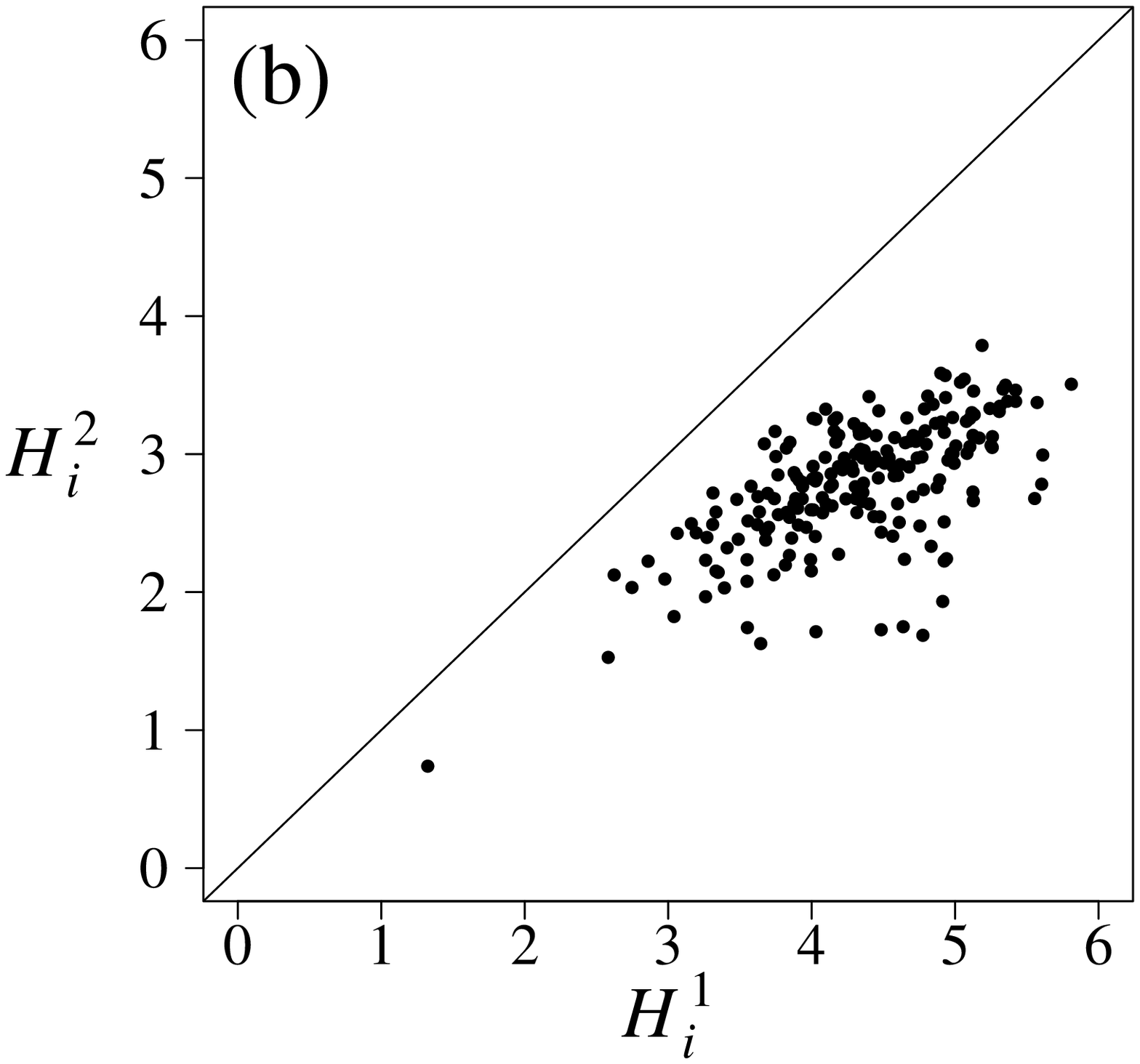}
\caption{
(a) Histograms of the entropies for $D_2$.
(b) Relationship between $H^1_i$ and $H^2_i$ in $D_2$. The solid line represents $H^1_i = H^2_i$.
}
\label{fig:hist_H_D2}
\end{figure}

\clearpage
\begin{figure}
\centering
\includegraphics[width=0.9\hsize]{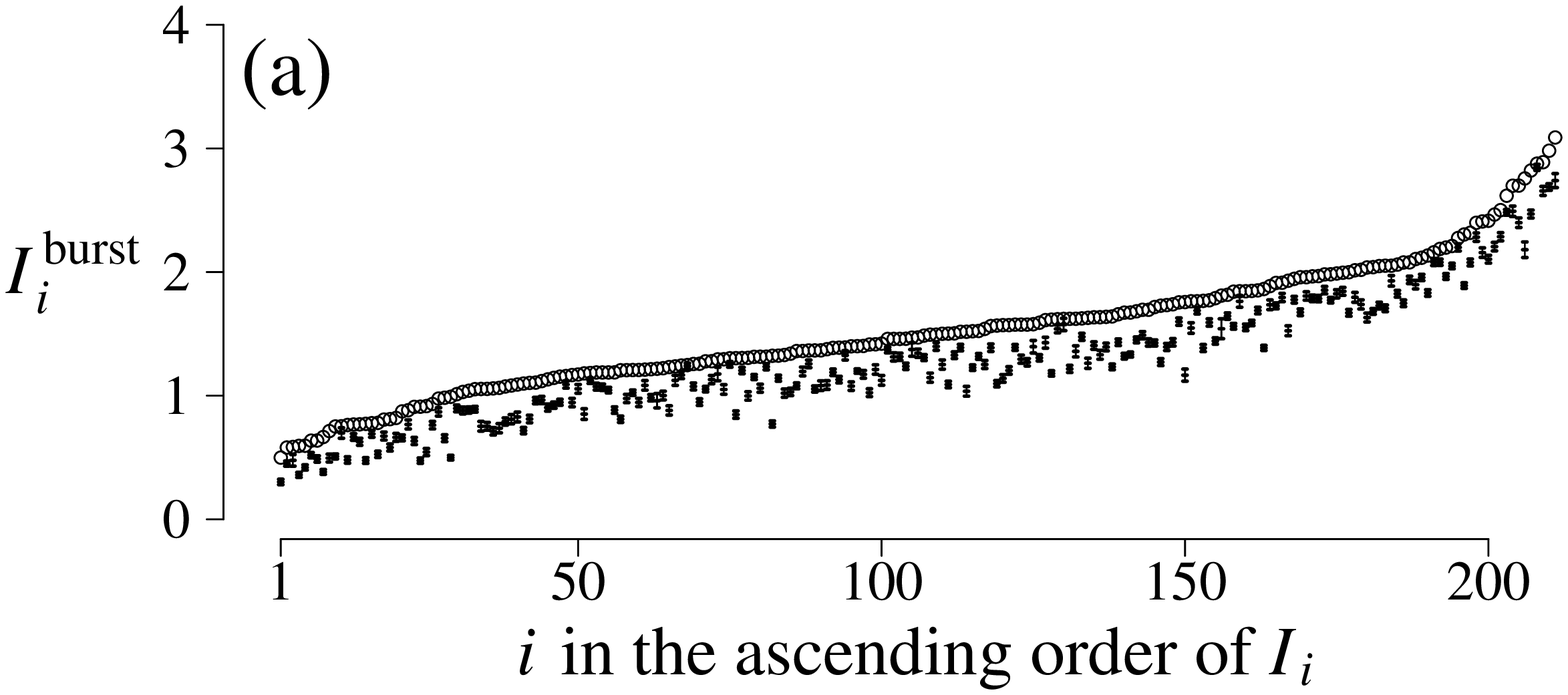}\\
\includegraphics[width=0.9\hsize]{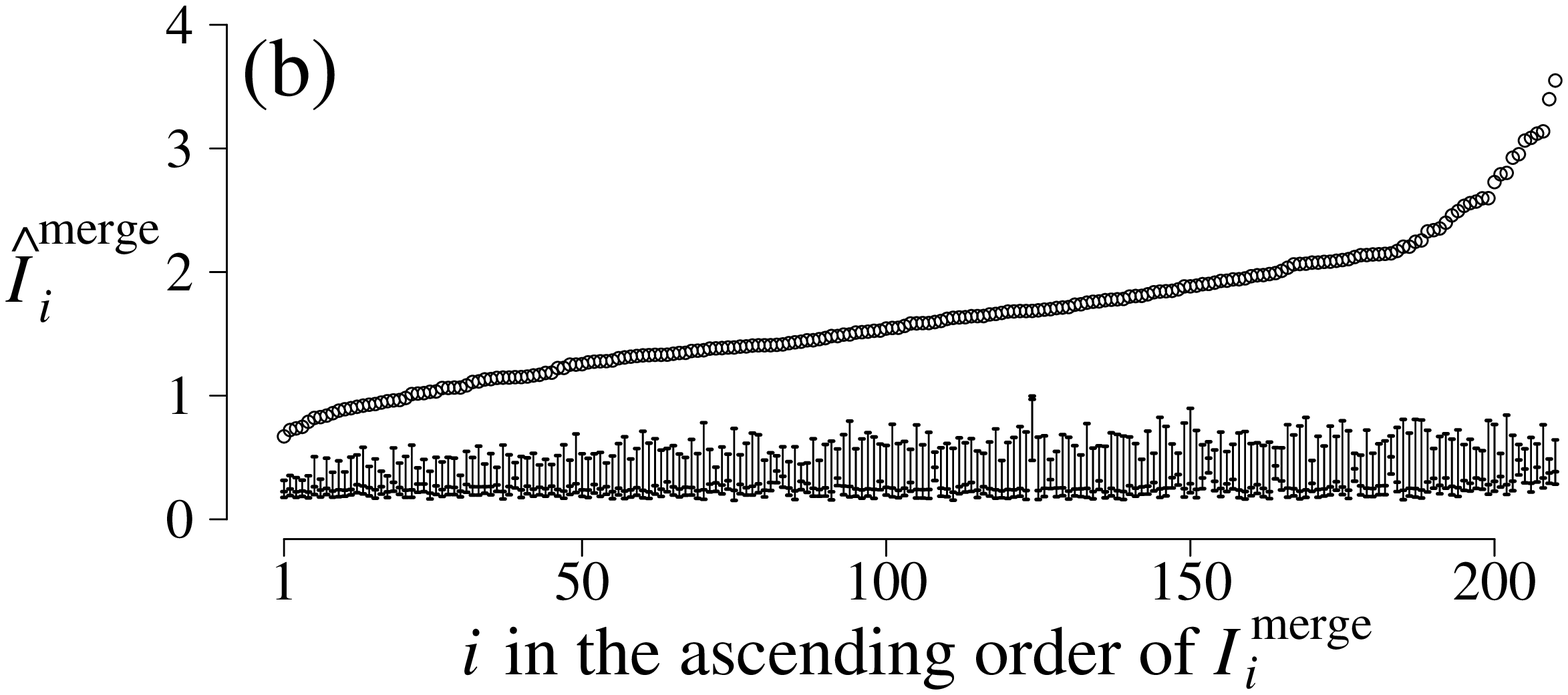}
\caption{
Results of the bootstrap tests for $D_2$ on the basis of (a) shuffling and (b) merging of the partner sequence.
See the caption of \FIG\ref{fig:IEI-SEtest} for legends.
}
\label{fig:IEI-SEtest_D2}
\end{figure}

\begin{figure}[h]
\centering
\includegraphics[width=0.45\hsize]{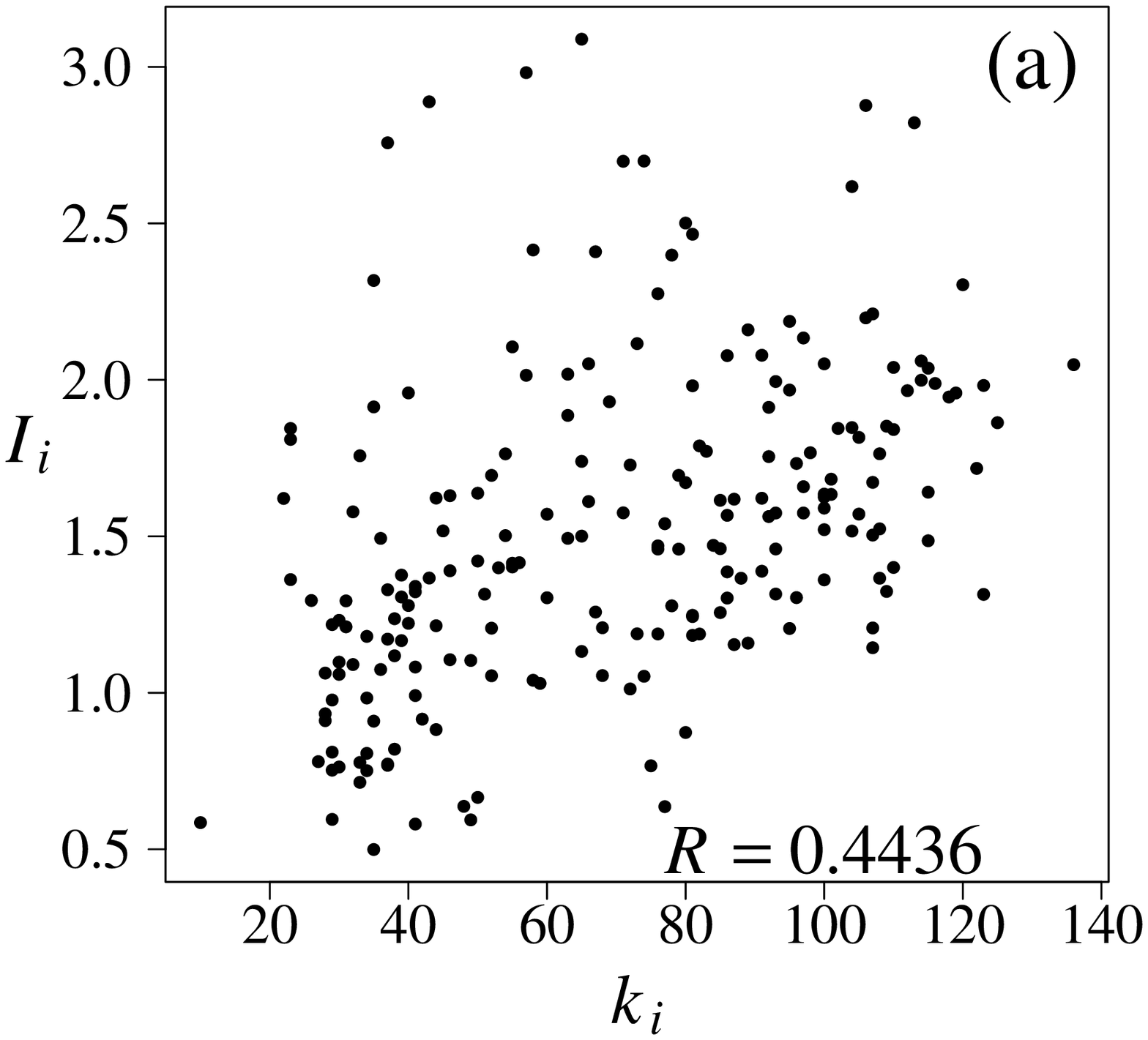}
\includegraphics[width=0.45\hsize]{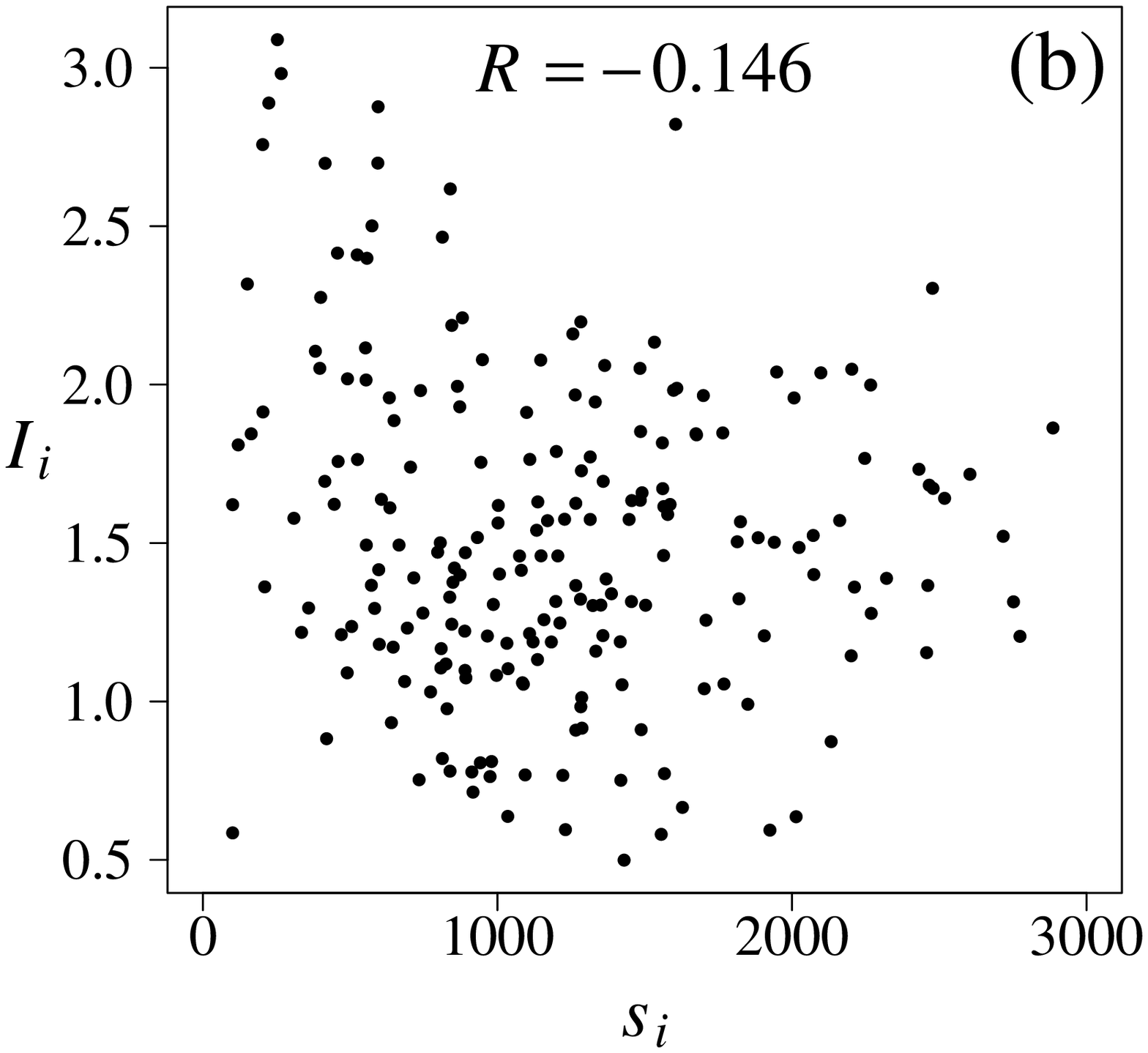}\\
\includegraphics[width=0.45\hsize]{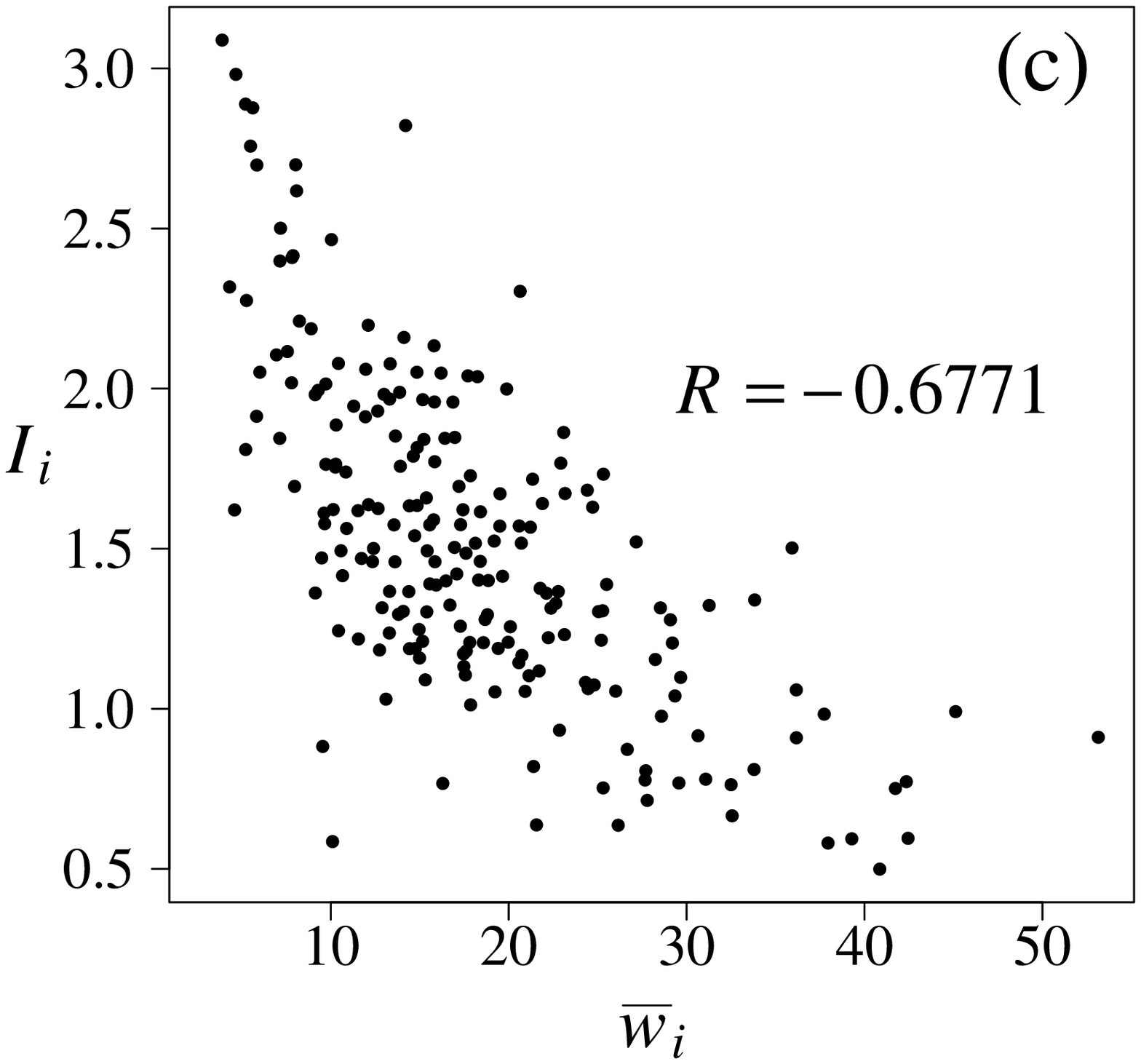}
\caption{
Mutual information $I_i$ is plotted against (a) degree $k_i$, (b) node strength $s_i$,
and (c) average node weight $\overline{w}_i$, for $D_2$.
The Pearson correlation coefficient $R$ between the plotted quantities is also shown.
}
\label{fig:corr_MI_D2}
\end{figure}

\begin{figure}[h]
\centering
\includegraphics[width=0.45\hsize]{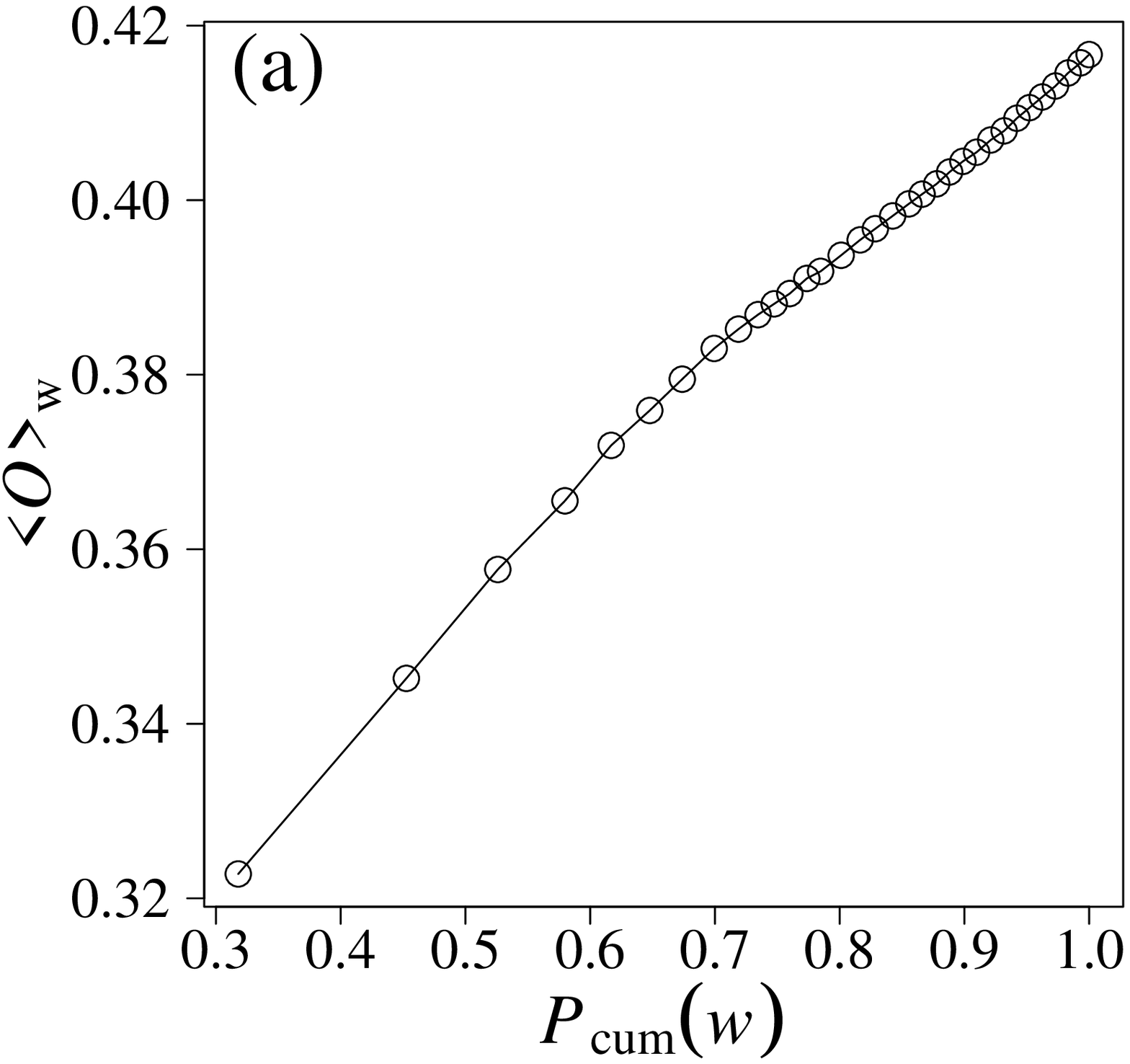}
\includegraphics[width=0.45\hsize]{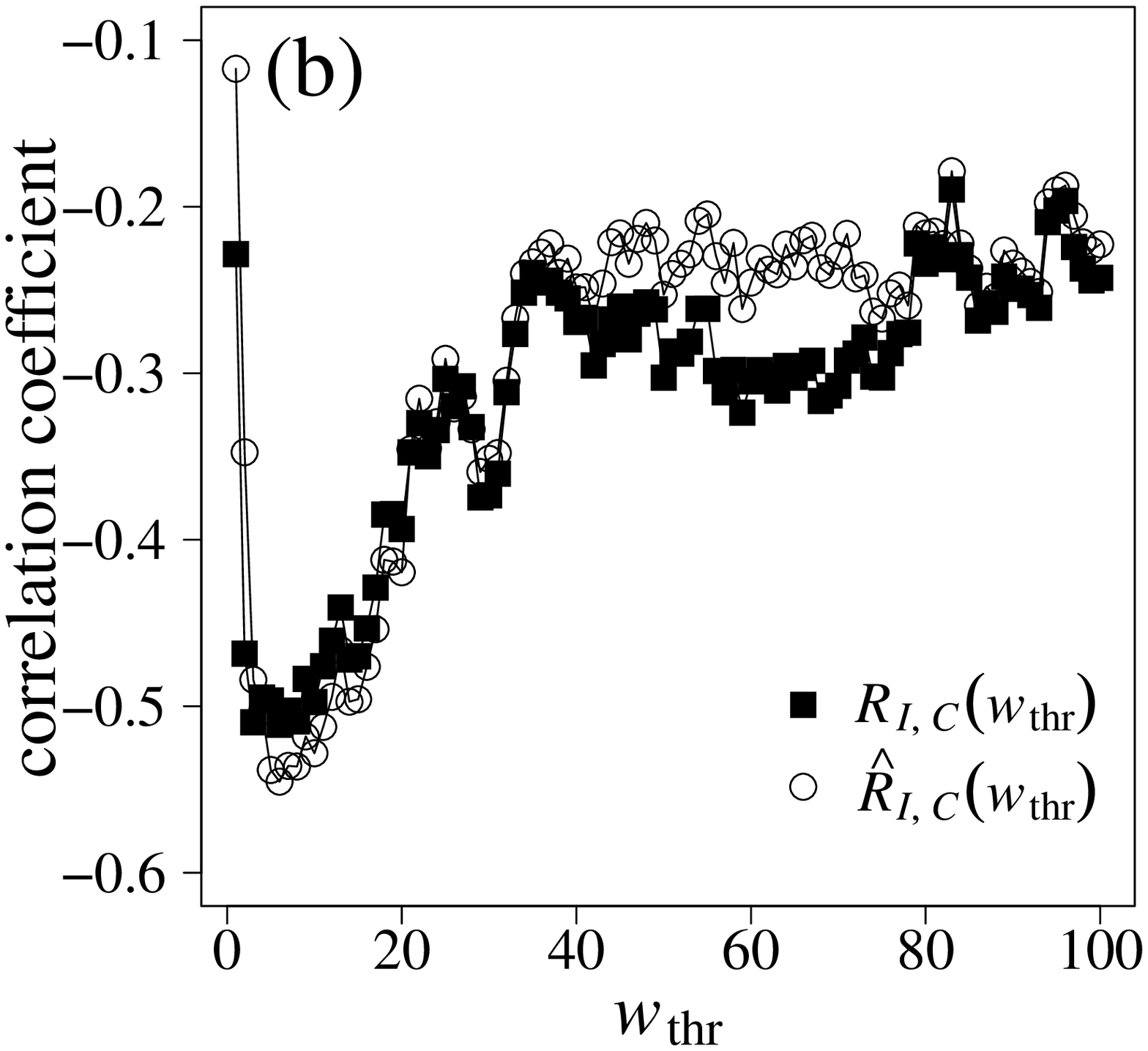}
\caption{
(a) Averaged neighborhood overlap $\langle O \rangle_w$ as a function of the fraction of links with weights smaller than $w$ for $D_2$.
(b) Pearson correlation coefficient between $I_i$ and $C_i(w_{\rm thr})$ (squares) and 
the partial correlation coefficient between them with $k_i(w_{\rm thr})$ and $s_i(w_{\rm thr})$ fixed (circles), for $D_2$.
}
\label{fig:overlap_MI.CC_D2}
\end{figure}

\begin{figure}
\centering
\includegraphics[width=0.9\hsize]{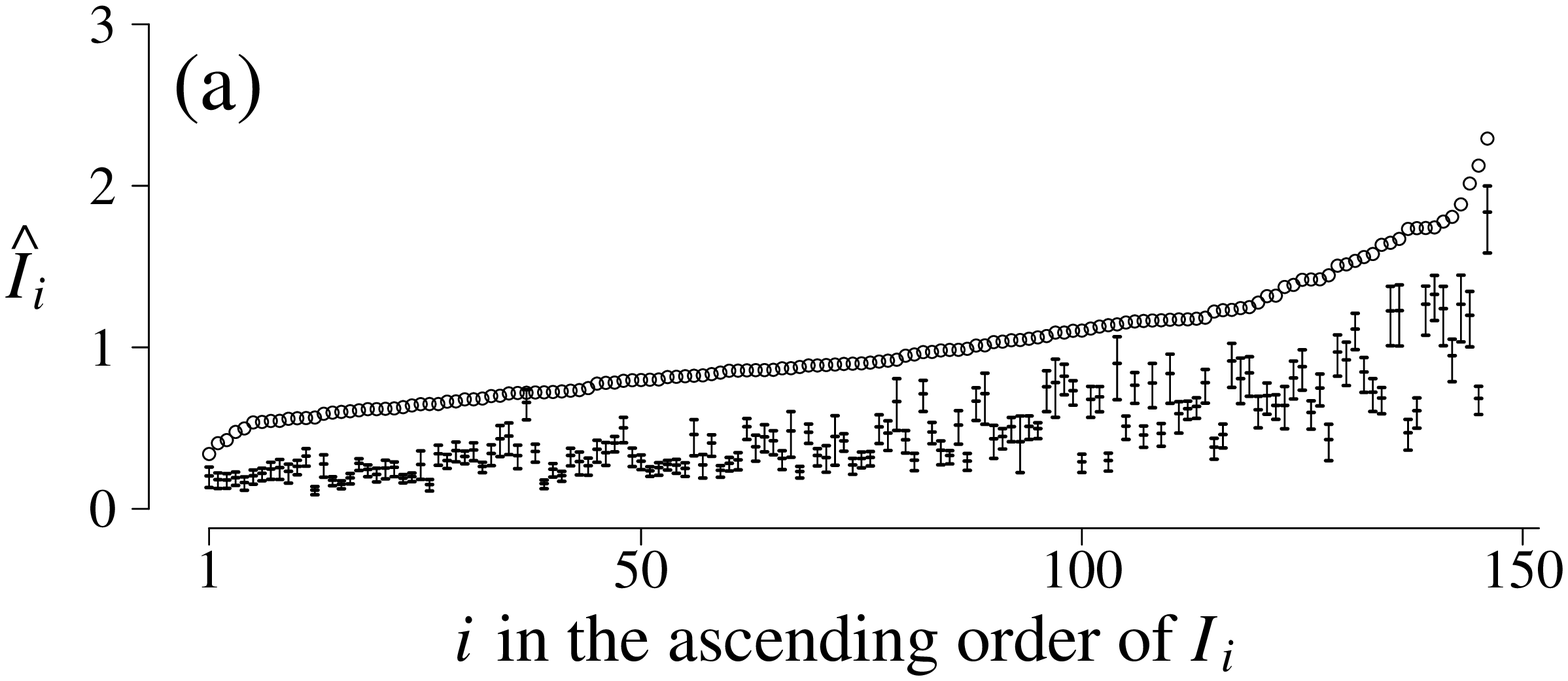}\\
\includegraphics[width=0.9\hsize]{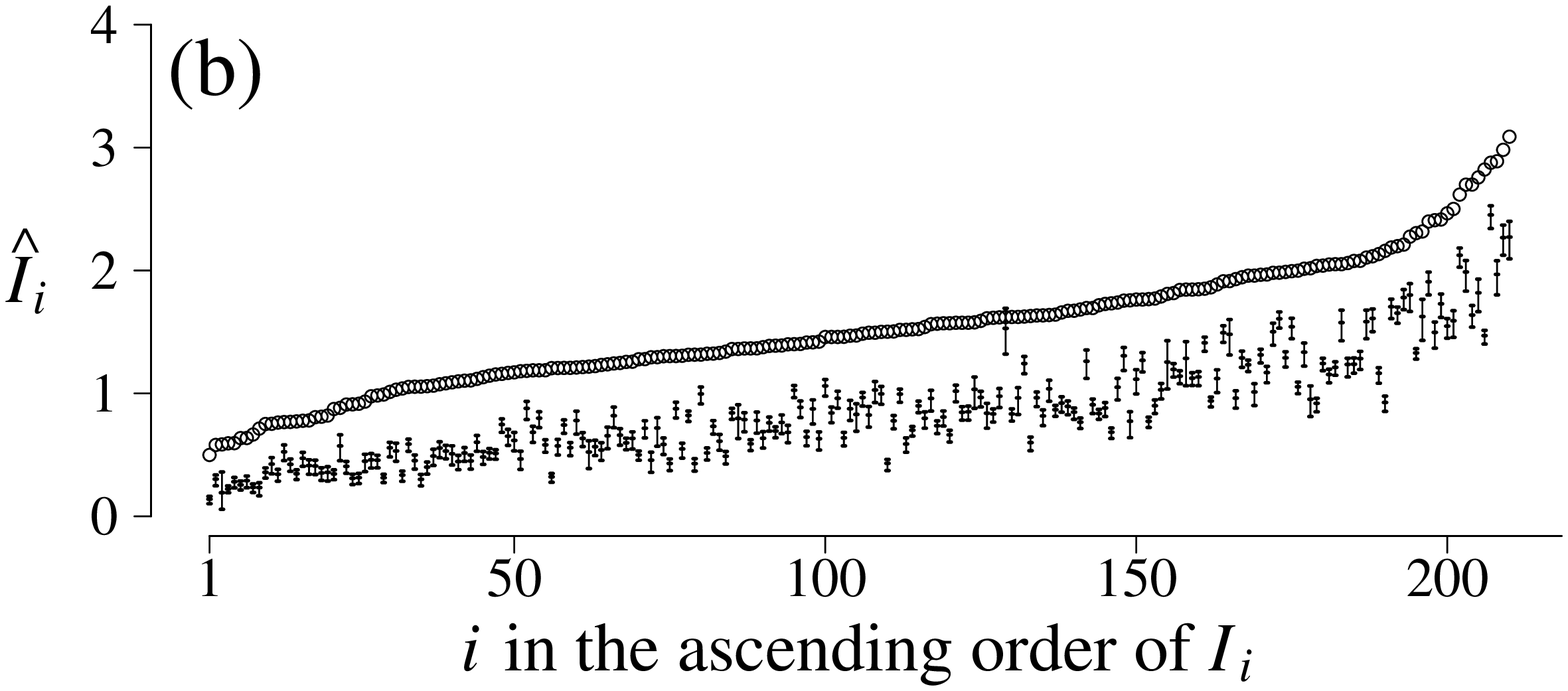}
\caption{Results of the bootstrap test of the finite size effect for (a) $D_1$ and (b) $D_2$.
$I_i$~(circles) and the confidential intervals (error bars) of individuals are plotted 
in the ascending order of $I_i$.
The lower and upper ends of the error bars represent $0$ and $99$ percentile points, respectively.
The ticks at the middle of the error bars indicate the mean.}
\label{fig:boottestMI}
\end{figure}

\clearpage
\begin{figure}[t]
\centering
\includegraphics[width=0.40\hsize]{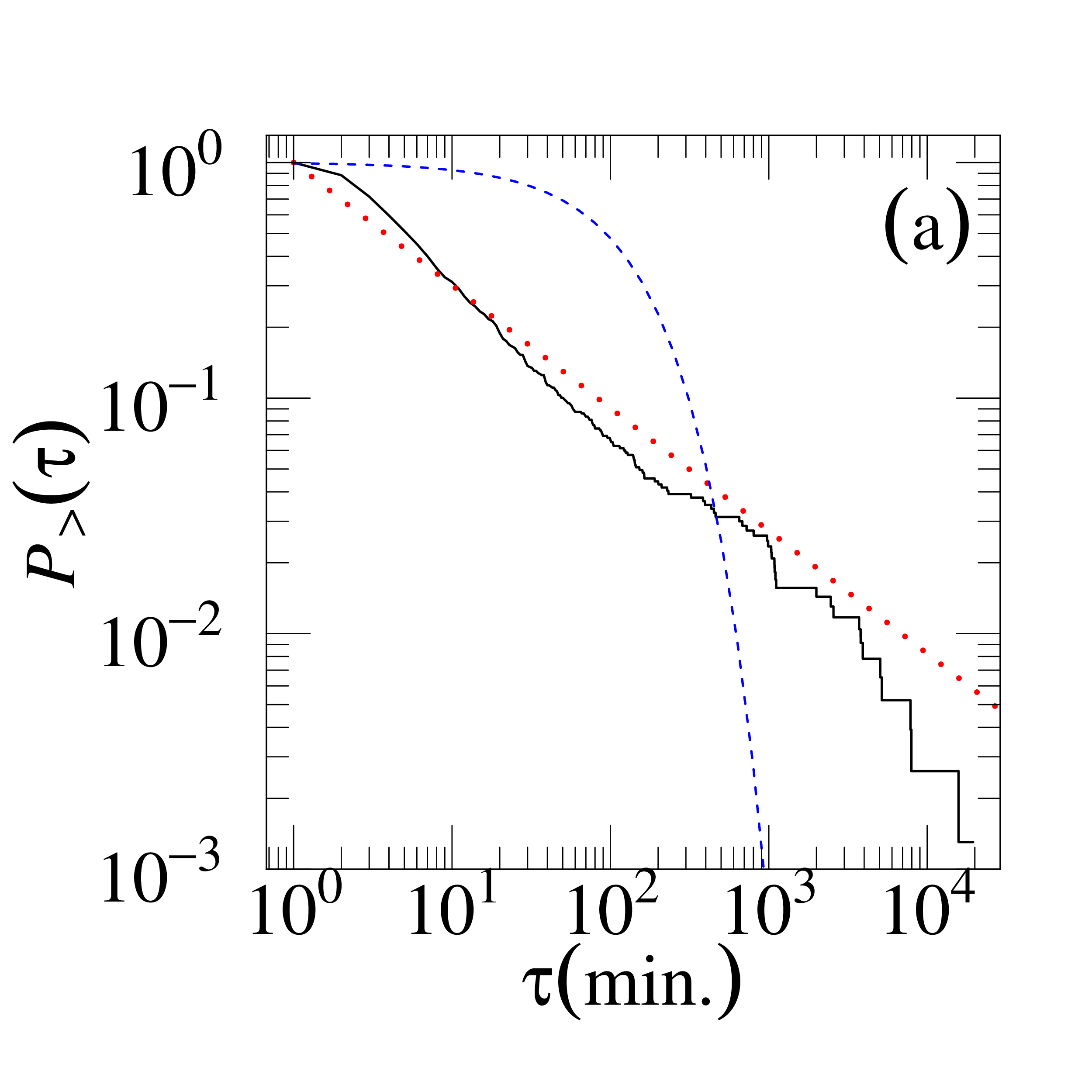}
\hspace*{10mm}
\includegraphics[width=0.43\hsize]{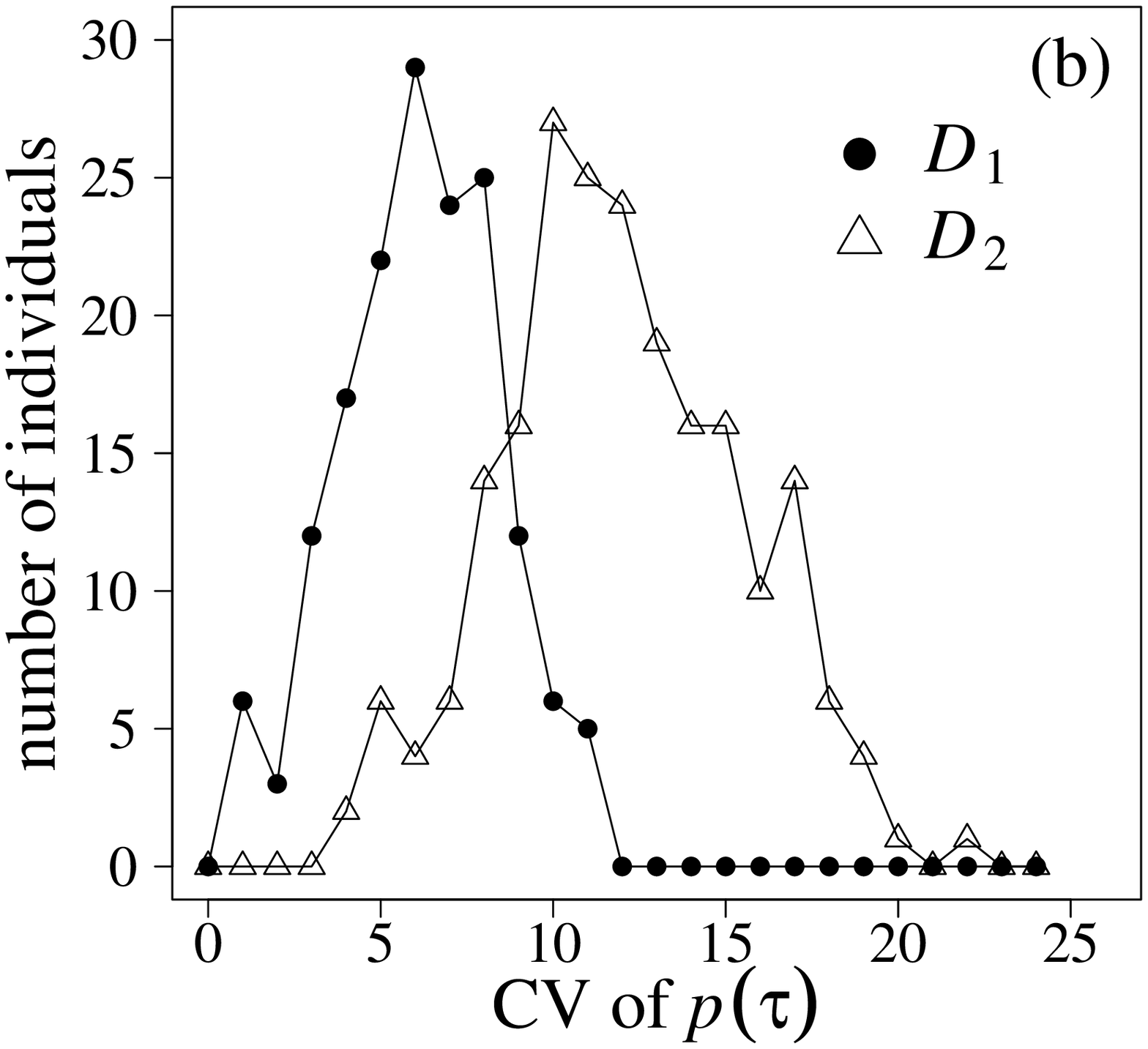}
\caption{
(a) Cumulative distribution of the interevent intervals of a typical individual in $D_1$ (solid line).
The dotted line represents the power-law fit with exponent~$-1.52$, which wa obtained from the maximum likelihood test~\cite{Clauset2009}.
The dashed line represents the exponential distribution with the same mean as that of the data.
(b) Distributions of the CV of $p(\tau)$ in $D_1$ and $D_2$.
}
\label{fig:taus}
\end{figure}

\begin{figure}
\centering
\includegraphics[width=0.45\hsize]{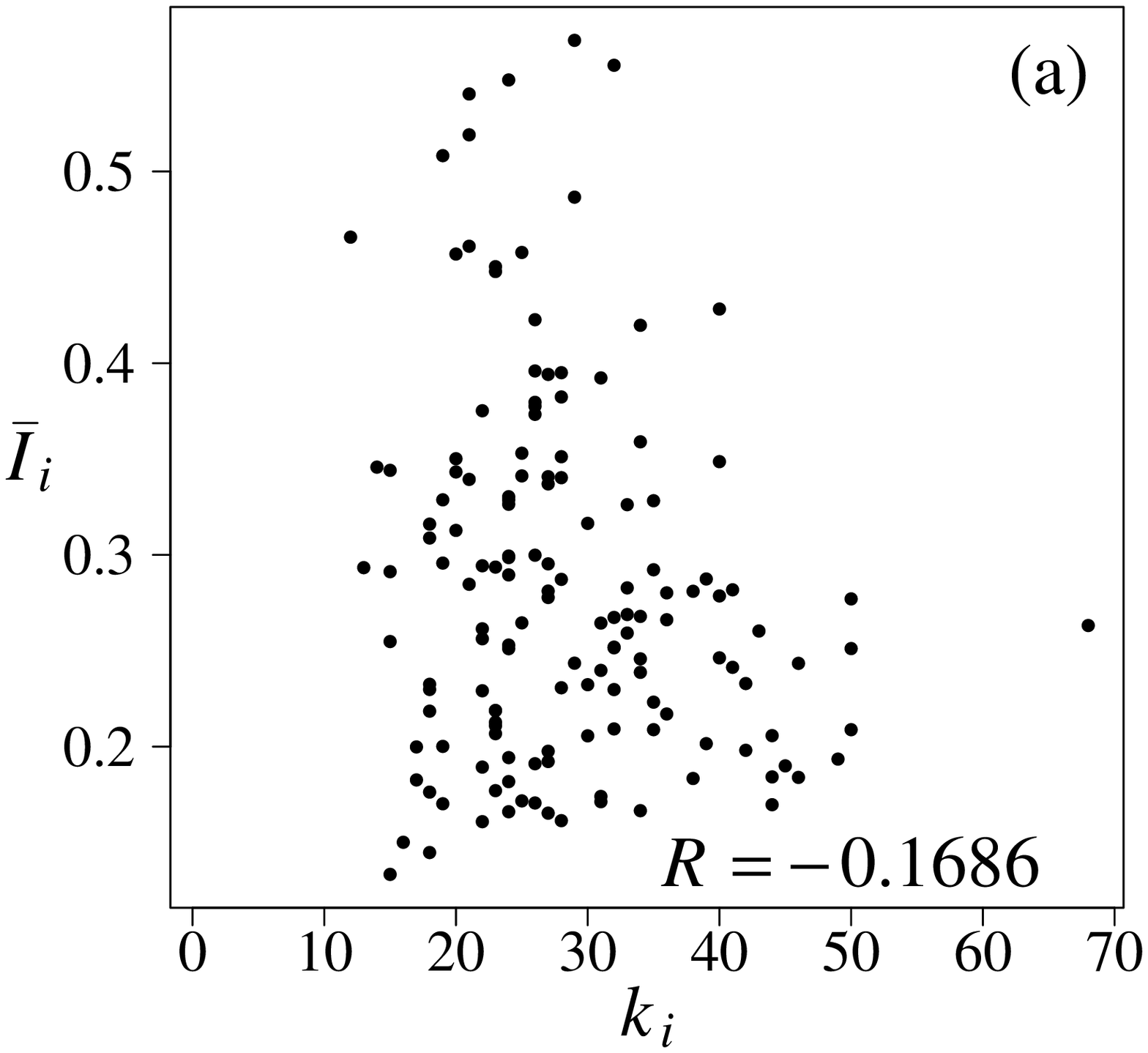}
\includegraphics[width=0.45\hsize]{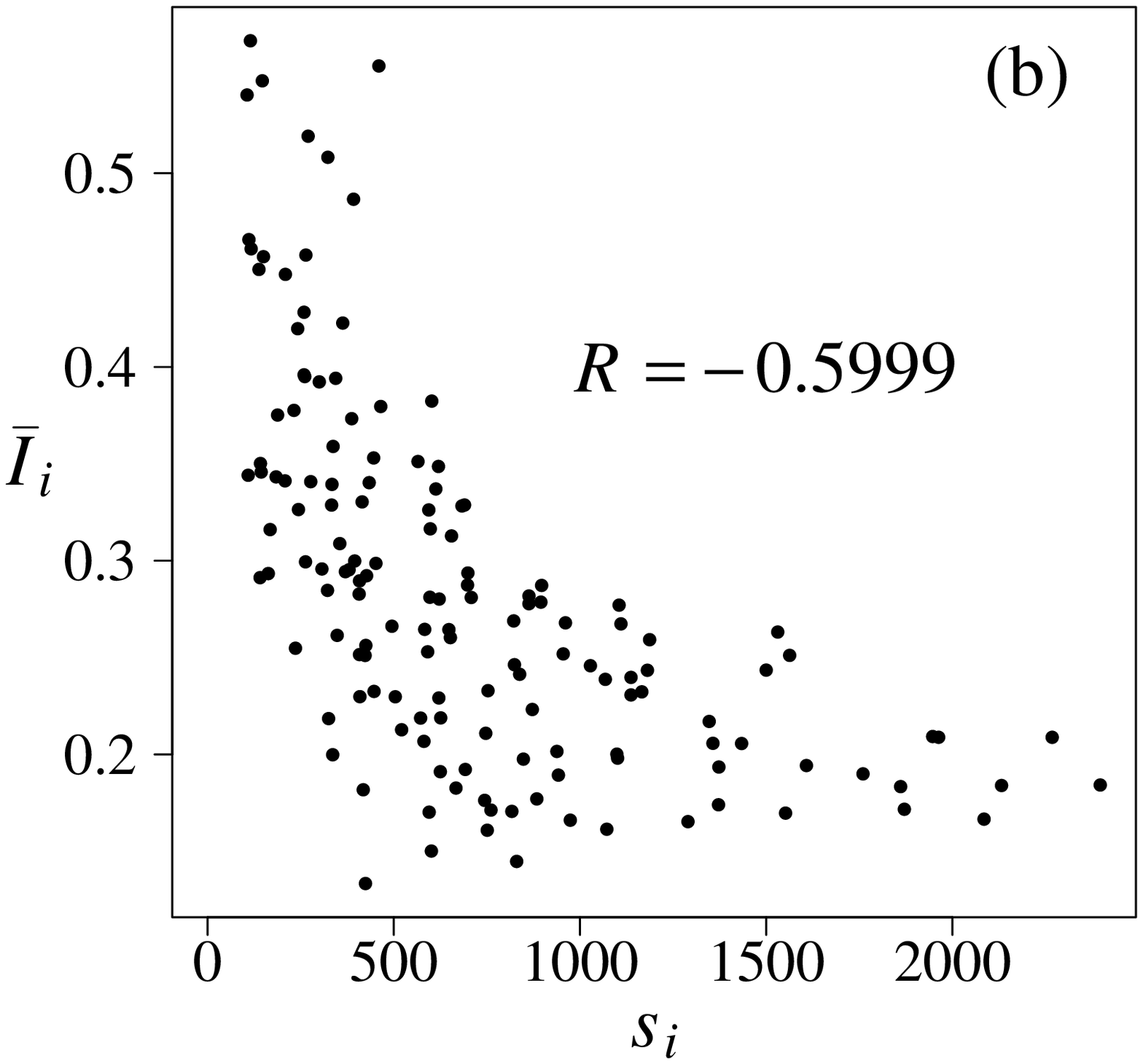}
\includegraphics[width=0.45\hsize]{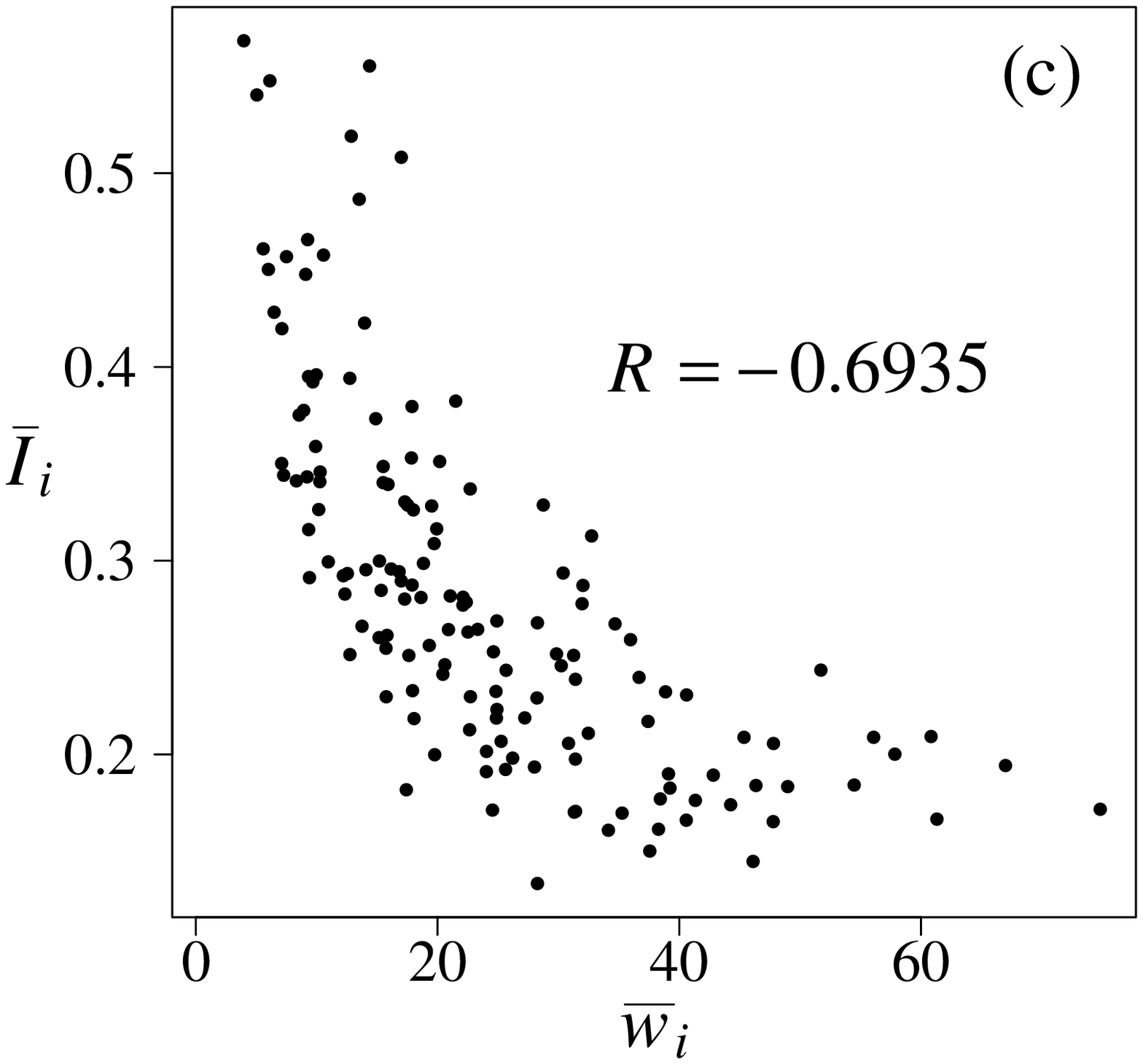}
\includegraphics[width=0.45\hsize]{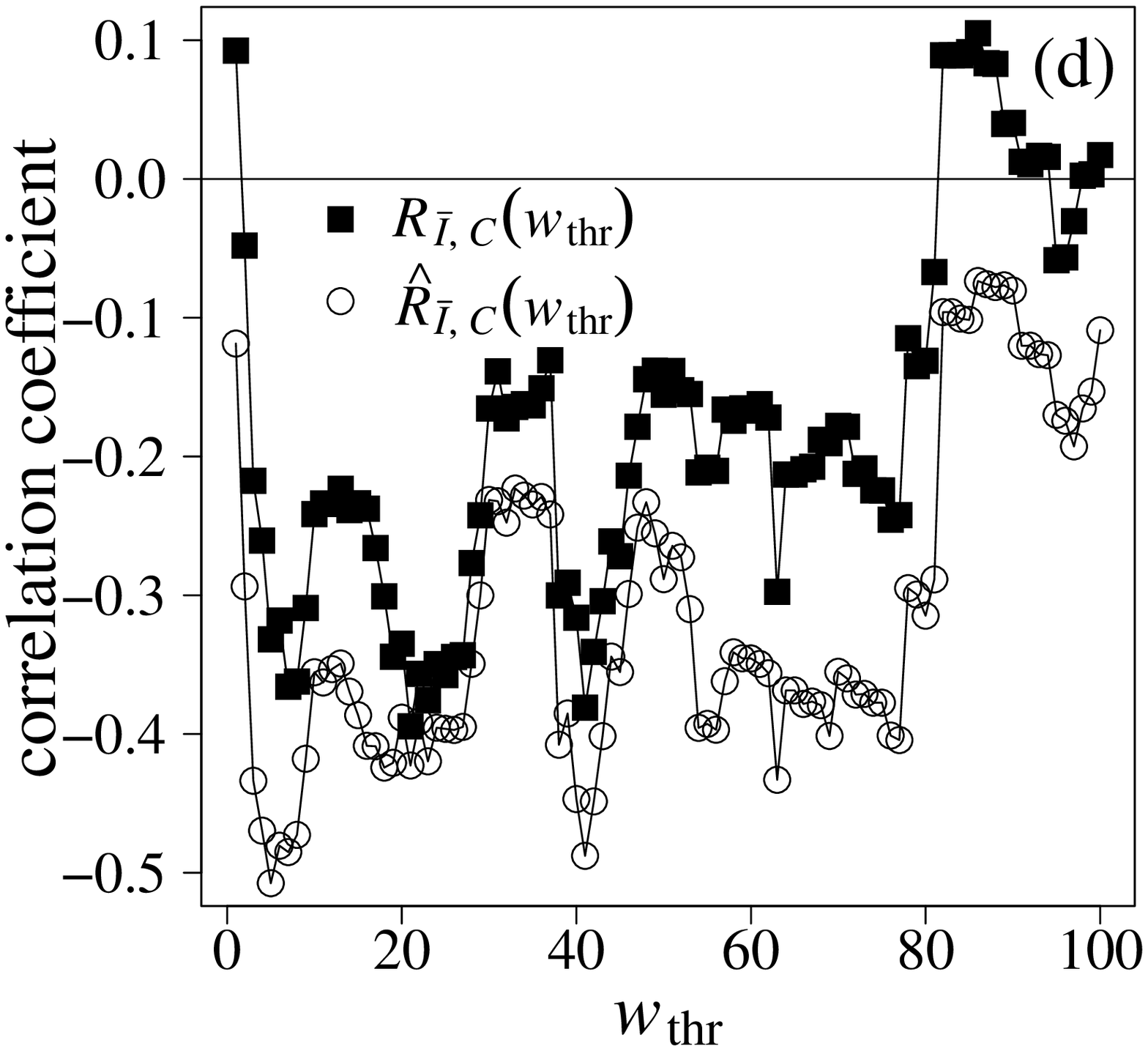}
\caption{
Normalized mutual information $\overline{I}_i$ is plotted against (a) degree $k_i$, (b) node strength $s_i$,
and (c) average node weight $\overline{w}_i$, for $D_1$.
The Pearson correlation coefficient $R$ between the plotted quantities is also shown.
(d) Pearson correlation coefficient between $\overline{I}_i$ and $C_i(w_{\rm thr})$ (squares) and 
the partial correlation coefficient between them with $k_i(w_{\rm thr})$ and $s_i(w_{\rm thr})$ fixed (circles).
}
\label{fig:normMI}
\end{figure}

\begin{figure}[h]
\centering
\includegraphics[width=0.45\hsize]{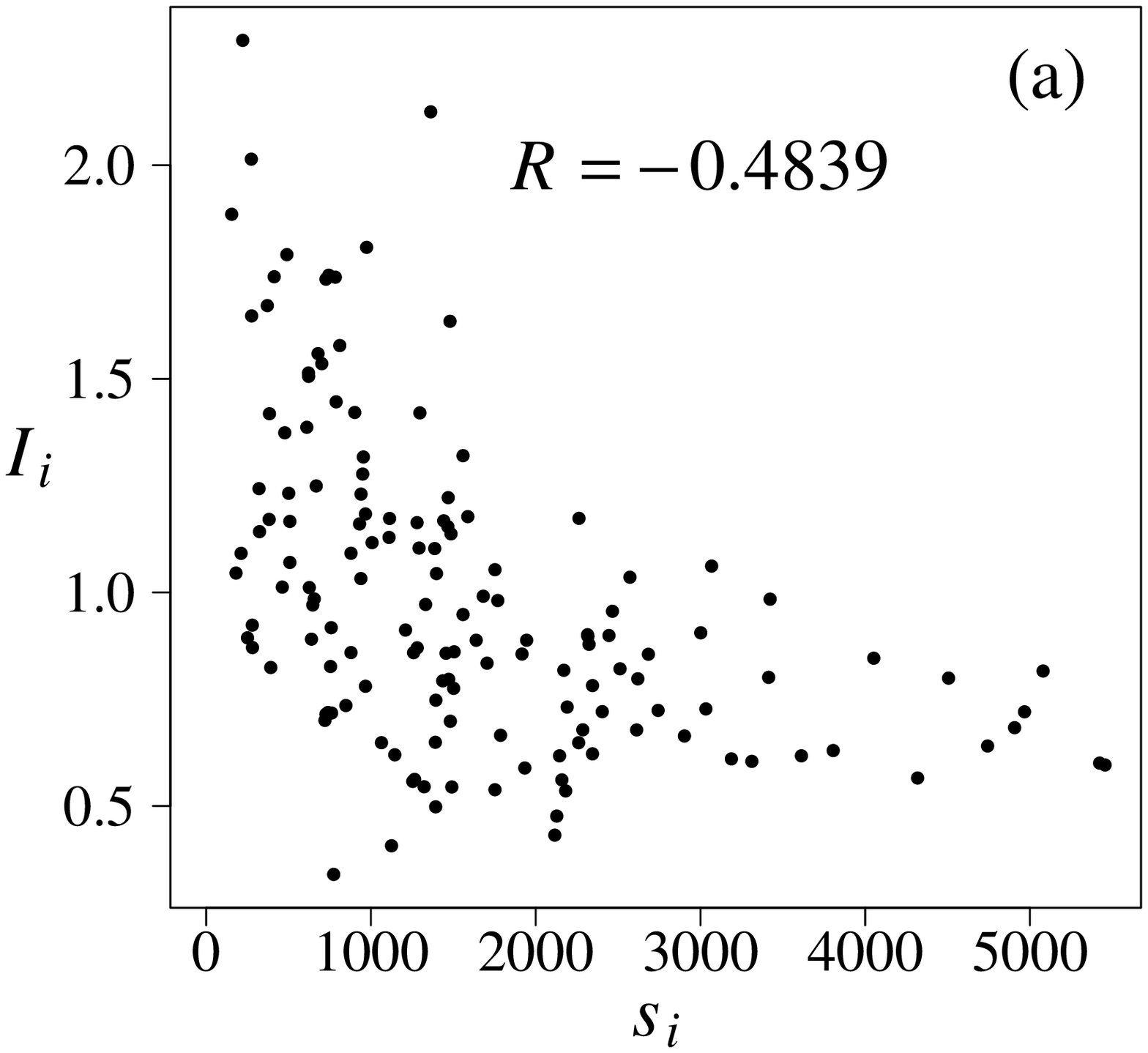}
\includegraphics[width=0.45\hsize]{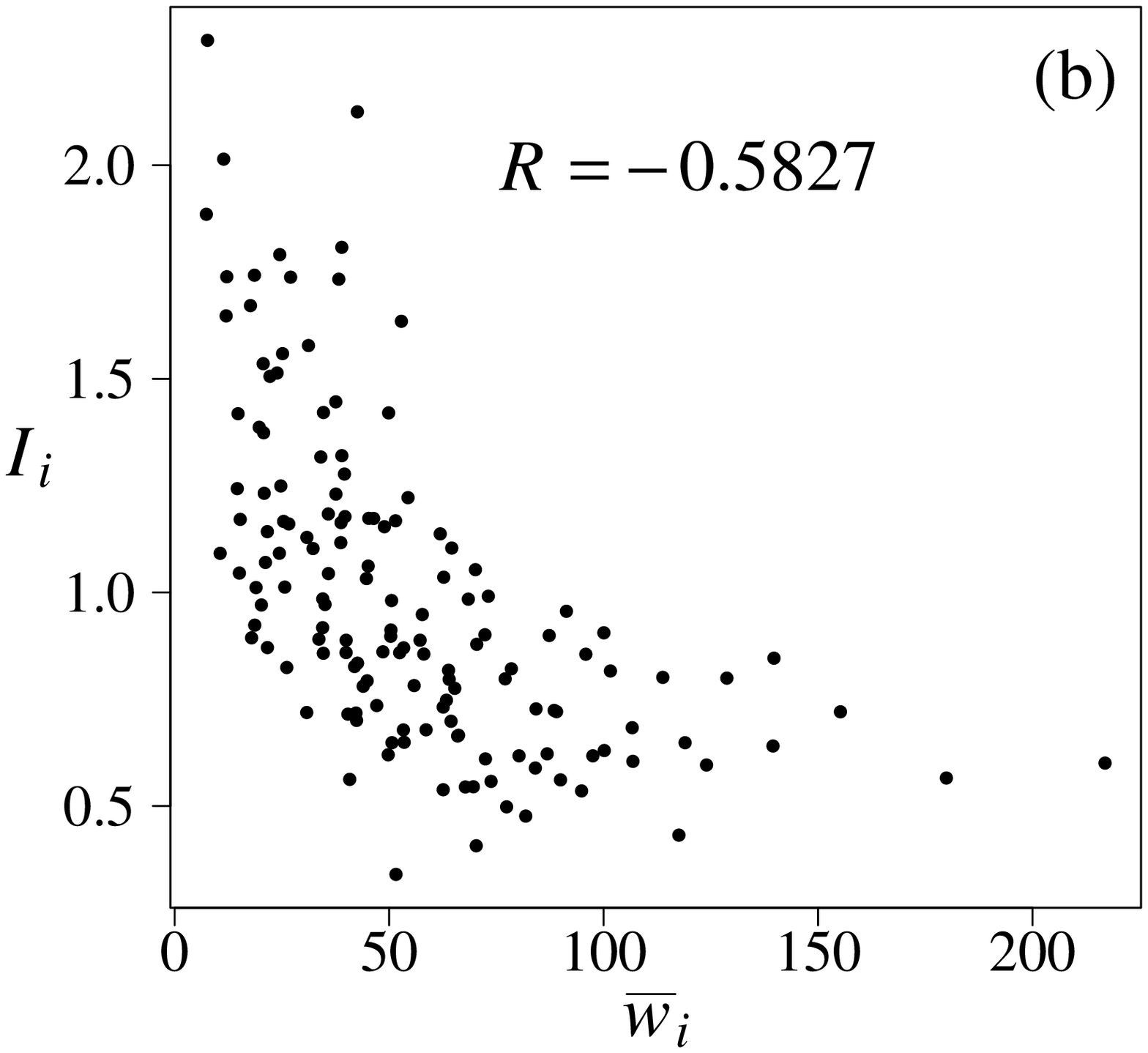}
\includegraphics[width=0.45\hsize]{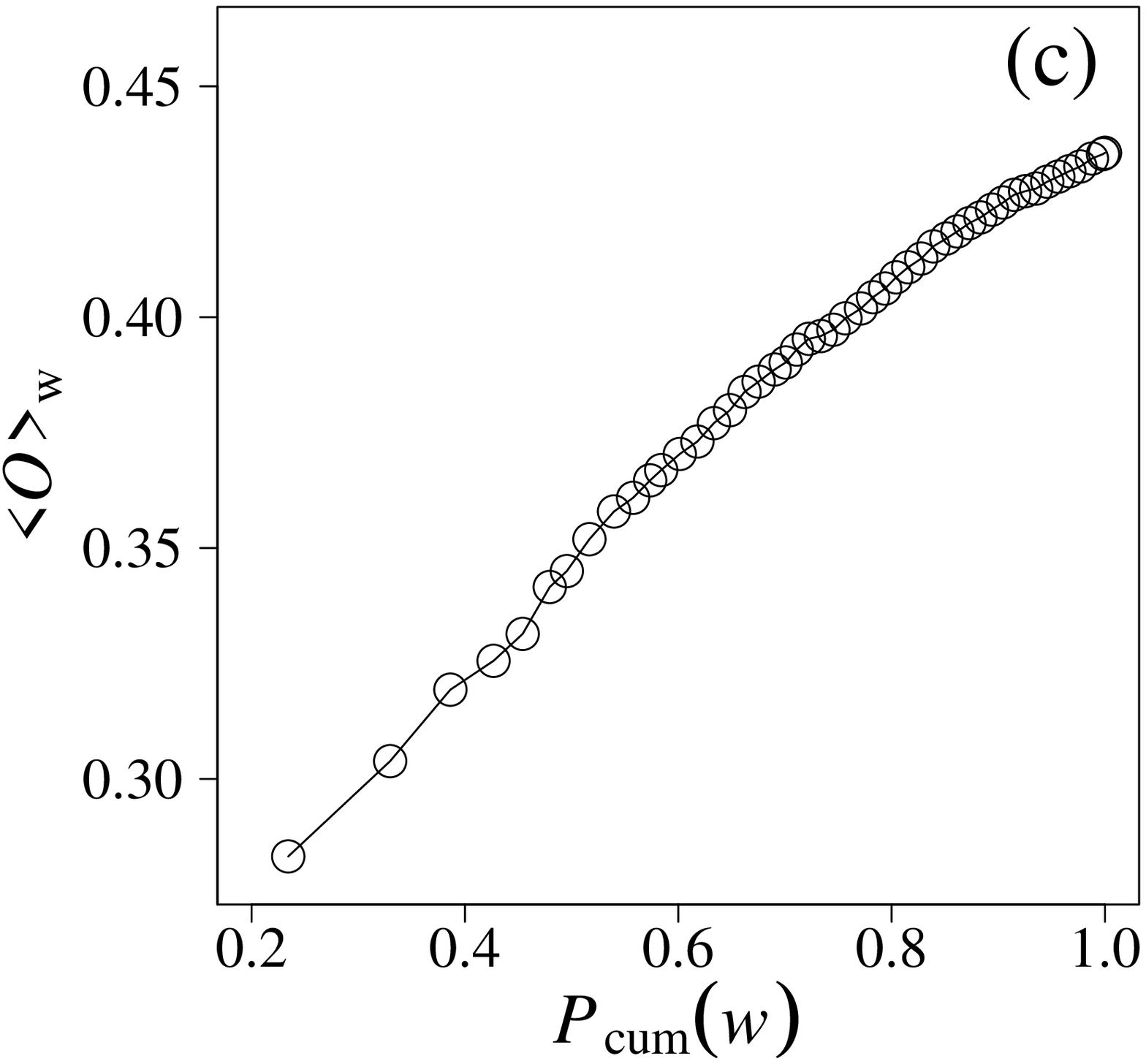}
\includegraphics[width=0.45\hsize]{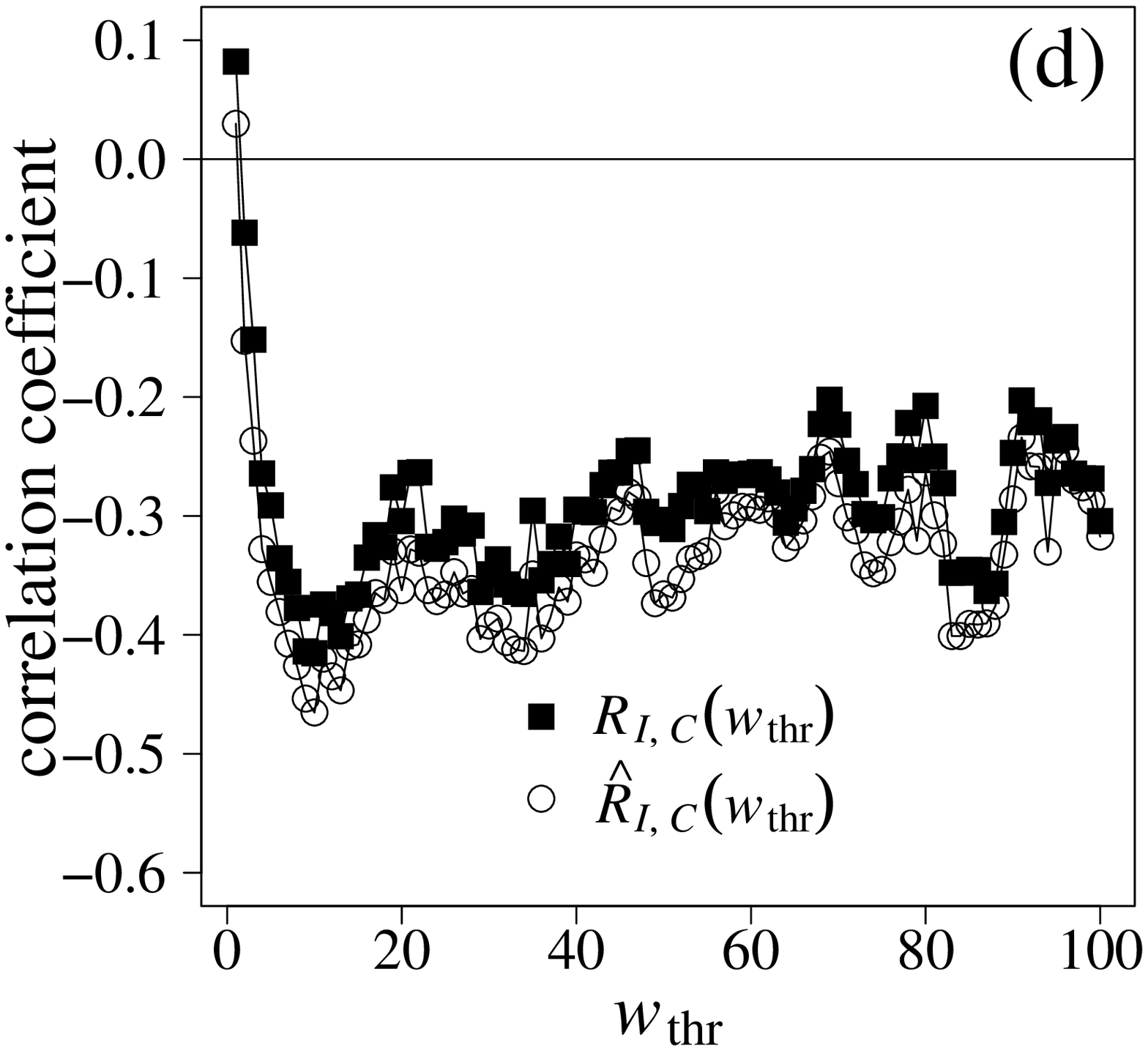}
\caption{Results when the link weight is defined by the total duration of the conversation events for each pair, for $D_1$.
The mutual information $I_i$ is plotted against (a) node strength $s_i$ and (b) average node weight $\overline{w}_i$.
The Pearson correlation coefficient $R$ between the plotted quantities is also shown.
(c) Averaged neighborhood overlap $\langle O \rangle_w$ as a function of the fraction of links with weights smaller than $w$.
(d) Pearson correlation coefficient between $I_i$ and $C_i(w_{\rm thr})$ (squares) and 
the partial correlation coefficient between them with $k_i(w_{\rm thr})$ and $s_i(w_{\rm thr})$ fixed (circles).
}
\label{fig:duration_weight}
\end{figure}

\begin{figure}[h]
\centering
\includegraphics[width=0.45\hsize]{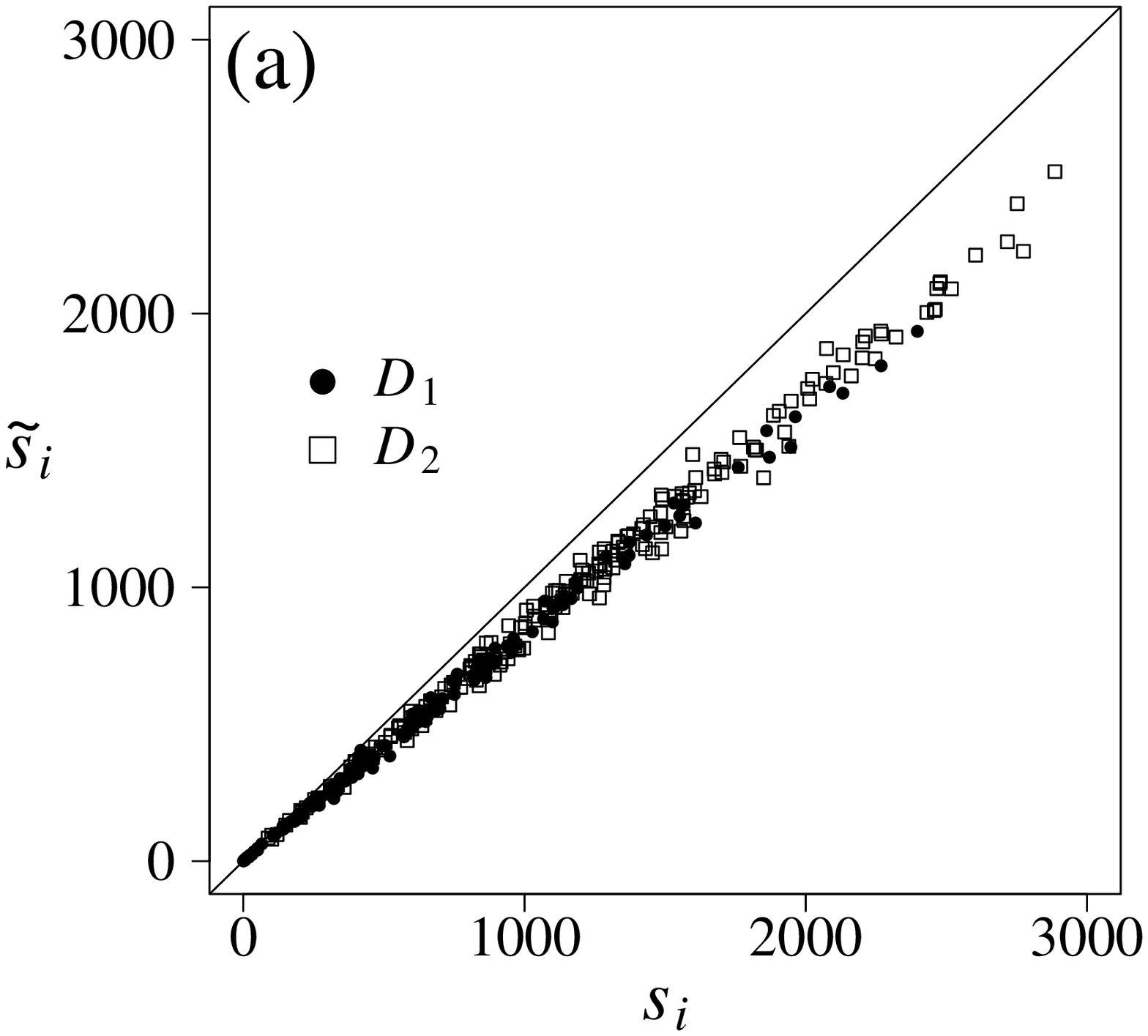}
\includegraphics[width=0.45\hsize]{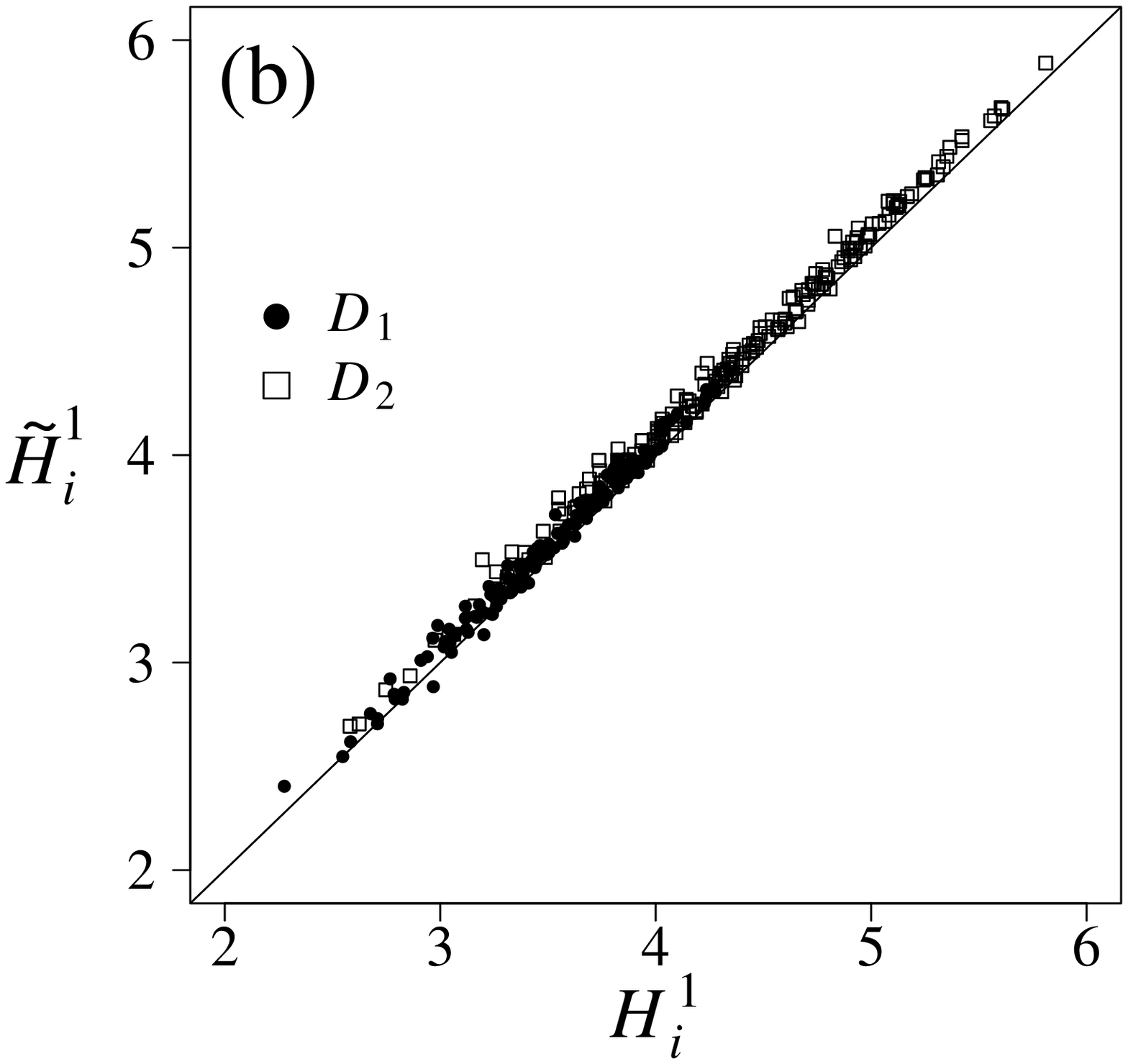}
\includegraphics[width=0.45\hsize]{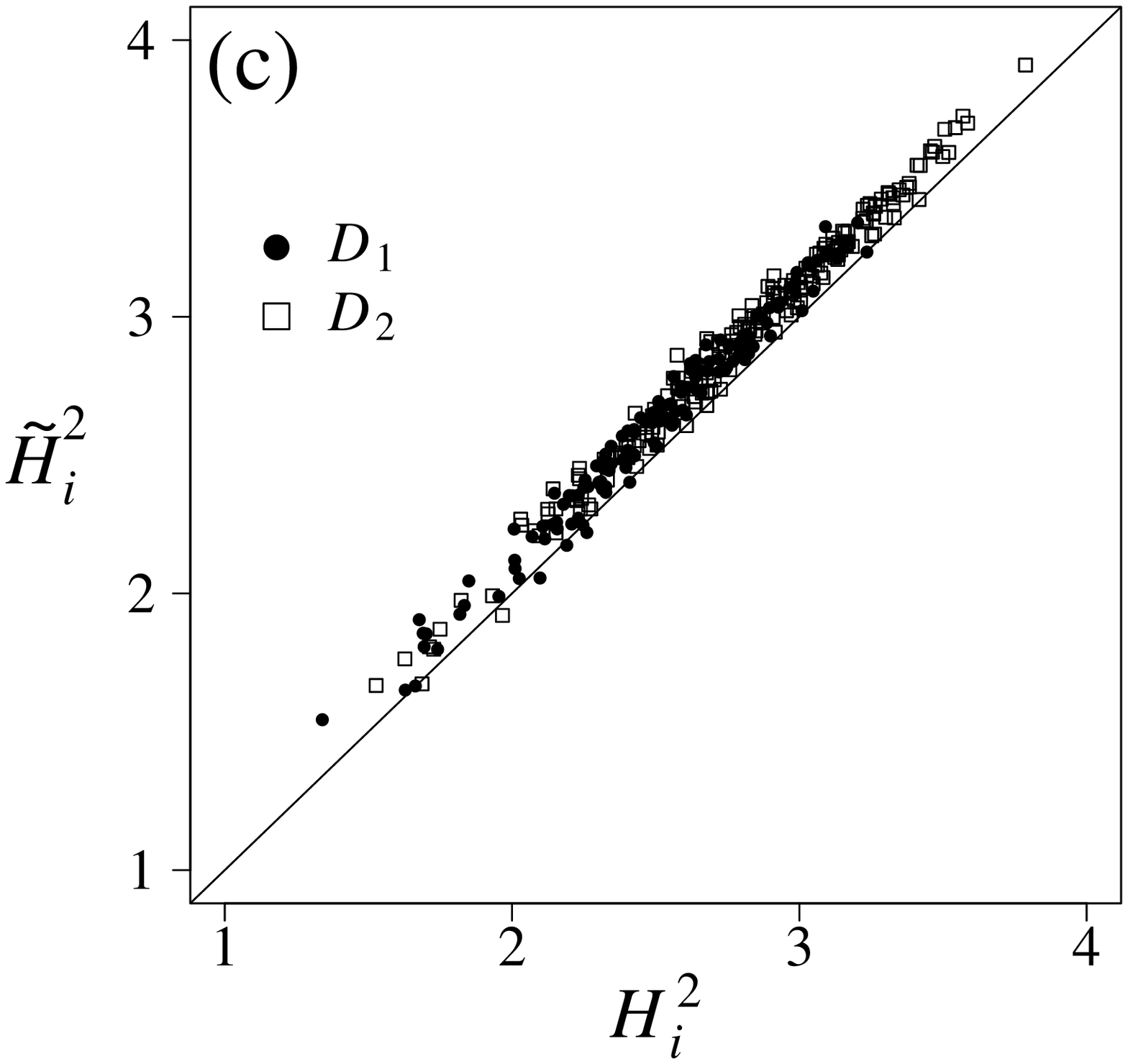}
\includegraphics[width=0.45\hsize]{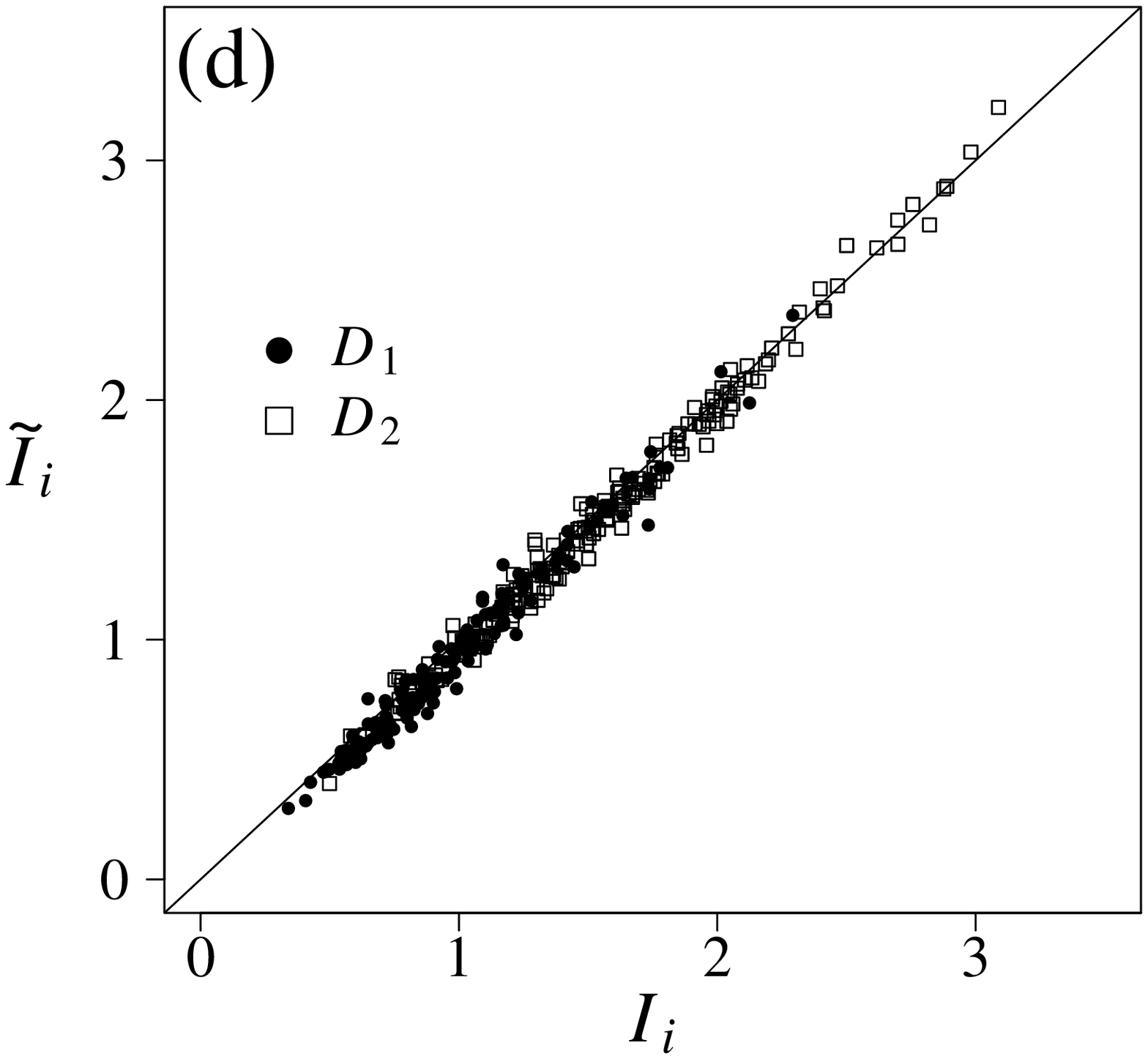}
\caption{Comparison between the interpolated and original data for data sets $D_1$ and $D_2$.
(a) Node strength, (b) uncorrelated entropy, (c) conditional entropy, and (d) mutual information 
for the interpolated data with $m=1$ are plotted against those without interpolation (\ie, original data sets). }
\label{fig:compare_m1}
\end{figure}

\begin{figure}
\centering
\includegraphics[width=0.45\hsize]{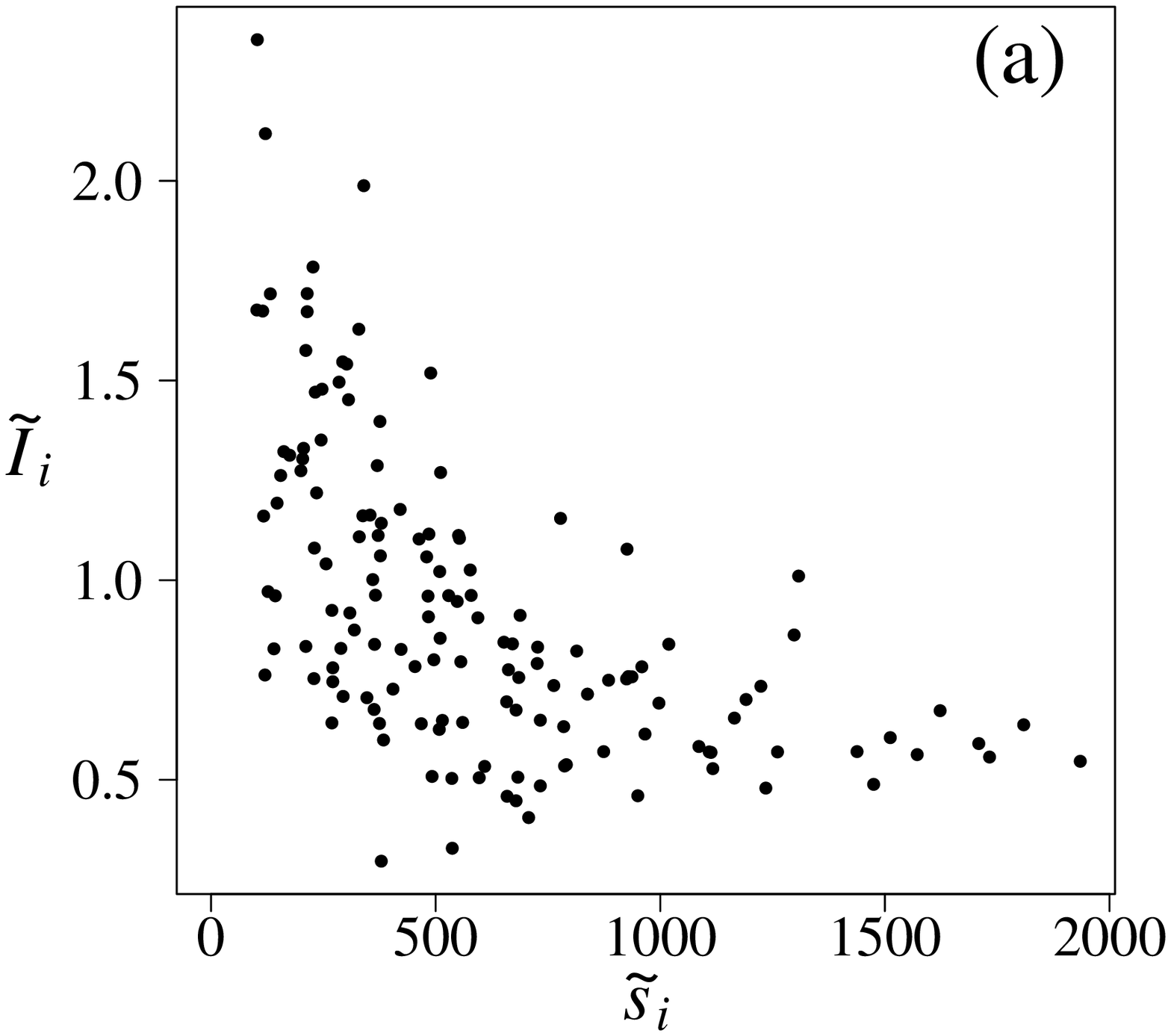}
\includegraphics[width=0.45\hsize]{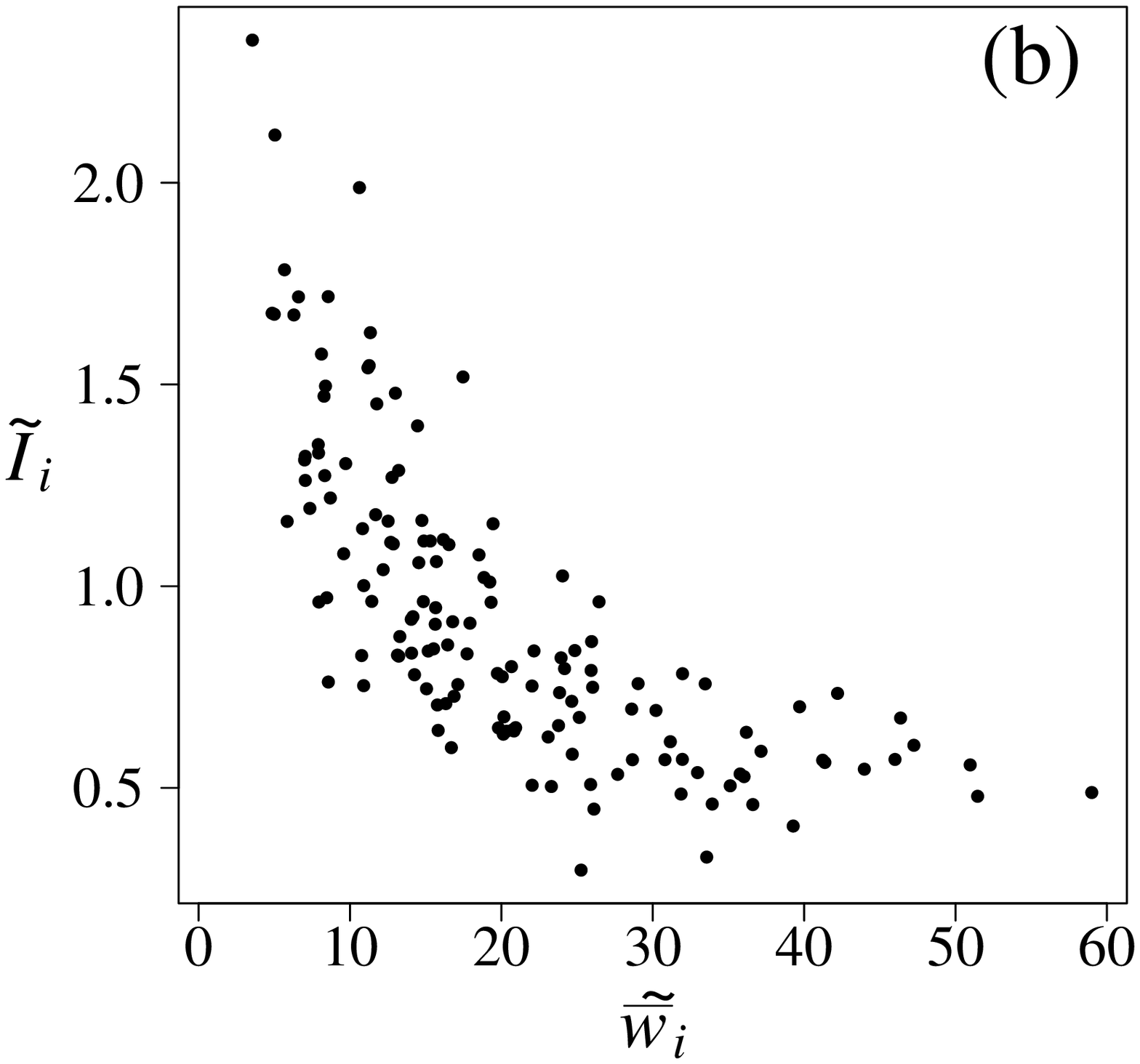}
\includegraphics[width=0.45\hsize]{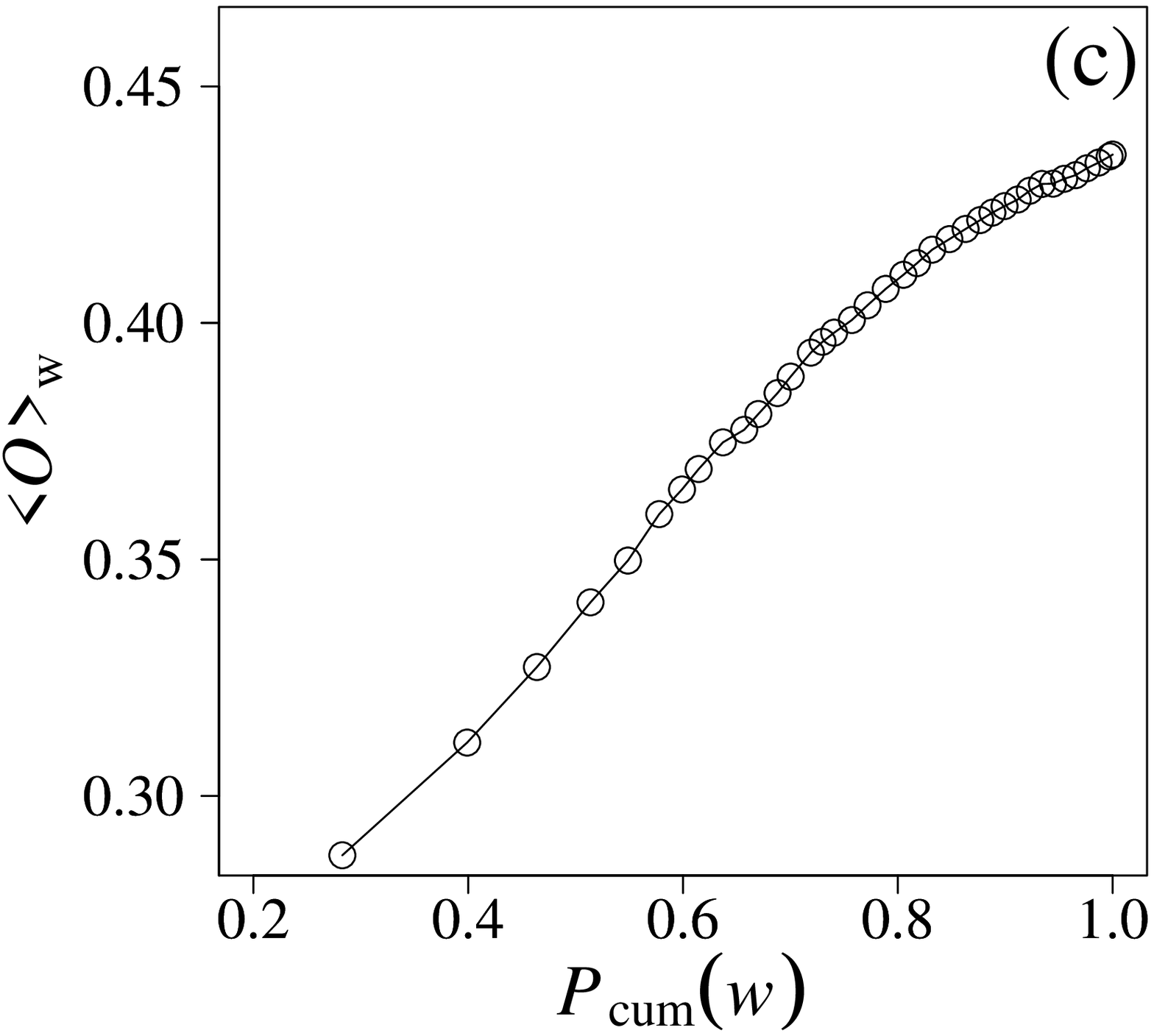}
\includegraphics[width=0.45\hsize]{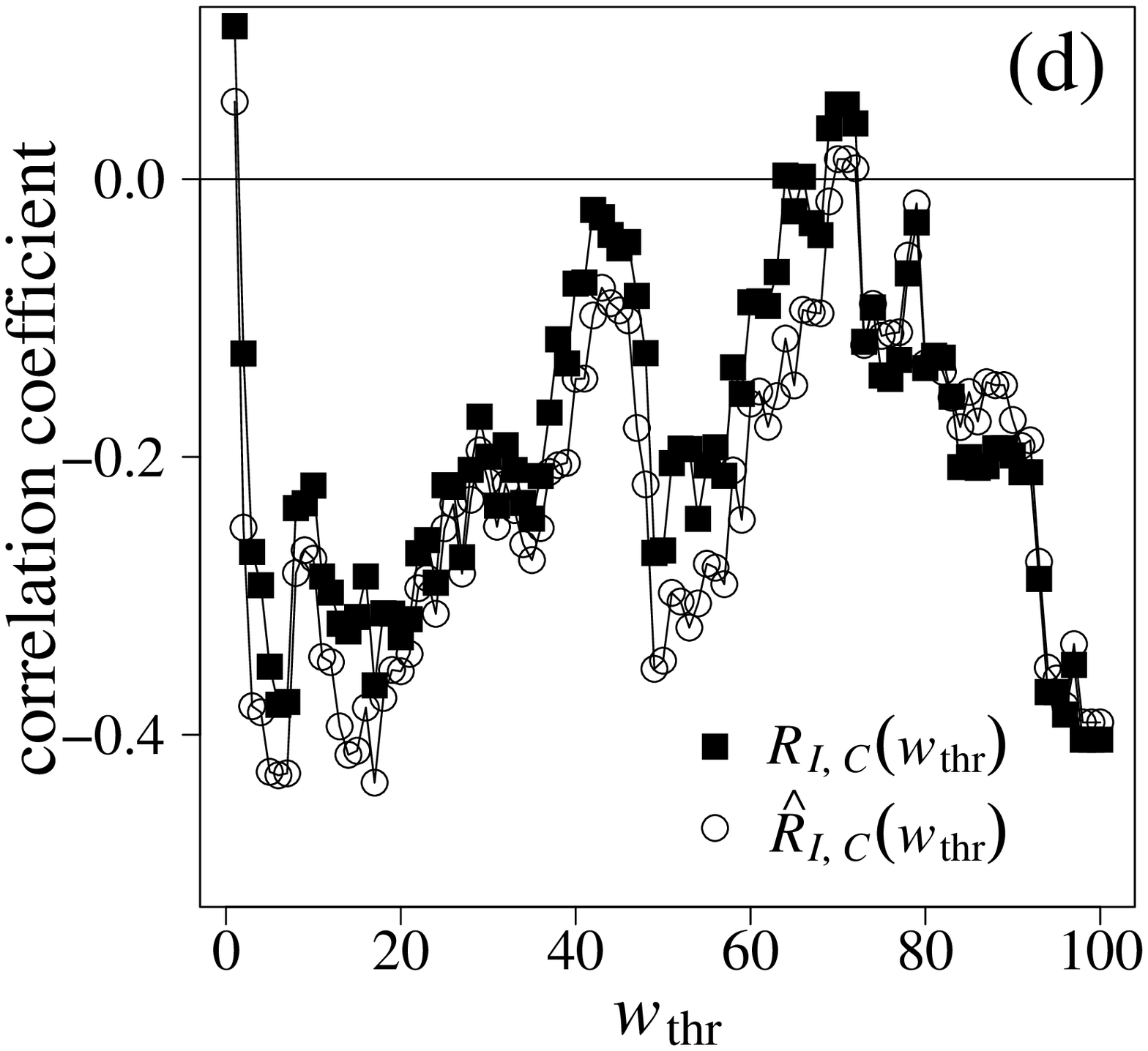}
\caption{
Results for the interpolated data with $m=1$.
We use data set~$D_1$.
The mutual information is plotted against (a) node strength and (b) mean weight.
(c) Averaged neighborhood overlap $\langle O \rangle_w$
as a function of the fraction of links with weights smaller than $w$.
(d) Pearson correlation coefficient (squares) and partial correlation coefficient (circles)
between $I_i$ and $C_i(w_{\rm thr})$.}
\label{fig:confirm_m1}
\end{figure}

\begin{figure}[h]
\centering
\includegraphics[width=0.45\hsize]{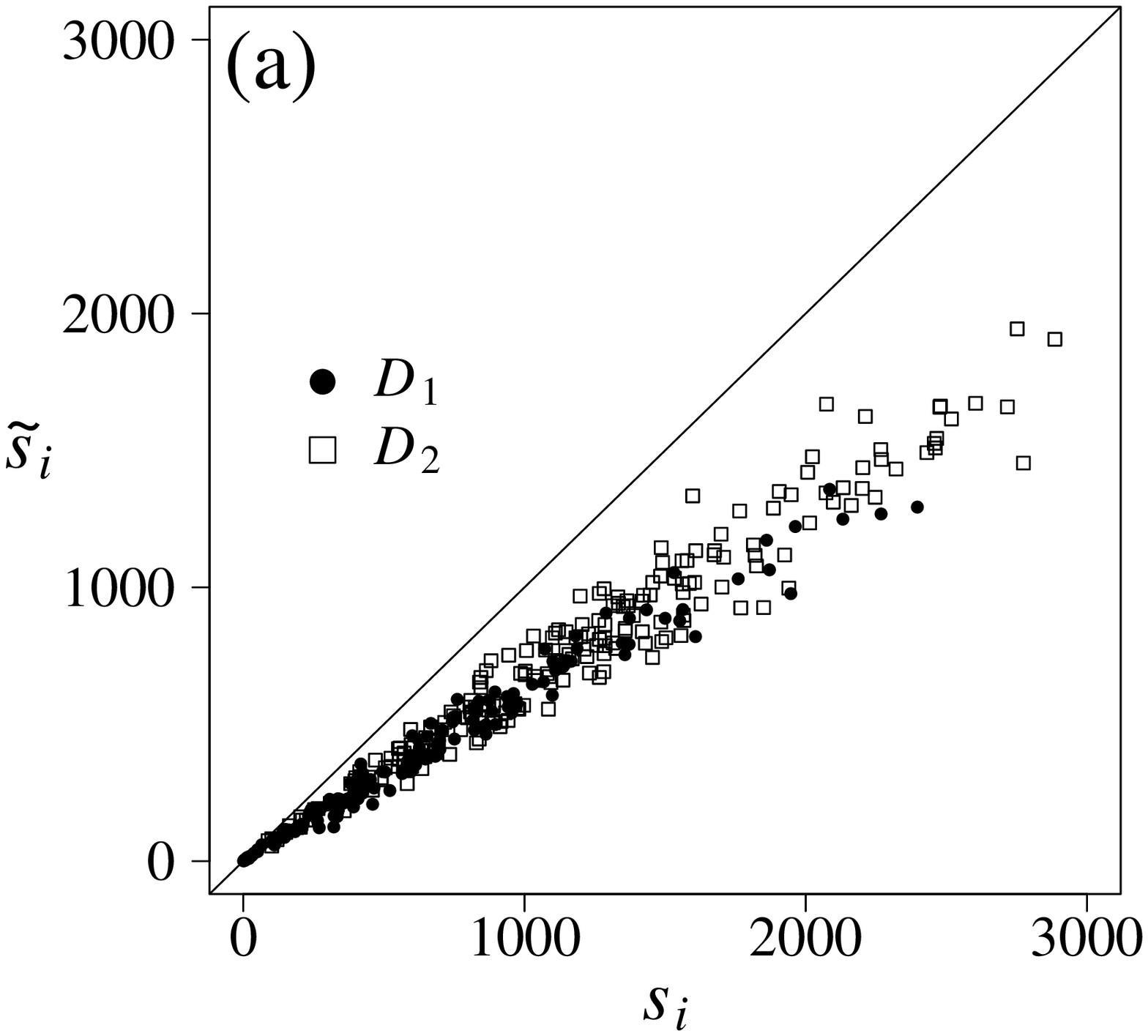}
\includegraphics[width=0.45\hsize]{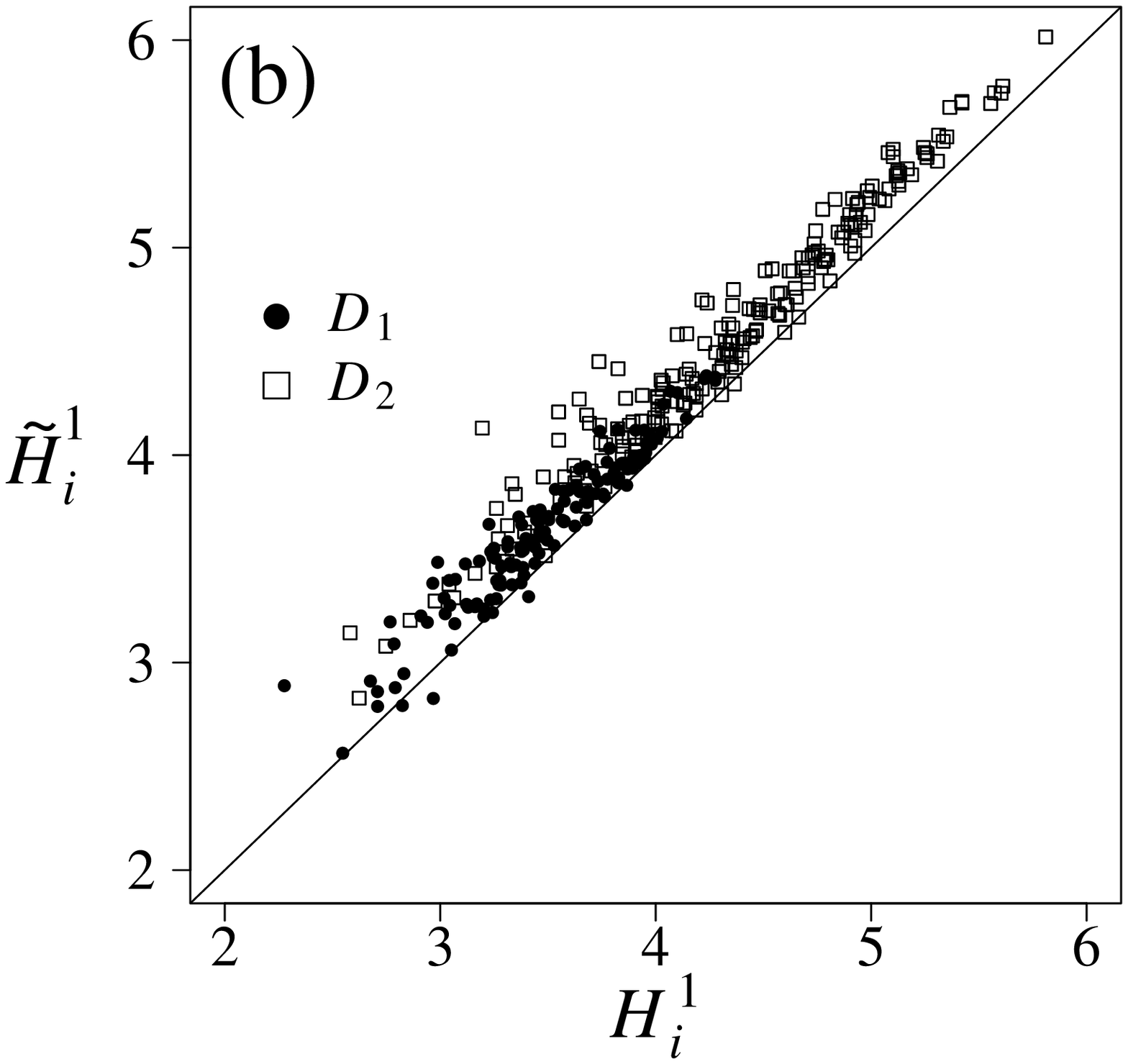}
\includegraphics[width=0.45\hsize]{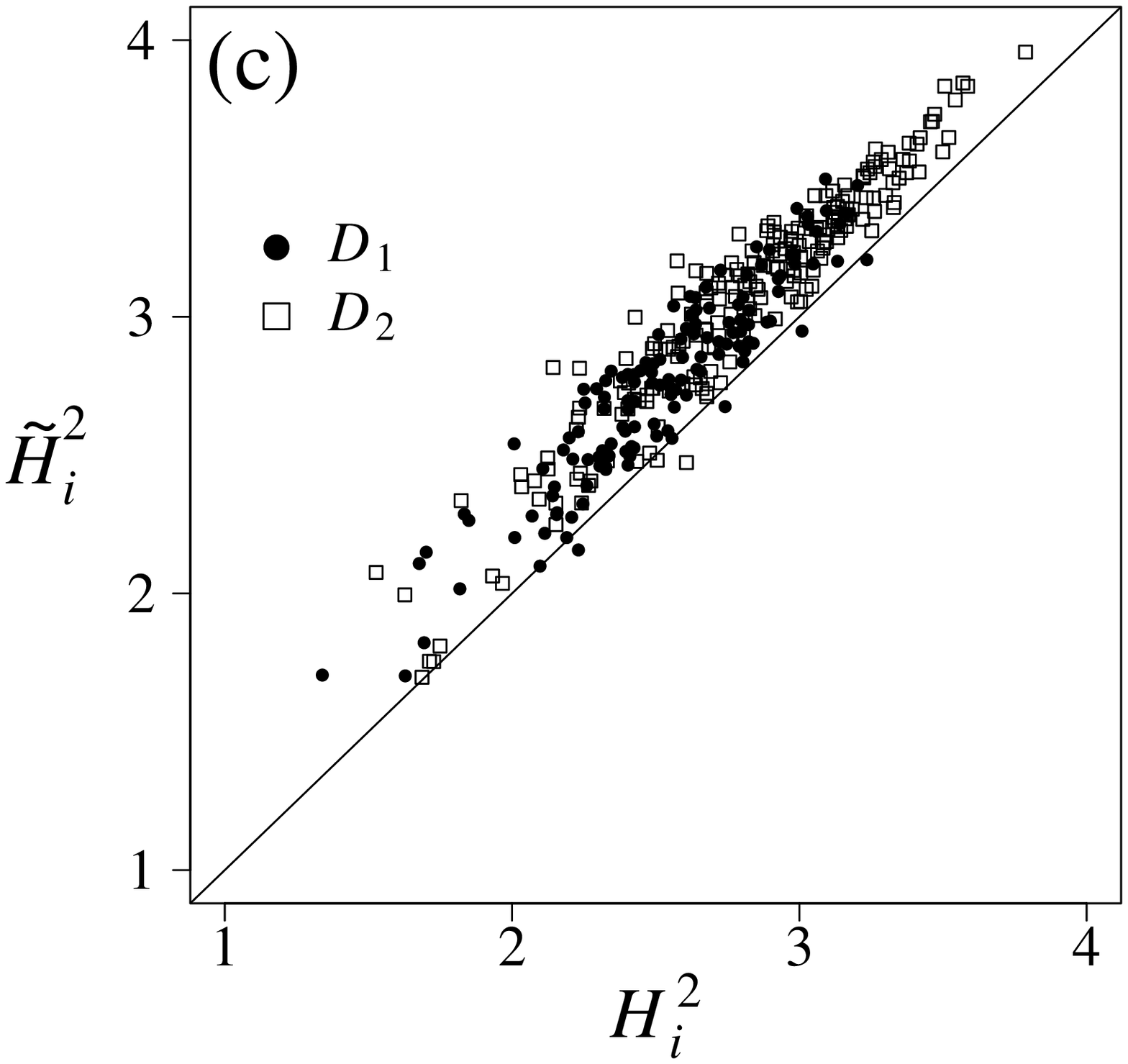}
\includegraphics[width=0.45\hsize]{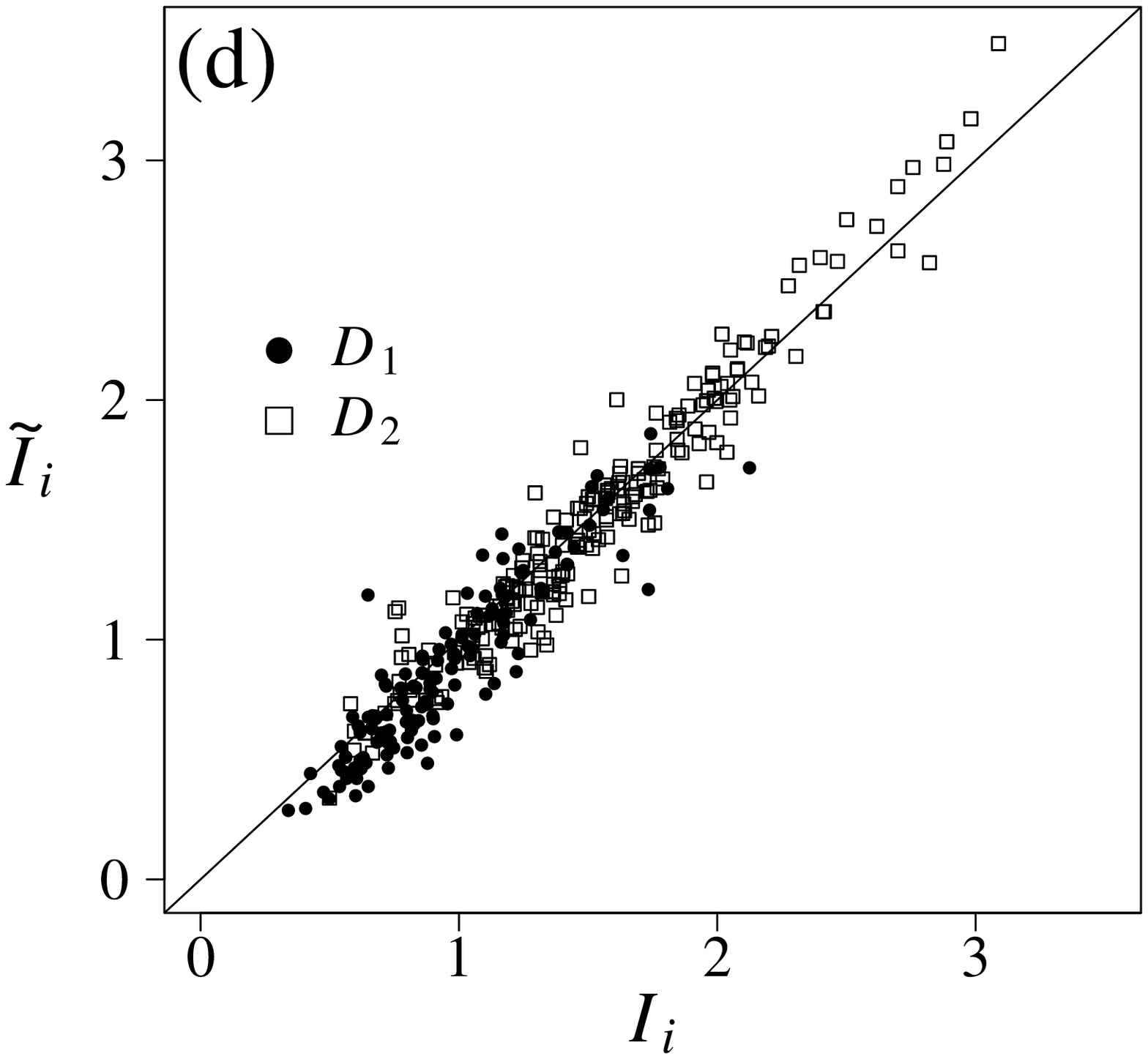}
\caption{Comparison between the interpolated and original data for data sets $D_1$ and $D_2$.
(a) Node strength, (b) uncorrelated entropy, (c) conditional entropy, and (d) mutual information 
for the interpolated data with $m=5$ are plotted against those without interpolation (\ie, original data sets). }
\label{fig:compare_m5}
\end{figure}

\begin{figure}
\centering
\includegraphics[width=0.45\hsize]{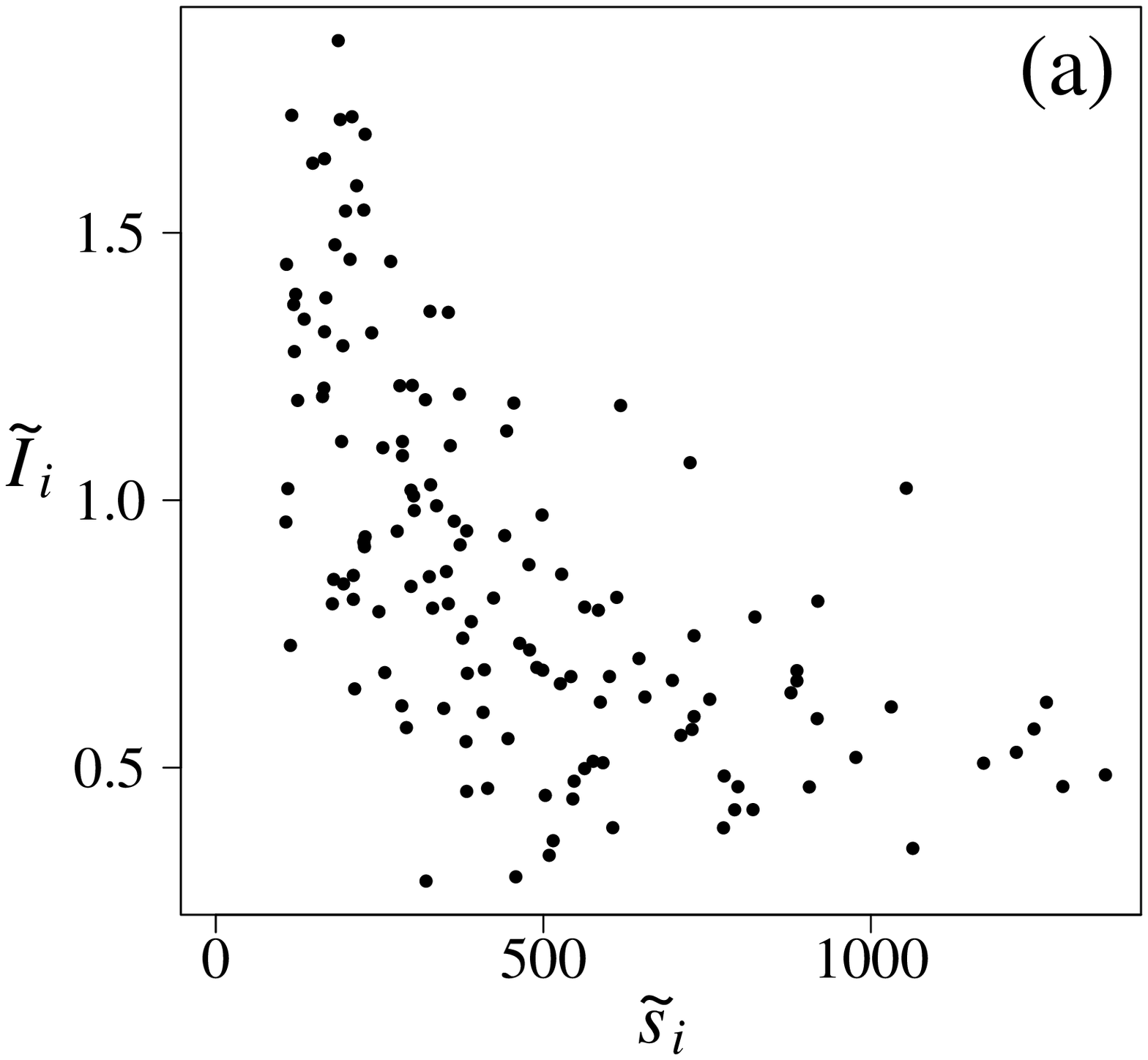}
\includegraphics[width=0.45\hsize]{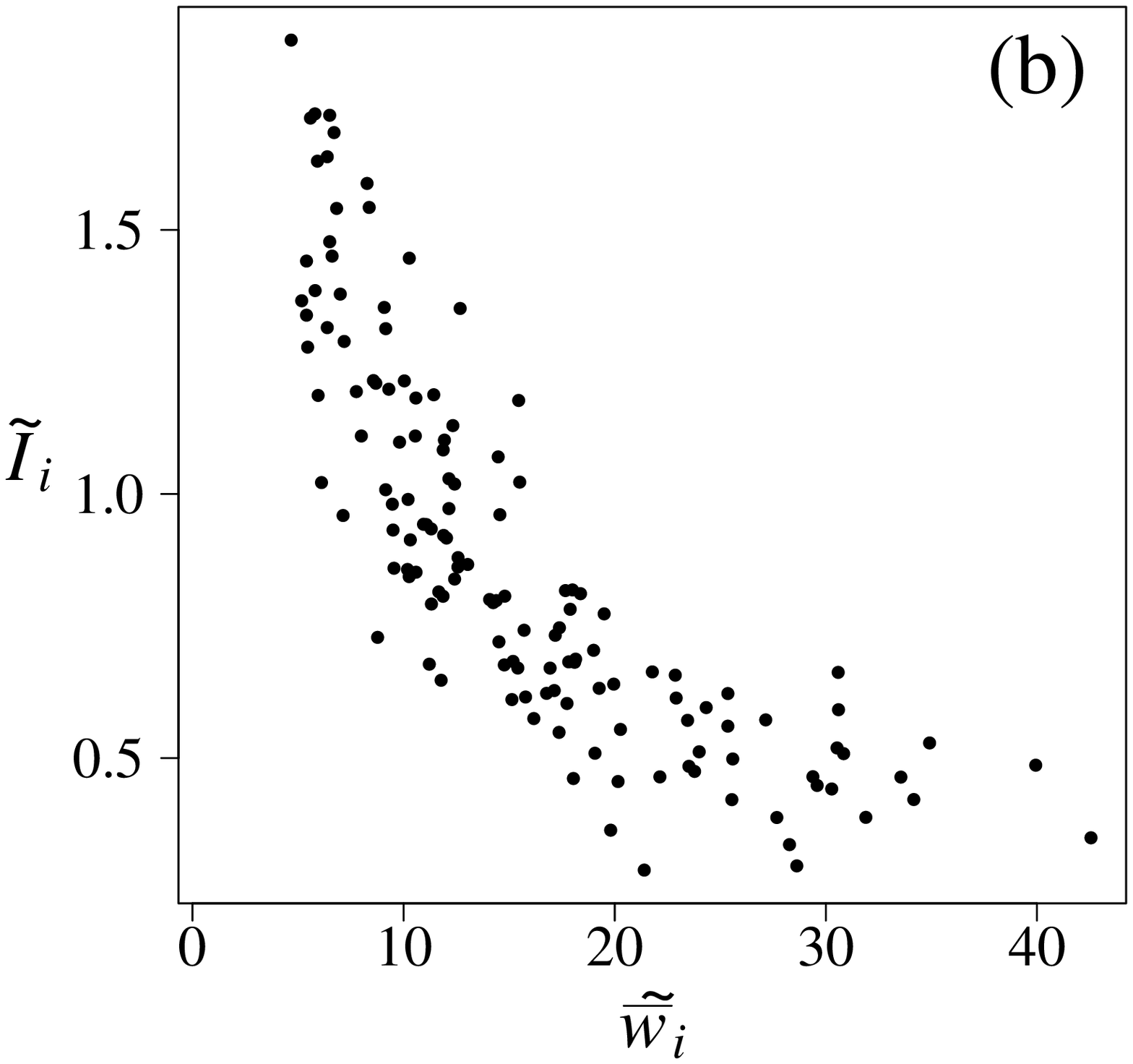}
\includegraphics[width=0.45\hsize]{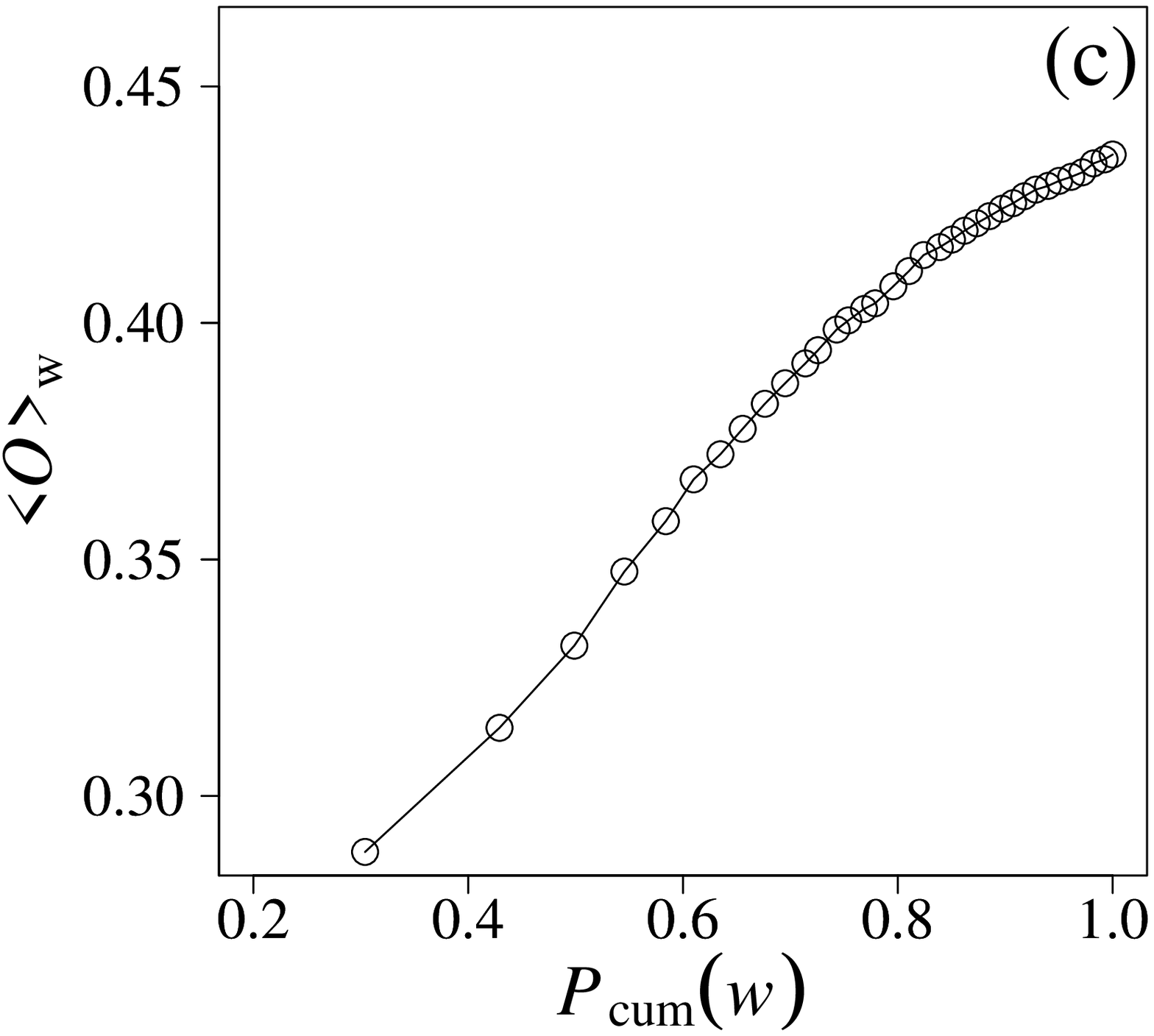}
\includegraphics[width=0.45\hsize]{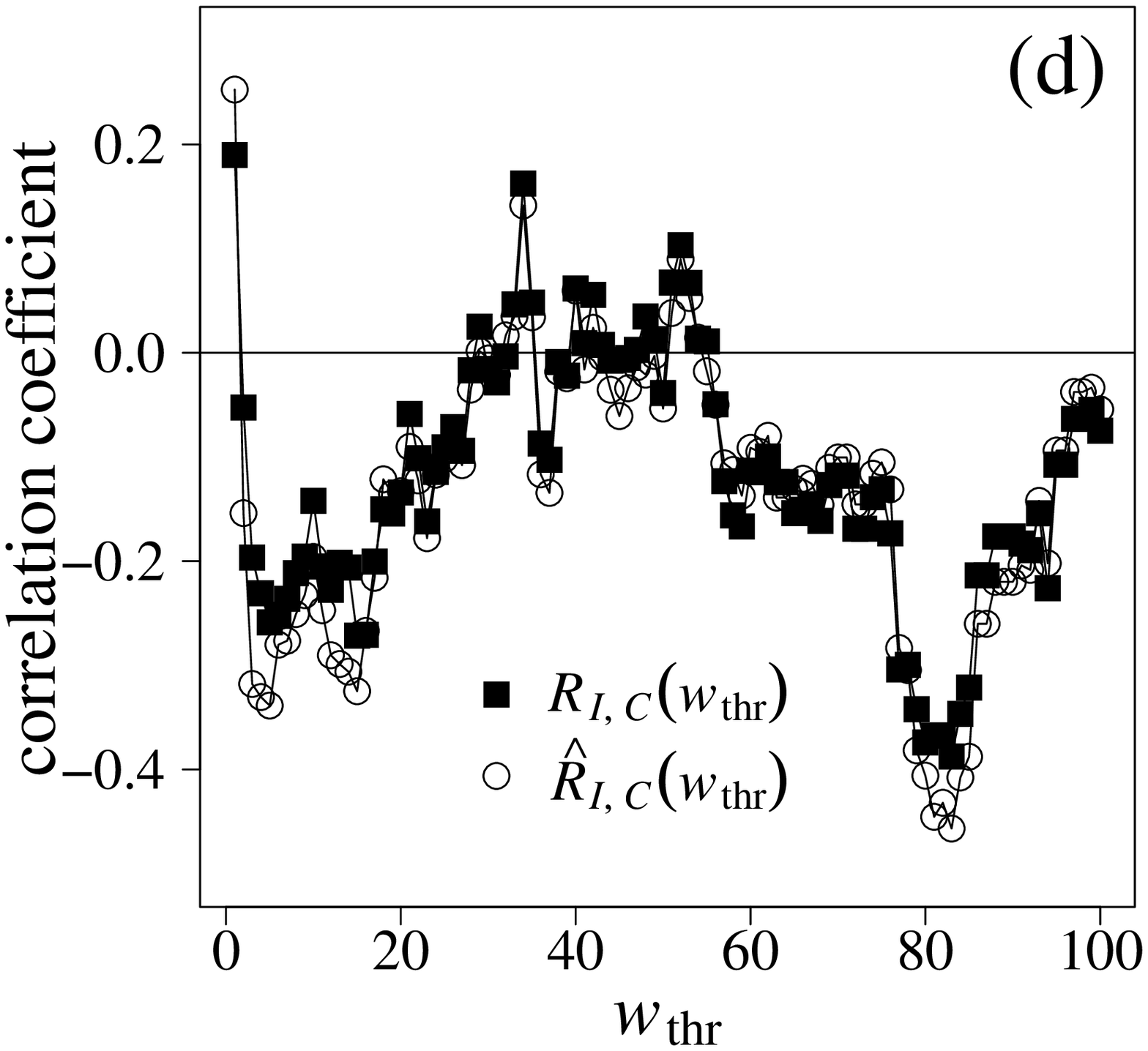}
\caption{
Results for the interpolated data with $m=5$.
We use data set~$D_1$.
See the caption of \FIG\ref{fig:confirm_m1} for legends.}
\label{fig:confirm_m5}
\end{figure}

\end{document}